\documentclass[a4paper,11pt]{article}
\usepackage[utf8]{inputenc}
\usepackage{fullpage}
\usepackage{amsmath,amssymb,amsfonts}
\usepackage{mathtools}
\usepackage{worldflags}
\usepackage[hyphenbreaks]{breakurl}

\usepackage[
colorlinks,
citecolor=blue,
pdftitle={Fredholm determinant for q-Painlevé III₃},
pdfauthor={Pavlo Gavrylenko}
]{hyperref}

\usepackage[
natbib=true,
sorting=nyt,
giveninits=true,
isbn=false,
backref=true,
style=alphabetic,
maxbibnames=99,
]{biblatex}

\DeclareFieldFormat[misc,online,proceedings,report]{title}{\mkbibquote{#1}}
\usepackage{tikz}
\usetikzlibrary{patterns}
\usetikzlibrary{shapes.geometric}

\usepackage{amsthm}
\newtheorem{theorem}{Theorem}[section]
\newtheorem{lemma}[theorem]{Lemma}
\newtheorem{definition}[theorem]{Definition}
\newtheorem{conjecture}[theorem]{Conjecture}
\newtheorem{remark}[theorem]{Remark}

\title{Riemann-Hilbert problems, Fredholm determinants, explicit combinatorial expansions, and connection formulas for the general $q$-Painlevé III$_3$ tau functions}
\author{Pavlo Gavrylenko\(^{\worldflag[width=3.3pt]{IT}\,\worldflag[width=3.3pt]{UA}}\)\thanks{\href{mailto:pasha.145@gmail.com}{pasha.145@gmail.com}}}
\date{\it\normalsize
\flushleft
\(^{\worldflag[width=3.3pt]{IT}}\)Scuola Internazionale Superiore di Studi Avanzati, Via Bonomea 265, Trieste 34136, Italy\\
\(^{\worldflag[width=3.3pt]{UA}}\)Bogolyubov Institute for Theoretical Physics, 14-b Metrolohichna str., Kyiv 03143, Ukraine\\
}

\newcommand{\tr}{\operatorname{tr}}
\newcommand{\diag}{\operatorname{diag}}
\newcommand{\Res}{\mathop{\operatorname{Res}}}
\numberwithin{equation}{section}

\begin{document}
\maketitle

\begin{abstract}
We reformulate the $q$-difference linear system corresponding to the $q$-Painlev\'e equation of type $A_7^{(1)'}$ as a Riemann-Hilbert problem on a circle. Then, we consider the Fredholm determinant built from the jump of this Riemann-Hilbert problem and prove that it satisfies bilinear relations equivalent to $P(A_7^{(1)'})$. We also find the minor expansion of this Fredholm determinant in explicit factorized form and prove that it coincides with the Fourier series in $q$-deformed conformal blocks, or partition functions of the pure $5d$ $\mathcal{N}=1$ $SU(2)$ gauge theory, including the cases with the Chern-Simons term. Finally, we solve the connection problem for these isomonodromic tau functions, finding in this way their global behavior.
\end{abstract}

\tableofcontents

\section{Introduction}

Painlev\'e equations were found originally as a result of a classification of equations of the form
\begin{equation}
y''(t)=F(y'(t), y(t), t)
\end{equation}
satisfying condition that all singularities with positions depending on initial conditions can only be poles.
This is called Painlev\'e condition or Painlev\'e property.
This condition is an analog of integrability in the non-autonomous setting.
Alternatively, Painlev\'e equations can be thought of as isomonodromic deformation problems that have 2-dimensional phase space.

When we switch to the d-difference or \(q\)-difference setting, Painlev\'e condition can be replaced by the so-called singularity confinement, stating that once discrete evolution leads to the situation where the solution becomes infinite, it should always return to the finite domain in a finite number of steps \cite{Grammaticos:1991zz}.
As in the differential case, it is possible to classify all second-order difference equations satisfying this property.
Hidetaka Sakai relates the singularity confinement property to the existence of the spaces of initial data for discrete equations and classifies such spaces in \cite{2001CMaPh.220..165S}\footnote{For a more detailed study of the relation between the singularity confinement and the spaces of initial data, see \cite{2002nlin......4070T,2019JPhA...52t5201M} and references therein.}.
As in the differential case, all \(q\)-Painlev\'e equations can be obtained as equations of \(q\)-isomonodromic deformations of some \(2\times 2\) \(q\)-difference systems \cite{2009JPhA...42k5201M,Kajiwara2017}.

The equation of our interest will be Painlev\'e~\(A_7^{(1)'}\), which is one of the two possible \(q\)-deformations of Painlev\'e~III\(_3\).
It was first obtained in \cite{Grammaticos:2021zhm,GTRCT:2002}\footnote{To be precise, the equation in the mentioned works is written for \(\mathsf{g}(t)^2\), not directly for \(\mathsf{g}(t)\). The precise form of the \(A_7^{(1)'}\) equation is written in \cite{SAKH:2007}, and in terms of the present work, it is a system of two equations on \(\mathsf{g}(t)^2\) and \(\mathsf{g}(t)\mathsf{g}(t/q)\). In this sense, both equations from \cite{Grammaticos:2021zhm} and from the present work are related to \cite{SAKH:2007} by some finite covers.} and has the form
\begin{equation}
\mathsf{g}(qt)\mathsf{g}(t/q)=\frac{\mathsf{g}(t)^2+t}{\mathsf{g}(t)^2+1}.
\end{equation}

General solutions of the differential and difference Painlev\'e equations are highly transcendental and were considered unintelligible for a long time.
However, the work \cite{Gamayun:2012ma} showed that the general solution (the tau function) of Painlev\'e~VI can be written as an explicit combinatorial series with factorized coefficients.
From a physical point of view, this expansion is a discrete Fourier series (also called Zak transform) in conformal blocks in \(c=1\) 2d CFT or in the instanton partition function in an \(\mathcal{N}=2\) \(4d\) gauge theory.
Such a formula is called the Gamayun-Iorgov-Lisovyy formula or Kyiv formula.

Generalizations of the Kyiv formula to other Painlev\'e equations that have an appropriate region of expansion were obtained in \cite{Gamayun:2013auu}.
Then it was also generalized to the \(P(A_7^{(1)'})\) case in \cite{Bershtein:2016aef}.

Currently, there exist four--five different types of proofs of the Kyiv formula.
One idea of a proof is to substitute the Fourier series into the equation and get some bilinear relations on conformal blocks/instanton partition functions, the so-called \(\mathbb{C}^2/\mathbb{Z}^2\) blow-up relations.
After that, one option is to prove these relations on the CFT side.
This was done for the differential case in \cite{Bershtein:2014yia}.
Another option is to derive them from the already known Nakajima-Yoshioka or \(\mathbb{C}^2\) blow-up relations.
This was done in the \(q\)-difference case in \cite{Shchechkin:2020ryb}.

A little bit similar, but different proof \cite{Jeong:2020uxz,Nekrasov:2020qcq} uses directly the Nakajima-Yoshioka blow-up relations with surface defect.
It establishes a relation between \(c=\infty\) conformal blocks with Gamayun-Iorgov-Lisovyy formula on one side, and with the classical action of the Painlev\'e equation on the other side, see also \cite{Litvinov:2013sxa}.
These blow-up relations and such kind of a proof can also be obtained in the CFT/representation theory framework \cite{Bershtein:2024kwe}.

Another idea of the proof is to study monodromies of degenerate fields viewed as operators acting on the space of conformal blocks and diagonalize them by Fourier transformation.
This was done in \cite{Iorgov:2014vla} for the \(2\times2\) isomonodromic problems.
A similar idea also works for the \(q\)-difference equations.
It was implemented in \cite{Jimbo:2017ael} for \(q\)-Painlev\'e~VI, which was later degenerated to lower Painlev\'e equations in \cite{2019SIGMA..15..074M}.

One by-product of the proof of \cite{Iorgov:2014vla} is the recovery of the free fermions from degenerate fields and identification of the CFT primary fields with the Jimbo-Miwa-Ueno monodromy fields.
One can take this observation and reverse the logic, namely, starting from the free fermions and monodromy fields prove that the latter are CFT primary fields and that they provide a solution to the isomonodromy problem \cite{Gavrylenko:2016moe}.
This free-fermionic approach gives a representation of the isomonodromic tau function as some explicit block Fredholm determinant.

In the last and the most mathematical and minimalistic proof of the Kyiv formula \cite{Gavrylenko:2016zlf}, we take a certain Fredholm determinant as the starting point and then prove that it is the isomonodromic tau function.
After that, we can compute the minor expansion of this determinant explicitly and identify it with the Kyiv formula.
Motivations for this Fredholm determinant can be different.
One of them is a free-fermionic computation from \cite{Gavrylenko:2016moe}.
Another one from \cite{Gavrylenko:2016zlf} involves some projection operators that first appeared in the study of \(\bar{\partial}\) and Dirac operators \cite{Palmer1990-kp,Palmer1993}.

Maybe the most simple and intuitive motivation for this Fredholm determinant comes from \cite{Cafasso:2017xgn}.
Namely, all simple enough isomonodromic problems can be effectively reformulated as the Riemann-Hilbert problems on a circle with jump \(J(z)\) given by solutions of elementary building block isomonodromic problems.
At the same time, there exists a known functional on the space of matrices on a circle, the so-called Widom determinant, defined by
\begin{equation}
\tau_W[J] = \det_{\mathcal{H}_+}\Pi_+J^{-1}\Pi_+J\Pi_+.
\end{equation}
It appeared initially in the study of matrix Toeplitz determinants and is a direct generalization of the Szeg\H{o} asymptotic formula \cite{WIDOM1974284,WIDOM19761}.
Surprisingly, it also coincides with the isomonodromic tau function if we take appropriate \(J(z)\).
We will use the same approach in the present paper.

Fredholm determinant representation of the isomonodromic tau function is important not only because it provides proof of the formula for solution.
As it was suggested by Oleg Lisovyy and later implemented in \cite{DelMonte:2022nem}, Fredholm determinant can also be used to express monodromy derivatives of the tau function in terms of data of the linear system giving the full closed 1-form \(d\log\tau\) on the space of times and monodromy parameters.
An alternative way to derive such formulas would be to guess this 1-form and then prove its closedness \cite{2010CMaPh.294..539B,2016arXiv160104790B}.
In this sense, the Fredholm determinant provides a more systematic approach, not based on any guesswork.

The latter approach is especially useful in the \(q\)-difference setting because we do not have a full understanding of what the analog of the formula \(\partial_t\log\tau = H\) is in this case.
In the differential setting, time derivatives of the tau function are expressed only in terms of the isomonodromic connection.
At the same time, monodromy derivatives also depend on the flat section of this connection (solution to the isomonodromic system).
This dependence also includes solutions to the auxiliary isomonodromic problems, see, e.g., \cite{2015arXiv150607485I,2010CMaPh.294..539B,DelMonte:2022nem}.
An important feature of the \(q\)-difference isomonodromic deformations is that times and monodromy parameters appearing as parameters of the equation are indistinguishable.
Therefore, it is natural to expect that the first difference derivative of the tau function with respect to time will depend on solutions of the original \(q\)-difference linear system and auxiliary linear system that describes one of the asymptotics.
We show that this is what actually happens, see Sections \ref{sec:detEvol}, \ref{sec:Yasymp}.

\

The present paper is devoted to the construction of the Fredholm determinant that describes the general solution of the \(q\)-Painlev\'e~\(A_7^{(1)'}\) equation.
We introduce such determinant as a Widom determinant of the appropriate Riemann-Hilbert problem on a circle.
Then, we compute the changes of this determinant under different transformations given by rational matrices and prove that it is indeed a \(q\)-Painlev\'e~\(A_7^{(1)'}\) tau function.
After this, we find the combinatorial expansion of the Fredholm determinant and express it in terms of Nekrasov functions.
This gives another rigorous proof of the Kyiv formula for the \(q\)-Painlev\'e~\(A_7^{(1)'}\) equation.
It is quite minimalistic in the sense that it contains only straightforward computations and does not rely on any additional tools, like representation theory, vertex operator algebras, or moduli spaces of instantons.

This paper mostly studies one of the two \(q\)-deformations of Painlev\'e~III\(_3\), the \(q\)-Painlev\'e equation of type \(A_7^{(1)'}\).
However, in Section \ref{sec:alternativeDet} we also provide a Fredholm determinant for \(q\)-Painlev\'e~\(A_7^{(1)}\).
This equation is also called \(q\)-Painlev\'e~I because it also has a limit to differential Painlev\'e~I\footnote{
The Fredholm determinant can be easily expanded in the regime that corresponds to the limit to Painlev\'e~III\(_3\).
Its limit to Painlev\'e~I is the open problem.
Determinants that will appear in this case can have the form similar to \cite{2021Nonli..34.6507D,2023arXiv231117051I}.
}.

\subsubsection*{Structure of the paper} The paper is organized as follows.
In Section \ref{sec:isomonodromy}, we introduce the \(q\)-difference linear system corresponding to the \(q\)-Painlev\'e~III\(_3\) equation.
We describe its solutions, their isomonodromic deformations, and local monodromy data.

In Section \ref{sec:parametrices}, we give a full description of the asymptotics of solution of the linear system around zero and infinity by approximating them with solutions of simpler auxiliary problems.
This provides us with an analog of the Stokes data in this case.
Moreover, these solutions of auxiliary problems describe the asymptotics of the full solution in different domains of the complex plane in terms of \(q\)-Bessel functions.

In Section \ref{sec:determinantVariation}, we formulate the Riemann-Hilbert problem for the \(q\)-Painlev\'e~III\(_3\) equation and define its tau function as the Widom determinant for this problem.
We also rewrite this Widom determinant as the block matrix Fredholm determinant with an integrable kernel.
After that, we compute the change of the tau function under the \(\mathbb{Z}_2\)-B\"acklund transformation and under the isomonodromic evolution.
In this way, we identify the ratio of the two tau functions with the \(q\)-Painlev\'e transcendent and prove that the tau functions themselves satisfy the bilinear form of the \(q\)-Painlev\'e equation of type \(A_7^{(1)'}\).

In Section \ref{sec:combinatorics}, we write the Fredholm determinant in the Laurent basis and compute its minor expansion explicitly.
Different terms in this expansion are naturally labeled by Maya diagrams.
Each term is given by an explicit factorized expression.
We also prove that these expressions can further be simplified and expressed in terms of Nekrasov functions.

In Section \ref{sec:alternativeDet}, we prove an equivalence between two slightly different combinatorial formulas for the tau function.
We also provide a Riemann-Hilbert formulation and the corresponding Fredholm determinant for a different \(q\)-deformation of Painlev\'e~III\(_3\), \(q\)-Painlev\'e~\(A_7^{(1)}\).

In Section \ref{sec:connection}, we find a relation between the local asymptotics and the tau functions.
We also solve the connection problem for the \(q\)-Painlev\'e III\(_3\) tau functions.
Namely, we find the mapping between the monodromy data and also find corresponding connection constant as some explicit combination of elliptic gamma functions.

In Section \ref{sec:discussion}, we present some open problems and directions of research.

\

All sections of this paper are organized in the following way.
We write a short summary of the results at the beginning of each section.
We also list all the main formulas that will be used later.
Then, in the subsections, we write the detailed technical computations.

\section*{Acknowledgements}
We would like to thank
Alexei Borodin,
Fabrizio Del Monte,
Harini Desiraju,
Alba Grassi,
and especially
Mikhail Bershtein and
Oleg Lisovyy
for the useful discussions and comments.

The research is partly supported by the INFN Iniziativa Specifica GAST, and by the MIUR PRIN Grant 2020KR4KN2 ``String Theory as a bridge between Gauge Theories and Quantum Gravity''.  
The author also acknowledges funding from the EU project Caligola (HORIZON-MSCA-2021-SE-01), Project ID: 101086123, and CA21109 - COST Action CaLISTA.

\section{Isomonodromic deformations and monodromy}
\label{sec:isomonodromy}
In this section, we introduce a linear system describing one of the two \(q\)-Painlev\'e III\(_3\) equations \eqref{eq:linear}, its B\"acklund transformation \eqref{eq:backlundg}, and the fundamental solution \eqref{eq:psidef} with prescribed monodromy.

\subsection{Linear system and isomonodromic deformations}
We realize the Painlev\'e~\(A_7^{(1)'}\) equation as a compatibility condition of the two equations:
\begin{equation}
\label{eq:linear}
Y(t,qz)=Y(t,z)L(t,z),
\end{equation}
\begin{equation}
\label{eq:Bmatrix}
Y(qt,z)=Y(t,z)B(t,z),
\end{equation}
where
\begin{multline}
\label{eq:Lax}
L(t,z)=
\begin{pmatrix}
1 & 0\\
\frac{z \mathsf{g}(t)}{\mathsf{g}(qt)} & 1-z
\end{pmatrix}
\begin{pmatrix}
1 & \frac{\mathsf{g}(qt)}{\mathsf{g}(t)}\\
0 & 1
\end{pmatrix}
\begin{pmatrix}
\frac{\mathsf{g}(t)}{\mathsf{g}(qt)} & 0\\
0 & \frac{\mathsf{g}(qt)}{\mathsf{g}(t)}
\end{pmatrix}
\begin{pmatrix}
1 & 0\\
\mathsf{g}(t)\mathsf{g}(qt) & 1
\end{pmatrix}
\begin{pmatrix}
1 & \frac{t}{z \mathsf{g}(qt)\mathsf{g}(t)}\\
0 & 1-\frac{t}{z}
\end{pmatrix}=\\=
\begin{pmatrix}
\frac{\mathsf{g}(qt)}{\mathsf{g}(t)}+\mathsf{g}(qt)\mathsf{g}(t) & 1+\frac{t}{z \mathsf{g}(t)^2}\\
z+\mathsf{g}(t)^2 & \frac{t}{\mathsf{g}(qt)\mathsf{g}(t)}+\frac{\mathsf{g}(t)}{\mathsf{g}(qt)}
\end{pmatrix}
\end{multline}
and
\begin{equation}
B(t,z)^{-1}=
\begin{pmatrix}
1 & 0\\
\mathsf{g}(qt)\mathsf{g}(t) & 1
\end{pmatrix}
\begin{pmatrix}
1 & \frac{qt}{z \mathsf{g}(qt)\mathsf{g}(t)}\\
0 & 1-\frac{qt}{z}
\end{pmatrix}=
\begin{pmatrix}
1 & \frac{qt}{z \mathsf{g}(qt)\mathsf{g}(t)}\\
\mathsf{g}(qt)\mathsf{g}(t) & 1
\end{pmatrix}.
\end{equation}
This compatibility condition reads
\begin{equation}
\label{eq:compatibility}
Y(qt,qz)=Y(t,z)B(t,z)L(qt,z)=Y(t,z)L(t,z)B(t,qz),
\end{equation}
or in other words,
\begin{equation}
L(qt,z)=B(t,z)^{-1}L(t,z)B(t,qz).
\end{equation}
The latter equation can be satisfied iff
\begin{equation}
\label{eq:Painleve}
\mathsf{g}(qt)\mathsf{g}(t/q)=\frac{\mathsf{g}(t)^2+t}{\mathsf{g}(t)^2+1},
\end{equation}
which is known as \(P(A_7^{(1)'})\) \cite{Grammaticos:2021zhm}.
Here and in most of the paper, we will consider the case \(|q|<1\).
Notice that the explicit form of \eqref{eq:compatibility} is
\begin{equation}
\label{eq:tzshift}
Y(qt,qz)=Y(t,z)
\begin{pmatrix}
\frac{\mathsf{g}(qt)}{\mathsf{g}(t)} & 1\\
z & \frac{\mathsf{g}(t)}{\mathsf{g}(qt)}
\end{pmatrix},
\end{equation}
which will also be used later.

\subsection{B\"acklund transformation}
Apart from the isomonodromic evolution, there is another important discrete transformation, the so-called B\"acklund transformation.
To construct it, consider the function
\begin{equation}
\label{eq:linBacklund}
\tilde{Y}(t,z)=C^bY(t,z)B^b(z),
\end{equation}
where
\begin{equation}
B^b(z)=
\begin{pmatrix}
0 & q^{1/4} z^{-1/2}\\
q^{-1/4} z^{1/2} & 0
\end{pmatrix}.
\end{equation}
It solves the equation
\begin{equation}
\tilde{Y}(t,qz)=\tilde{Y}(t,z)B^b(z)L(t,z)B^b(qz)=\tilde{Y}(t,z)\tilde{L}(t,z),
\end{equation}
where
\(\tilde{L}(t,z)\) is given by \eqref{eq:Lax}, but with \(\mathsf{g}(t)\) replaced by \(\tilde{\mathsf{g}}(t)\):
\begin{equation}
\label{eq:backlundg}
\tilde{\mathsf{g}}(t)=\frac{\sqrt{t}}{\mathsf{g}(t)}.
\end{equation}
We will later use tilde to denote the action of this involution on different objects.

\subsection{Monodromy}

\tikzset{
poleb/.pic = {\draw[fill, radius=0.03] circle;},
zerob/.pic = {\draw[radius=0.03] circle;},
poler/.pic = {\draw[fill,red] (-0.035,0)--(0,0.035)--(0.035,0)--(0,-0.035)--cycle;},
zeror/.pic = {\draw[red] (-0.035,0)--(0,0.035)--(0.035,0)--(0,-0.035)--cycle;},
}
\newcommand\zeror{{\protect\tikz[baseline=-.5ex]\protect\draw(0,0)pic{zeror};}}
\newcommand\poler{{\protect\tikz[baseline=-.5ex]\protect\draw(0,0)pic{poler};}}
\newcommand\zerob{{\protect\tikz[baseline=-.5ex]\protect\draw(0,0)pic{zerob};}}
\newcommand\poleb{{\protect\tikz[baseline=-.5ex]\protect\draw(0,0)pic{poleb};}}

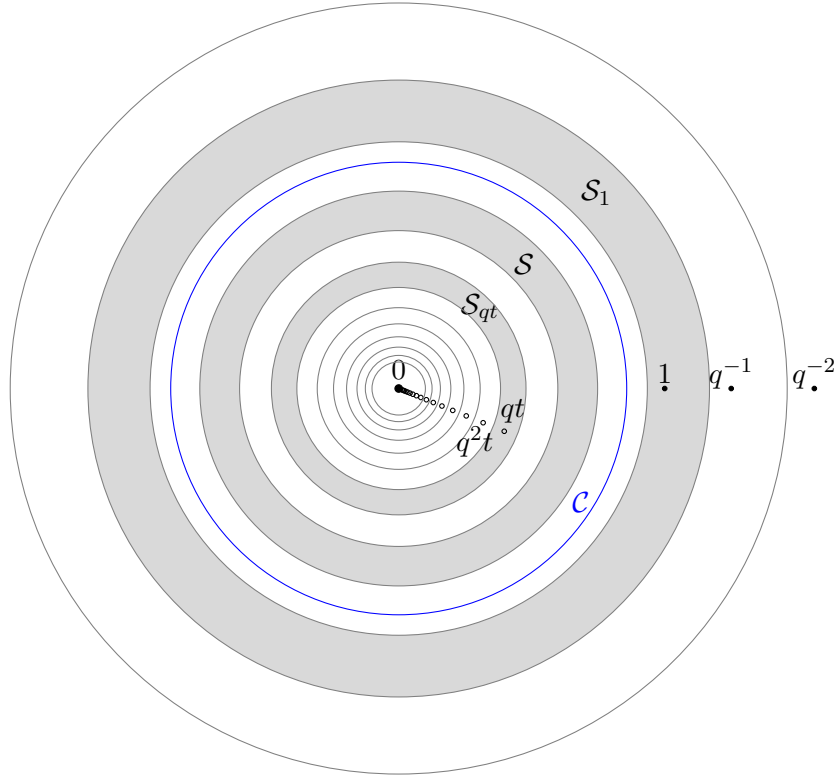
\begin{figure}[ht!]
\begin{center}
\begin{tikzpicture}[scale=1]
\draw[radius=3,start angle=0,end angle=-360,color=blue](3,0)arc;
\path[fill=gray!30!white] (2.61719,0) arc [radius=2.61719,start angle=0,end angle=360] -- (2.09375,0) arc[radius=2.09375,start angle=0,end angle=-360] -- cycle;
\node at (1.65,1.65) {\(\mathcal{S}\)};
\path[fill=gray!30!white] (1.34,0) arc [radius=1.34,start angle=0,end angle=360] -- (1.675,0) arc[radius=1.675,start angle=0,end angle=-360] -- cycle;
\node at (1.07,1.07) {\(\mathcal{S}_{qt}\)};
\path[fill=gray!30!white] (3.27148,0) arc [radius=3.27148,start angle=0,end angle=360] -- (4.08936,0) arc[radius=4.08936,start angle=0,end angle=-360] -- cycle;
\node at (2.6,2.6) {\(\mathcal{S}_1\)};
\node at (2.4,-1.5){\color{blue}\(\mathcal{C}\)};
\draw[fill, radius=0.05](0,0) circle;
\node at (0,0.25){\(0\)};
\foreach \x in {1.5, 1.2, 0.96, 0.768, 0.6144, 0.49152, 0.393216, 0.314573, 0.251658, 0.201327, 0.161061, 0.128849, 0.103079, 0.0824634, 0.0659707, 0.0527766} {\draw ({\x*sqrt(6/7)},{-\x*sqrt(1/7)}) pic{zerob};};
\foreach \x in {5.11169, 4.08936, 3.27148, 2.61719, 2.09375, 1.675, 1.34, 1.072, 0.8576, 0.68608, 0.548864, 0.439091, 0.351273} {\draw[radius=\x, thin,gray](0,0)circle;};
\node at (1.5,-0.3) {\(qt\)};
\node at (1,-0.7) {\(q^2t\)};
\foreach \x in {3.5, 4.375, 5.46875} {\draw (\x,0) pic{poleb};};
\node at (3.5,0.2) {\(1\)};
\node at (4.375,0.2) {\(q^{-1}\)};
\node at (5.46875,0.2) {\(q^{-2}\)};
\end{tikzpicture}
\end{center}
\caption{\label{fig:contours} Different domains of \(\mathbb{C}\mathrm{P}^1\).}
\end{figure}

To define monodromy data, we will use Krichever's approach \cite{Krichever:2004bb}\footnote{The original paper contains only the d-difference and elliptic equations, but the \(q\)-difference case can obviously be obtained by interpolation.
Such monodromies for the \(q\)-difference linear systems were already studied in \cite{Jimbo:2017ael,AFST2020629511190,Roffelsen2024,joshi2024segresurfacesgeometrypainleve,Joshi:2022kud}.}.
Namely, the basic object will be the solution \(Y(t,z)\) of the linear system \eqref{eq:linear}, analytic invertible, but not single-valued inside a circular strip \(\mathcal{S}_R^{\epsilon}\):
\begin{equation}
\label{eq:SRepsdef}
\mathcal{S}^{\epsilon}_R=\big\{\big.z\in \mathbb{C}\,\big|\,R|q|^{\epsilon}\le|z|\le R |q|^{1-\epsilon}\big\}, \qquad 0<\epsilon<1.
\end{equation}
We will omit \(\epsilon\) for simplicity, writing just \(\mathcal{S}_R\) and assuming that \(\epsilon\) can be any number between 0 and 1.

Multivaluedness of \(Y(t,z)\) can be encoded by a single monodromy matrix \(M\):
\begin{equation}
Y(t,e^{2\pi i}z)=MY(t,z).
\end{equation}
To establish some useful properties of \(M\), let us first study \(\det Y(t,z)\).
It satisfies the linear system
\begin{equation}
\label{eq:detsystem}
\det Y(t,qz) = \det L(t,z) \det Y(t,z),
\end{equation}
where
\begin{equation}
\det L(t,z)=(1-z)\left(1-\frac{t}{z}\right).
\end{equation}
The solutions of \eqref{eq:detsystem} that are analytic and invertible inside \(\mathcal{S}\) are given by
\begin{equation}
\label{eq:detsolution}
\det Y(t,z)=r^2\frac{\left(qt/z,q \right)_{\infty}}{(z;q)_{\infty}}, \qquad r\in \mathbb{C},
\end{equation}
where \((z;q)_n\) is the \(q\)-Pochhammer symbol
\begin{equation}
(z;q)_n=\prod_{i=0}^{n-1}\left(1-zq^i\right).
\end{equation}
In the case \(n=\infty\), we can rewrite this definition for \(|z|<1\):
\begin{equation}
(z;q)_{\infty}=\exp \left(-\sum_{n=1}^{\infty}\frac{z^n}{n(1-q^n)} \right).
\end{equation}
This definition does not work for \(|q|=1\), but remains valid for \(|q|>1\).
This case is related to \(|q|<1\) by
\begin{equation}
(z,1/q)_{\infty}=\frac1{(qz,q)_{\infty}}.
\end{equation}

We see that the solution \eqref{eq:detsolution} is actually single-valued in \(\mathcal{S}\), which means that
\begin{equation}
\det M=1.
\end{equation}
Therefore, \(M\) can be diagonalized by an appropriate choice of basis:
\begin{equation}
M=
e^{2\pi i \boldsymbol{\sigma}_{\mathbf{3}}\sigma},
\end{equation}
where
\begin{equation}
\boldsymbol{\sigma}_{\mathbf{3}}=
\begin{pmatrix}
1 & 0\\
0 & -1
\end{pmatrix}.
\end{equation}
Sometimes, we will also use the notation
\begin{equation}
q^{\sigma}=u.
\end{equation}

Thanks to the known monodromy properties, \(Y(t,z)\) can be written as
\begin{equation}
\label{eq:psidef}
Y(t,z)=z^{\boldsymbol{\sigma}_{\mathbf{3}}}\Psi(t,z),
\end{equation}
where \(\Psi(t,z)\) is single-valued in \(\mathcal{S}\).

We also see that the solution \(Y(t,z)\) becomes non-invertible at the points \(z=tq^n\), \(n>0\) and has singularities at \(z=q^{-n}\), \(n\ge 0\).
It is holomorphic and invertible away from these points.
Zero and infinity are essential singularities.

Such a behavior has a clear interpretation. Originally we started from \(Y(t,z)\) defined in \(\mathcal{S}_R\).
Then, we continued it to the entire plane using \eqref{eq:linear}.
The matrix \(L(t,z)\) degenerates at the points \(z=t\) and \(z=1\), therefore our recursive procedure necessarily produces a singularity in \(\mathcal{S}_1\) and a degeneration point in \(\mathcal{S}_{qt}\).
Further continuation only reproduces these singularities since \(t\) and \(1\) were the only singularities of \(L(t,z)\) in the finite domain.

So far, we have only one monodromy parameter, while the phase space is two-dimensional.
Another parameter is hidden in the relation between \(z\to 0\) and \(z\to \infty\) asymptotics of \(Y(t,z)\). 
Instead of finding these asymptotics explicitly, we introduce this parameter through the \emph{parametrices} in the next section.

\section{Parametrices}
\label{sec:parametrices}
In this section, we provide the description of the asymptotics and analytic behavior of the solution of the \(q\)-Painlev\'e III\(_3\) linear system \eqref{eq:psidef}.
The main objects are solutions of the two auxiliary linear systems \eqref{eq:Y0def}, \eqref{eq:Yinfdef}.
The main result of this section is that the ratios of these solutions with the solution of the full system \eqref{eq:phidef}, \(\Phi_+\) and \(\Phi_-\), are analytic invertible inside and outside the circle \(\mathcal{C}\), respectively, c.f. Figure~\ref{fig:contours}.
It also contains the equations \eqref{eq:fors1}, \eqref{eq:fors2} that relate the solution of the linear system to monodromy data encoded by the parametrices.
Another result is the extension of the original isomonodromic system \eqref{eq:Painleve} with the equation \eqref{eq:betaeq}.

\subsection{Auxiliary systems}
In the differential case, full monodromy data of \eqref{eq:linear} includes the asymptotics of \(Y(t,z)\) as \(z\to 0\) and \(z\to\infty\).
However, in our case, these monodromies are more complicated, and we prefer not to study them explicitly.
Instead, the global analytic behavior of \(Y(t,z)\) inside and outside the contour \(\mathcal{C}\) will be captured using the \(q\)-Bessel functions.
Namely, let us study the two auxiliary systems that approximate the behavior of \(L(t,z)\) \eqref{eq:Lax} around zero and infinity:
\begin{equation}
\label{eq:linearzero}
Y_0(t,qz)=Y_0(t,z)L_0(t,z),\qquad
L_0(t,z)=
\begin{pmatrix}
q^{\sigma} & \frac{\sqrt{t}}{z}\\
\sqrt{t} & q^{-\sigma}
\end{pmatrix},
\end{equation}
and 
\begin{equation}
\label{eq:linearinf}
Y_\infty(t,qz)=Y_\infty(t,z)L_\infty(t,z),\qquad
L_\infty(t,z)=
\begin{pmatrix}
q^{\sigma} & 1\\
z & q^{-\sigma}
\end{pmatrix}.
\end{equation}
Their solutions are given by
\begin{multline}
\label{eq:Y0def}
Y_0(t,z)=\\=
r_0
(z/(qt))^{\boldsymbol{\sigma}_{\mathbf{3}}\sigma}
(s_0)^{\boldsymbol{\sigma}_{\mathbf{3}}}
\begin{pmatrix}
j_2(q^{1-2\sigma},qt/z) & \frac{qt/z}{q^{\sigma}-q^{1-\sigma}}j_2(q^{2-2\sigma},qt/z)\\
\frac{-1}{q^{\sigma}-q^{-\sigma}}j_2(q^{1+2\sigma},qt/z) & j_2(q^{2\sigma},qt/z)
\end{pmatrix}t^{\frac1{4}\boldsymbol{\sigma}_{\mathbf{3}}}=\\=
r_0(s_0)^{\boldsymbol{\sigma}_{\mathbf{3}}}\mathcal{Y}_0(qt/z)t^{\frac1{4}\boldsymbol{\sigma}_{\mathbf{3}}}=
z^{\boldsymbol{\sigma}_{\mathbf{3}}}\Psi_-(t,z),
\end{multline}
\begin{multline}
\label{eq:Yinfdef}
Y_\infty(t,z)=
r_\infty z^{\boldsymbol{\sigma}_{\mathbf{3}}\sigma}
(s_\infty)^{\boldsymbol{\sigma}_{\mathbf{3}}}
\begin{pmatrix}
j_0(q^{2\sigma},z) & \frac{-1}{q^{\sigma}-q^{-\sigma}}j_0(q^{1+2\sigma},z)\\
\frac{-z}{q^{\sigma}-q^{1-\sigma}}j_0(q^{2-2\sigma},z) & j_0(q^{1-2\sigma},z)
\end{pmatrix}
q^{-\frac1{4}\boldsymbol{\sigma}_{\mathbf{3}}}=\\=
r_{\infty}(s_\infty)^{\boldsymbol{\sigma}_{\mathbf{3}}}
\mathcal{Y}_{\infty}(z)=
z^{\boldsymbol{\sigma}_{\mathbf{3}}\sigma}\Psi_+(t,z),
\end{multline}
where the \(q\)-Bessel functions are given by
\begin{equation}
j_k(u,z)=\sum_{n=0}^{\infty}\frac{q^{k n(n-1)/2}u^{kn/2}z^k}{(u;q)_n(q;q)_n}.
\end{equation}
The determinants of these solutions are equal to
\begin{equation}
\label{eq:detY0inf}
\det Y_0(t,z)=r_0^2\cdot(qt/z;q)_{\infty},\qquad
\det Y_{\infty}(t,z)=r_{\infty}^2\frac1{(z;q)_{\infty}},
\end{equation}
and their values at zero and infinity are
\begin{align}
\label{eq:psimval}
&\Psi_-(t,\infty)=
r_0\cdot
(qt)^{-\boldsymbol{\sigma}_{\mathbf{3}}\sigma}
(s_0)^{\boldsymbol{\sigma}_{\mathbf{3}}}
\begin{pmatrix}
1 & 0\\
\frac{-1}{q^{\sigma}-q^{-\sigma}} & 1
\end{pmatrix}t^{\frac1{4}\boldsymbol{\sigma}_{\mathbf{3}}},\\
\label{eq:psipval}
&\Psi_+(t,0)=
r_\infty\cdot(s_\infty)^{\boldsymbol{\sigma}_{\mathbf{3}}}
\begin{pmatrix}
1 & \frac{-1}{q^{\sigma}-q^{-\sigma}}\\
0 & 1
\end{pmatrix}q^{-\frac1{4}\boldsymbol{\sigma}_{\mathbf{3}}}.
\end{align}
Such description is called Mano decomposition after the work \cite{Mano2010}, where a similar analysis was carried out for the \(q\)-Painlev\'e VI equation.

The action of the B\"acklund transformation on the solutions of the auxiliary problems is described by
\begin{equation}
\label{eq:auxBacklund}
\tilde{Y}_{0,\infty}(s_{0,\infty},\sigma;t,z) = B^b(q) Y_{0,\infty}(s_{0,\infty},\sigma;t,z)B^b(z) = Y_{0,\infty}(s^{-1}_{0,\infty},1/2-\sigma;t,z),
\end{equation}
therefore the action of the B\"acklund transformation on the \(q\)-monodromy data\footnote{We will see later that \(s_0/s_{\infty}\) actually describes the \(q\)-monodromy data.
It corresponds to the Stokes data in the differential limit, but we will not distinguish such details and simply call them generalized monodromy data.}
is given by
\begin{equation}
\label{eq:backlundsigma}
\tilde{\sigma}=1/2-\sigma, \qquad \tilde{s}_0=1/s_0, \qquad \tilde{s}_{\infty}=1/s_{\infty}.
\end{equation}
The monodromies of \eqref{eq:Y0def}, \eqref{eq:Yinfdef} are equal to \(M\).
Consider now the two ratios,
\begin{equation}
\label{eq:phidef}
\Phi_+(t,z)=Y_0(t,z)^{-1}Y(t,z), \qquad
\Phi_-(t,z)=Y_\infty(t,z)^{-1}Y(t,z).
\end{equation}
These functions are single-valued in \(\mathcal{S}\) since their monodromies cancel.
Their global behavior, and consequently the global behavior of \(Y(t,z)\), is described by the following
\begin{theorem}
\label{thm:parametrices}
For generic \(q\)-monodromy data, the following holds:
\begin{enumerate}
\item\label{item:1} There exist \(s_0, s_{\infty}\in \mathbb{C}\), s.t. \(\Phi_+(t,z)\) is regular at \(z=qt\) and \(\Phi_-(t,z)\) is non-degenerate at \(z=1\).
If \(s_0\) and \(s_{\infty}\) are chosen in this way, \(\Phi_+\) is holomorphic invertible inside \(\mathcal{C}\) except at \(z=0\), while \(\Phi_-\) is holomorphic invertible outside \(\mathcal{C}\) except at \(z=\infty\).
\item\label{item:2} This choice of \(s_0\) and \(s_{\infty}\) also implies that \(\Phi_+\) is holomorphic invertible at \(0\), while \(\Phi_-\) is holomorphic invertible at \(\infty\).
\end{enumerate}
\end{theorem}
Such solutions possess an extra property, given by
\begin{lemma}
\label{thm:speriodicity}
The parameters \(s_{0,\infty}\) of parametrices are \(q\)-periodic functions of \(t\).
Namely, \(s_{0,\infty}(qt)^2 = s_{0,\infty}(t)^2\), and the signs can further be adjusted so that \(s_{0,\infty}(qt) = s_{0,\infty}(t)\).
\end{lemma}
In this way, we see that \(s_0\) and \(s_{\infty}\) together with \(\sigma\) describe the (generalized) monodromy data of the \(q\)-difference system.

These statements will be proved in the next two subsections.

\subsection{Regularity conditions for \(\Phi_{\pm}\)}
To study the analytic properties of \(\Phi_{\pm}(t,z)\) we use \eqref{eq:linear}, \eqref{eq:linearzero}, \eqref{eq:linearinf} and derive the difference equations they satisfy:
\begin{equation}
\label{eq:phiequations}
\Phi_+(t,qz)=L_0(t,z)^{-1}\Phi_+(t,z)L(t,z),\qquad
\Phi_-(t,z/q)=L_{\infty}(t,z/q)\Phi_-(t,z)L(t,z/q)^{-1}.
\end{equation}
The first singularity of \(\Phi_+(t,z)\) can appear at \(z=qt\), since
\(L_0(t,z)^{-1}\) has a pole at \(z=t\):
\begin{equation}
L_0(t,z)^{-1}=\frac1{1-\frac{t}{z}}
\begin{pmatrix}
q^{-\sigma} & \frac{-\sqrt{t}}{z}\\
-\sqrt{t} & q^{\sigma}
\end{pmatrix}
\approx
\frac1{1-\frac{t}{z}}
\begin{pmatrix}
q^{-\sigma}\sqrt[4]{t} \\
-\sqrt[4]{t}
\end{pmatrix}
\otimes
\begin{pmatrix}
\sqrt[4]{t} & -q^{\sigma}/\sqrt[4]{t}
\end{pmatrix}.
\end{equation}
If we suppose that \(\Phi_+(t,z)\) nevertheless remains regular at \(qt\), we get an equation
\begin{equation}
\label{eq:pole}
\Phi_+(t,z)=L_0(t,z/q)^{-1}\Phi_+(t,t)L(t,t)+O(z-qt)^0=O(z-qt)^0.
\end{equation}
In order to write it in a useful form, we also compute
\begin{equation}
\label{eq:Ltt}
L(t,t)=
\begin{pmatrix}
\frac{\mathsf{g}(t)^2+1}{\mathsf{g}(t)}\\
\frac{\mathsf{g}(t)^2+t}{\mathsf{g}(qt)}
\end{pmatrix}\otimes
\begin{pmatrix}
\mathsf{g}(qt) & \frac1{\mathsf{g}(t)}
\end{pmatrix}.
\end{equation}
To cancel the pole in \eqref{eq:pole}, we should have
\begin{equation}
\label{eq:fors}
\begin{pmatrix}
\sqrt[4]{t} & -q^{\sigma}/\sqrt[4]{t}
\end{pmatrix}
\Phi_+(t,t)
\begin{pmatrix}
\frac{\mathsf{g}(t)^2+1}{\mathsf{g}(t)} \\
\frac{\mathsf{g}(t)^2+t}{\mathsf{g}(qt)} 
\end{pmatrix}=0,
\end{equation}
or using the definition \eqref{eq:Y0def}
\begin{equation}
\label{eq:fors1}
\begin{pmatrix}
1 & -q^{\sigma}
\end{pmatrix}
\mathcal{Y}_0(q)^{-1}
\begin{pmatrix}
s_0^{-1}(t) & 0\\
0 & s_0(t)
\end{pmatrix}
Y(t,t)
\begin{pmatrix}
\frac{\mathsf{g}(t)^2+1}{\mathsf{g}(t)} \\
\frac{\mathsf{g}(t)^2+t}{\mathsf{g}(qt)} 
\end{pmatrix}=s_0(t)(\ldots)+s_0^{-1}(t)(\ldots)=0.
\end{equation}
This equation allows us to find \(s_0(t)^2\).
Naively, this quantity is \(t\)-dependent, but we will later show that it is actually \(q\)-periodic.

After fixing \(s_0(t)\) in such a way, \(\Phi_+(t,z)\) becomes a function holomorphic and invertible\footnote{Since \(\det \Phi_+(t,z)\) is holomorphic invertible at \(qt\), the absence of poles in the matrix elements implies the absence of non-trivial kernel.} at \(qt\), hence holomorphic and invertible in \(\mathcal{S}_{qt}\), the strip containing \(qt\).
We can then use the linear system \eqref{eq:linear} to continue this function to the whole interior of \(\mathcal{C}\) for the moment except \(0\).
It will be analytic and single-valued in this region.
The same computation can be done for the exterior of \(\mathcal{C}\), which proves the first part of Theorem~\ref{thm:parametrices}.

\subsubsection*{Extra properties}
It is also useful to look at the equation \eqref{eq:phiequations} at \(z=t\):
\begin{equation}
L_0(t,t)\Phi_+(t,qt)=\Phi_+(t,t)L(t,t),
\end{equation}
or explicitly
\begin{equation}
\begin{pmatrix}
q^{\sigma}/\sqrt[4]{t} \\
\sqrt[4]{t}
\end{pmatrix}\otimes
\begin{pmatrix}
\sqrt[4]{t} & q^{-\sigma}/\sqrt[4]{t}
\end{pmatrix}
\Phi_+(t,qt)
=
\Phi_+(t,t)
\begin{pmatrix}
\frac{\mathsf{g}(t)^2+1}{\mathsf{g}(t)}\\
\frac{\mathsf{g}(t)^2+t}{\mathsf{g}(qt)}
\end{pmatrix}\otimes
\begin{pmatrix}
\mathsf{g}(qt) & \frac1{\mathsf{g}(t)}
\end{pmatrix},
\end{equation}
which can be either understood ``as is'', or as a pair of equations:
\begin{equation}
\label{eq:PhiPlusOmegat}
\Phi_+(t,t)\begin{pmatrix}
\frac{\mathsf{g}(t)^2+1}{\mathsf{g}(t)}\\
\frac{\mathsf{g}(t)^2+t}{\mathsf{g}(qt)}
\end{pmatrix}
=
\alpha(t) \begin{pmatrix}
q^{\sigma} /\sqrt[4]{t}\\
\sqrt[4]{t}
\end{pmatrix},\quad
\begin{pmatrix}
\sqrt[4]{t} & q^{-\sigma}/\sqrt[4]{t}
\end{pmatrix}\Phi_+(t,qt)=
\alpha(t) \begin{pmatrix}
\mathsf{g}(qt) & \frac{1}{\mathsf{g}(t)}
\end{pmatrix}
\end{equation}
for some scalar \(\alpha(t)\).
Notice that \eqref{eq:fors} follows from this pair of equations.
Another consequence of this equation is the following identity:
\begin{equation}
\begin{pmatrix}
\mathsf{g}(qt) & \frac1{\mathsf{g}(t)}
\end{pmatrix}
\Phi_+(t,qt)^{-1}
\begin{pmatrix}
-q^{-\sigma} /\sqrt[4]{t}\\
\sqrt[4]{t}
\end{pmatrix}=0,
\end{equation}
or equivalently
\begin{equation}
\label{eq:xidef}
\Xi(t)=\frac{r}{r_0}\Phi_+(t,qt)^{-1}\begin{pmatrix}
-q^{-\sigma} /\sqrt[4]{t}\\
\sqrt[4]{t}
\end{pmatrix}=
\beta(t)\begin{pmatrix}
\frac{-1}{\mathsf{g}(qt)}\\ \mathsf{g}(t)
\end{pmatrix}.
\end{equation}

\subsubsection*{\(q\)-periodicity of \(s_{0,\infty}(t)\)}
Now we study the \(t\)-dependence of \(s_0(t)\) defined by the equation \eqref{eq:fors1}.
The shifted equation for the shifted \(s_0(qt)\) has the form
\begin{equation}
\begin{pmatrix}
1 & -q^{\sigma}
\end{pmatrix}
\mathcal{Y}_0(q)^{-1}
\begin{pmatrix}
s_0^{-1}(qt) & 0\\
0 & s_0(qt)
\end{pmatrix}
Y(qt,qt)
\begin{pmatrix}
\frac{\mathsf{g}(qt)^2+1}{\mathsf{g}(qt)} \\
\frac{\mathsf{g}(qt)^2+qt}{\mathsf{g}(q^2t)}
\end{pmatrix}=0.
\end{equation}
Now we use the \(q\)-Painlev\'e equation \eqref{eq:Painleve} and \eqref{eq:tzshift} for \(z=t\):
\begin{equation}
\begin{pmatrix}
1 & -q^{\sigma}
\end{pmatrix}
\mathcal{Y}_0(q)^{-1}
\begin{pmatrix}
s_0^{-1}(qt) & 0\\
0 & s_0(qt)
\end{pmatrix}
Y(t,t)
\begin{pmatrix}
\frac{\mathsf{g}(qt)}{\mathsf{g}(t)} & 1\\
t & \frac{\mathsf{g}(t)}{\mathsf{g}(qt)}
\end{pmatrix}
\begin{pmatrix}
\frac{\mathsf{g}(qt)^2+1}{\mathsf{g}(qt)} \\
\mathsf{g}(t)(\mathsf{g}(qt)^2+1)
\end{pmatrix}=0,
\end{equation}
so that finally
\begin{equation}
\begin{pmatrix}
1 & -q^{\sigma}
\end{pmatrix}
\mathcal{Y}_0(q)^{-1}
\begin{pmatrix}
s_0^{-1}(qt) & 0\\
0 & s_0(qt)
\end{pmatrix}
Y(t,t)
\begin{pmatrix}
\frac{\mathsf{g}(t)^2+1}{\mathsf{g}(t)} \\
\frac{\mathsf{g}(t)^2+t}{\mathsf{g}(qt)}
\end{pmatrix}=0.
\end{equation}
In this way we see that \(s_0(qt)\) solves the same equation \eqref{eq:fors1} as \(s_0(t)\), which means that \(s_0(qt)^2=s_0(t)^2\), and therefore \(s_0(t)\) is a quasi-constant, which is in perfect agreement with the differential case.
Since the only requirement for \(Y_0\) was the regularity condition \eqref{eq:pole}, it is still defined up to an overall constant for any \(t\), and therefore we can fix this constant to make \(s_0(t)\) \(q\)-periodic.
This proves Lemma~\ref{thm:speriodicity}.

\subsubsection*{Equations at \(z=1\)}
We can now repeat literally the same arguments for the other point, \(z=1\), and get another equation:
\begin{equation}
\Phi_-(t,1)L(t,1)=L_{\infty}(t,1)\Phi_-(t,q),
\end{equation}
and therefore
\begin{equation}
L_{\infty}(t,1)\Phi_-(t,q)=\Phi_-(t,1)L(t,1),
\end{equation}
or explicitly
\begin{equation}
\label{eq:intertwinerInf}
\begin{pmatrix}
q^{\sigma} \\
1
\end{pmatrix}\otimes
\begin{pmatrix}
1 & q^{-\sigma}
\end{pmatrix}\Phi_-(t,q)=
\Phi_-(t,1)
\begin{pmatrix}
\frac{\mathsf{g}(qt)}{\mathsf{g}(t)} \\
1
\end{pmatrix}\otimes
\begin{pmatrix}
\mathsf{g}(t)^2+1 & \frac{\mathsf{g}(t)^2+t}{\mathsf{g}(qt)\mathsf{g}(t)}
\end{pmatrix}.
\end{equation}
It can be shown in the same manner that \(s_{\infty}(t)\) is also invariant under the isomonodromic flow.
We can also write down an explicit equation for \(s_{\infty}(t)\) using \eqref{eq:Yinfdef}:
\begin{equation}
\label{eq:fors2}
\begin{pmatrix}
1 & q^{-\sigma}
\end{pmatrix}
\mathcal{Y}_{\infty}(q)^{-1}
\begin{pmatrix}
s_{\infty}(t)^{-1} & 0\\
0 & s_{\infty}(t)
\end{pmatrix}
Y(t,q)
\begin{pmatrix}
-\frac{\mathsf{g}(t)^2+t}{\mathsf{g}(qt)\mathsf{g}(t)} \\ \mathsf{g}(t)^2+1
\end{pmatrix}=0.
\end{equation}

\subsection{Behavior of \(\Phi_{\pm}\) around zero and infinity}
By now we know that \(\Phi_+\) and \(\Phi_-\) are single-valued, holomorphic and invertible inside and outside \(\mathcal{C}\), excluding \(0\) and \(\infty\), respectively.
To prove the last part of Theorem~\ref{thm:parametrices}, it remains to study their asymptotics at \(0\) and \(\infty\).

To do this, we introduce the following operators acting on the space of matrices
\begin{equation}
\mathcal{L}_+(t,z)[A]=L_0(t,z)^{-1}A L(t,z),\qquad
\mathcal{L}_-(t,z)[A]=L_{\infty}(t,z/q)A L(t,z/q)^{-1}.
\end{equation}
In this notation, \eqref{eq:phiequations} can be rewritten as
\begin{equation}
\Phi_\pm(t,q^{\pm1}z)=\mathcal{L}_\pm(t,z)[\Phi_\pm(t,z)].
\end{equation}
We also define the double shift:
\begin{equation}
\label{eq:doubleshift}
\mathcal{L}\!\mathcal{L}_{\pm}(t,z)=\mathcal{L}_{\pm}(t,q^{\pm1}z)\circ \mathcal{L}_{\pm}(t,z),\qquad
\Phi_\pm(t,q^{\pm2}z)=\mathcal{L}\!\mathcal{L}_\pm(t,z)[\Phi_\pm(t,z)].
\end{equation}
The series expansions of \(\mathcal{L}_{\pm}\) operators have the form
\begin{equation}
\label{eq:Lcaldef}
\mathcal{L}_{\pm}(t,z)=\sum_{n=-1}^{\infty} z^{\pm n} \mathcal{L}_{\pm,n},
\qquad
\mathcal{L}\!\mathcal{L}_{\pm}(t,z)=\sum_{n=0}^{\infty} z^{\pm n} \mathcal{L}\!\mathcal{L}_{\pm,n}.
\end{equation}
It is much more convenient to study the double shift since \(\mathcal{L}\!\mathcal{L}_{\pm}\) are regular at \(0\) and \(\infty\).
The spectra of the leading terms of these operators are
\begin{equation}
\mathcal{L}\!\mathcal{L}_{\pm,0}\sim \diag(1,1,q,q^{-1}).
\end{equation}
These eigenvalues define the asymptotics of \(\Phi_+\) and \(\Phi_-\) as \(z\to 0\) and \(z\to \infty\), respectively:
\begin{equation}
\Phi_{\pm}(t,z)=\phi_{\pm,0}+\phi_{\pm,1/2}z^{1/2}+\phi_{\pm,-1/2} z^{-1/2}+\phi_{\pm,1}z^{\pm 1}+\ldots.
\end{equation}
Since \(\Phi_{\pm}\) are single-valued, the terms \(\phi_{\pm,1/2}\), \(\phi_{\pm,-1/2}\) vanish, and therefore we see that \(\Phi_+(t,z)\) and \(\Phi_-(t,z)\) have limits as \(z\to 0\) and \(z\to \infty\), respectively.
This completes the proof of the last part of Theorem~\ref{thm:parametrices}.

\subsubsection*{Limiting values}
We can also find the limits of \(\Phi_+\) and \(\Phi_-\) at \(0\) and at \(\infty\), respectively.
They are defined by the equations
\begin{equation}
\mathcal{L}\!\mathcal{L}_{\pm,0}[\phi_{\pm,0}]=\phi_{\pm,0},\qquad
\mathcal{L}_{\pm,-1}[\phi_{\pm,0}]=0,\qquad
\mathcal{L}_{\pm,0}[\phi_{\pm,0}]-\phi_{\pm,0}=-\mathcal{L}_{\pm,-1}[\phi_{\pm,1}].
\end{equation}
The solutions of these equations are given by
\begin{align}
\label{eq:phipval}
&\Phi_+(t,0)=\phi_{+,0}=
\frac{r}{r_0}\begin{pmatrix}
1 & c(t)/\sqrt{t}\\
0 & 1
\end{pmatrix} t^{-\frac1{4}\boldsymbol{\sigma}_{\mathbf{3}}}\mathsf{g}(t)^{\boldsymbol{\sigma}_{\mathbf{3}}},\\
\label{eq:phimval}
&\Phi_-(t,\infty)=\phi_{-,0}=
\frac{r}{r_{\infty}}\begin{pmatrix}
1 & 0\\
-c(t) & 1
\end{pmatrix},
\end{align}
where
\begin{equation}
c(t)=\frac{qt+\mathsf{g}(qt)^2+q\mathsf{g}(t)^2+\mathsf{g}(qt)^2\mathsf{g}(t)^2}{(q-1)\mathsf{g}(qt)\mathsf{g}(t)}-
\frac{q^{1/2-\sigma}+q^{\sigma-1/2}}{q^{1/2} - q^{-1/2}}.
\end{equation}
We also fix the overall factors to make them consistent with \eqref{eq:detsolution} and \eqref{eq:detY0inf}.

\subsection{Isomonodromic evolution and B\"acklund transformations of \(\Phi_{\pm}\)}
It is also useful to study the behavior of \(\Phi_{\pm}\) under certain discrete transformations.
It follows from \eqref{eq:Bmatrix}, \eqref{eq:linearzero}, \eqref{eq:linearinf}, \eqref{eq:Y0def}, \eqref{eq:Yinfdef} that
\begin{equation}
\Phi_-(qt,z)=\Phi_-(t,z)B(t,z),
\end{equation}
\begin{equation}
\Phi_+(qt,z)=q^{-\frac1{4}\boldsymbol{\sigma}_{\mathbf{3}}}L_0(t,z/q)\Phi_+(t,z)B(t,z),
\end{equation}
\begin{equation}
\tilde{\Phi}_{\pm}(t,z)=B^b(t,z)^{-1}\Phi_{\pm}(t,z)B^b(t,z).
\end{equation}
We can also compute
\begin{multline}
\Phi_+(qt,qz) = q^{-\frac1{4}\boldsymbol{\sigma}_{\mathbf{3}}}L_0(t,z)\Phi_+(t,qz)B(t,qz)
= q^{-\frac1{4}\boldsymbol{\sigma}_{\mathbf{3}}}\Phi_+(t,z)L(t,z)B(t,qz)
= \\ = q^{-\frac1{4}\boldsymbol{\sigma}_{\mathbf{3}}}\Phi_+(t,z)
\begin{pmatrix}
\mathsf{g}(qt)/\mathsf{g}(t) & 1\\
z & \mathsf{g}(t)/\mathsf{g}(qt)
\end{pmatrix}.
\end{multline}
This equation also makes sense for \(z=qt\):
\begin{equation}
\Phi_+(qt,q^2t) = q^{-\frac1{4}\boldsymbol{\sigma}_{\mathbf{3}}}\Phi_+(t,qt)
\begin{pmatrix}
\mathsf{g}(qt)/\mathsf{g}(t) & 1\\
qt & \mathsf{g}(t)/\mathsf{g}(qt)
\end{pmatrix}.
\end{equation}
Let us write analogous equation for \(\Xi(t)\) defined in \eqref{eq:xidef}:
\begin{multline}
\Xi(qt)=
\frac{r}{r_0}\Phi_+(qt,q^2t)^{-1}\begin{pmatrix}
-q^{-\sigma}/\sqrt[4]{qt}\\
\sqrt[4]{qt}
\end{pmatrix}=\\=
\frac1{1-qt}\frac{r}{r_0} 
\begin{pmatrix}
\mathsf{g}(t)/\mathsf{g}(qt) & -1\\
-qt & \mathsf{g}(qt)/\mathsf{g}(t)
\end{pmatrix}
\Phi_+(t,qt)^{-1}\begin{pmatrix}
-q^{-\sigma}/\sqrt[4]{t}\\
\sqrt[4]{t}
\end{pmatrix}=\\=
\frac{\beta(t)}{1-qt} 
\begin{pmatrix}
\mathsf{g}(t)/\mathsf{g}(qt) & -1\\
-qt & \mathsf{g}(qt)/\mathsf{g}(t)
\end{pmatrix}
\begin{pmatrix}
\frac{-1}{\mathsf{g}(qt)}\\ \mathsf{g}(t)
\end{pmatrix}=\\=
\frac{\beta(t)}{1-qt}
\begin{pmatrix}
-\frac{\mathsf{g}(t)(1+\mathsf{g}(qt)^2)}{\mathsf{g}(qt)^2}\\
\frac{qt+\mathsf{g}(qt)^2}{\mathsf{g}(qt)}
\end{pmatrix}=
\frac{\beta(t)}{1-qt}
\begin{pmatrix}
-\frac{qt+\mathsf{g}(qt)^2}{\mathsf{g}(q^2t)\mathsf{g}(qt)^2}\\
\frac{qt+\mathsf{g}(qt)^2}{\mathsf{g}(qt)}
\end{pmatrix}=\\=
\frac{\beta(t)}{1-qt}\frac{qt+\mathsf{g}(qt)^2}{\mathsf{g}(qt)^2}
\begin{pmatrix}
\frac{-1}{\mathsf{g}(q^2t)} \\
\mathsf{g}(qt)
\end{pmatrix}=
\beta(qt)
\begin{pmatrix}
\frac{-1}{\mathsf{g}(q^2t)}\\ \mathsf{g}(qt)
\end{pmatrix}.
\end{multline}
The function \(\beta(t)\) therefore satisfies the linear equation
\begin{equation}
\label{eq:betaeq}
\beta(qt)=\frac{1+\mathsf{g}(qt)^{-2}qt}{1-qt}\beta(t),
\end{equation}
and its solution cannot be expressed in terms of the Painlev\'e transcendent, so \(\beta(t)\) is an additional coordinate in the \(q\)-isomonodromic system.

\subsection{Algebraic solution}
We can check that the following expression solves \eqref{eq:linear}:
\begin{equation}
Y^{alg.}(t,z)=T^{alg.}
\begin{pmatrix}
\frac{\left(-\sqrt{qt/z};\sqrt{q}\right)_{\infty}}{\left(-\sqrt{z};\sqrt{q}\right)_{\infty}} & 
\frac{\left(-\sqrt{qt/z};\sqrt{q}\right)_{\infty}}{\left(-\sqrt{z};\sqrt{q}\right)_{\infty}}\\
-\frac{\left(\sqrt{qt/z};\sqrt{q}\right)_{\infty}}{\left(\sqrt{z};\sqrt{q}\right)_{\infty}} & 
\frac{\left(\sqrt{qt/z};\sqrt{q}\right)_{\infty}}{\left(\sqrt{z};\sqrt{q}\right)_{\infty}}
\end{pmatrix}
\left(z/\sqrt{q}\right)^{1/4},
\end{equation}
where
\begin{equation}
T^{alg.}=\frac12 q^{-\frac18 \boldsymbol{\sigma}_{\mathbf{3}}}\begin{pmatrix} 1 & -1\\ 1 & 1\end{pmatrix}.
\end{equation}
This solution corresponds to
\begin{equation}
\label{eq:gAlgebraic}
\mathsf{g}^{alg.}(t)=\pm t^{1/4}.
\end{equation}

Auxiliary systems \eqref{eq:linearzero}, \eqref{eq:linearinf} have analogous solutions:
\begin{equation}
\label{eq:Y0exact}
Y^{alg.}_0(t,z)=T^{alg.}
\begin{pmatrix}
\left(-\sqrt{qt/z};\sqrt{q}\right)_{\infty} & 
\left(-\sqrt{qt/z};\sqrt{q}\right)_{\infty}\\
-\left(\sqrt{qt/z};\sqrt{q}\right)_{\infty} & 
\left(\sqrt{qt/z};\sqrt{q}\right)_{\infty}
\end{pmatrix}
\left(z/\sqrt{q}\right)^{1/4}
\end{equation}
and
\begin{equation}
\label{eq:Yinfexact}
Y_{\infty}^{alg.}(t,z)=T^{alg.}
\begin{pmatrix}
\frac1{\left(-\sqrt{z};\sqrt{q}\right)_{\infty}} & 
\frac1{\left(-\sqrt{z};\sqrt{q}\right)_{\infty}}\\
\frac{-1}{\left(\sqrt{z};\sqrt{q}\right)_{\infty}} & 
\frac1{\left(\sqrt{z};\sqrt{q}\right)_{\infty}}
\end{pmatrix}
\left(z/\sqrt{q}\right)^{1/4}.
\end{equation}
These solutions have the same monodromy matrices
\begin{equation}
M=
\begin{pmatrix}
i & 0\\
0 & -i
\end{pmatrix},
\end{equation}
therefore they correspond to \(\sigma=\frac1{4}\).
By comparing these solutions with \eqref{eq:Y0def} and \eqref{eq:Yinfdef} we find
\begin{equation}
s_0^{alg.}=s_{\infty}^{alg.}=\pm1.
\end{equation}
The determinants of these solutions are
\begin{equation}
\det Y^{alg.}(t,z)=\frac{(qt/z;q)_{\infty}}{(z;q)_{\infty}},\quad
\det Y_0^{alg.}(t,z)=(qt/z;q)_{\infty},\quad
\det Y_{\infty}^{alg.}(t,z)=\frac1{(z;q)_{\infty}}.
\end{equation}

\section{Determinants and their transformations}
\label{sec:determinantVariation}
Now we would like to switch from the original \(q\)-isomonodromic system \eqref{eq:linear} to a Riemann-Hilbert description.
Define the following jump matrix
\begin{equation}
\label{eq:Jfactorizations}
J(z)=\Phi_+(t,z)\Phi_-(t,z)^{-1}=\Psi_-(t,z)^{-1}\Psi_+(t,z)=Y_0(t,z)^{-1}Y_{\infty}(t,z).
\end{equation}
Equivalence of its different representations follows from \eqref{eq:Y0def}, \eqref{eq:Yinfdef}, \eqref{eq:phidef}.
Such jump matrix defines the Riemann-Hilbert problem
\begin{equation}
\Phi_+(t,z)=J(z)\Phi_-(t,z),
\end{equation}
which is equivalent to the original system \eqref{eq:linear} once we know the solutions of the auxiliary systems.

We suggest the following definition, inspired by a similar construction in the differential case~\cite{Cafasso:2017xgn}:
\begin{definition}
The \(q\)-isomonodromic tau function of \eqref{eq:linear} is defined as the Widom determinant
\begin{equation}
\label{eq:widom}
\tau(t)=\tau_W[J]=\det_{\mathcal{H}_+}\Pi_+J^{-1}\Pi_+J\Pi_+,
\end{equation}
where \(\Pi_+\) is the projector onto
\begin{equation}
\mathcal{H}_+=\mathbb{C}[z]\otimes \mathbb{C}^2,
\end{equation}
the space of non-negative Laurent modes, while \(\Pi_-\) is the projector onto
\begin{equation}
\mathcal{H}_-=z^{-1}\mathbb{C}[z^{-1}]\otimes \mathbb{C}^2,
\end{equation}
the space of the negative Laurent modes.
\end{definition}

The projectors \(\Pi_{\pm}\) satisfy the obvious relations
\begin{equation}
\Pi_{\pm}^2=\Pi_{\pm}, \qquad \Pi_+\Pi_-=\Pi_-\Pi_+=0.
\end{equation}

In this section, we study the changes of \eqref{eq:widom} under different transformations and prove that (i) the ratio of this tau function and its B\"acklund transformation gives the Painlev\'e transcendent \eqref{eq:gformula}, and (ii) it satisfies the bilinear equation \eqref{eq:bilin1}, which can also be rewritten as \eqref{eq:bilin2}, \eqref{eq:bilin3}.
This proves that the tau function \eqref{eq:widom} is actually the tau function of \(q\)-Painlev\'e III\(_3\).

\subsection{Inversion formulas}
To work with the Fredholm determinants, we first check the following relations
\begin{equation}
\Pi_+ \Psi_-^{-1}\Psi_+ \cdot \Psi_+^{-1} \Pi_+ \Psi_- = \Pi_+ \Psi_-^{-1} (\mathbb{I}-\Pi_-)\Psi_-=\Pi_+,
\end{equation}
\begin{equation}
\Pi_+ \Phi_-\Phi_+^{-1} \cdot \Phi_+ \Pi_+ \Phi_-^{-1}=\Pi_+\Phi_-(\mathbb{I}-\Pi_-)\Phi_-^{-1}=\Pi_+,
\end{equation}
which means that
\begin{equation}
\label{eq:Jinverse}
\left(\Pi_+ J\right)^{-1}= \Psi_+^{-1}\Pi_+\Psi_-,\qquad
\left( \Pi_+J^{-1}\right)^{-1}=\Phi_+\Pi_+ \Phi_-^{-1}.
\end{equation}
In this way, we are able to invert the operator in \eqref{eq:widom}.
This knowledge will be used in subsequent computations.



\subsection{B\"acklund transformation of the determinant}
We first compute the change of the determinant \eqref{eq:widom} under the B\"acklund transformation described by \eqref{eq:linBacklund} and \eqref{eq:auxBacklund}:
\begin{multline}
\tilde{J}(t,z)=B^b(t,z)^{-1}J(t,z)B^b(t,z) = \\=
\begin{pmatrix}
0 & -i q^{1/4}z^{-1/2}\\
-i q^{-1/4}z^{1/2} & 
\end{pmatrix}
J(t,z)
\begin{pmatrix}
0 & i q^{1/4}z^{-1/2}\\
i q^{-1/4}z^{1/2} & 
\end{pmatrix}=\\=
\begin{pmatrix}
0 & -i q^{1/4}\\
-i q^{-1/4} & 
\end{pmatrix}
\Lambda_2^{-1}
J(t,z)
\Lambda_2
\begin{pmatrix}
0 & i q^{1/4}\\
i q^{-1/4} & 
\end{pmatrix},
\end{multline}
where
\begin{equation}
\Lambda_2=
\begin{pmatrix}
1 & 0\\
0 & z
\end{pmatrix}.
\end{equation}
Therefore, the transformed determinant has the form
\begin{multline}
\label{eq:ToeplitzBacklund1}
\tilde{\tau}=\det_{\mathcal{H}_+}\Pi_+ \left(B^b\right)^{-1}J^{-1} \left(B^b\right)\Pi_+ \left(B^b\right)^{-1}J \left(B^b\right)\Pi_+=
\det_{\mathcal{H}_+}\Pi_+\Lambda_2^{-1}J^{-1}\Lambda_2\Pi_+\Lambda_2^{-1}J\Lambda_2\Pi_+.
\end{multline}
We can easily check that
\begin{equation}
\label{eq:projectorsConj}
\Lambda_2 \Pi_+ \Lambda_2^{-1}=\Pi_+ - v_0\otimes \bar{v}_0,
\end{equation}
where 
\begin{equation}
v_0=
\begin{pmatrix}
0 \\
1 
\end{pmatrix},\quad
\bar{v}_0\cdot 
\begin{pmatrix}
f_1 \\
f_2 
\end{pmatrix}=\oint f_2(z)\frac{dz}{2\pi iz} = (\Pi_+ f_2)(0) \eqcolon \operatorname{ev}_0 \Pi_+ f_2.
\end{equation}
The operator \(\operatorname{ev}_0\) thus computes the value of its argument at zero.
We also introduce the following identity
\begin{equation}
\det_{\mathcal{H}_+}(X) = \det_{\mathcal{H}_+}\left(v_0\otimes \bar{v}_0 + \Lambda_2\Pi_+ X\Pi_+ \Lambda_2^{-1}\right),
\end{equation}
stating simply that we can realize the action of \(A\) on the subspace of \(\mathcal{H}_+\) spanned by \(z^n \otimes e_1\) and \(z^{n+1}e_2\) for \(n\ge 0\) and complete it with the unit action on \(z^0\otimes e_2\).

This identity allows us to rewrite the determinant \eqref{eq:ToeplitzBacklund1} as follows:
\begin{equation}
\tilde{\tau}=\det_{\mathcal{H}_+}\left(v_0\otimes \bar{v}_0+
\Lambda_2\Pi_+ \Lambda_2^{-1}J^{-1}\Lambda_2\Pi_+ \Lambda_2^{-1}J \Lambda_2\Pi_+\Lambda_2^{-1}\right) = \det_{\mathcal{H}_+} \mathcal{X}_1.
\end{equation}
Now, we transform the operator \(\mathcal{X}_1\) using \eqref{eq:projectorsConj}:
\begin{multline}
\mathcal{X}_1=v_0\otimes \bar{v}_0 + \left( \Pi_+-v_0\otimes \bar{v}_0  \right) J^{-1} \left( \Pi_+-v_0\otimes \bar{v}_0  \right) J \left( \Pi_+-v_0\otimes \bar{v}_0  \right)=\\=
\Pi_+J^{-1}\Pi_+J\Pi_++v_0\otimes \bar{v}_0 \cdot\left( 1+ \bar{v}_0 J^{-1} \Pi_+ J v_0 - \bar{v}_0J^{-1}v_0 \cdot \bar{v}_0J v_0\right)-\\-
\Pi_+J^{-1}\Pi_+Jv_0\otimes \bar{v}_0-v_0\otimes \bar{v}_0 J^{-1}\Pi_+J\Pi_+-
\Pi_+J^{-1}v_0\otimes \bar{v}_0 J\Pi_++\\+
v_0\otimes \bar{v}_0J\Pi_+ \cdot \bar{v}_0J^{-1}v_0+
\Pi_+J^{-1}v_0\otimes \bar{v}_0 \cdot \bar{v}_0J v_0.
\end{multline}
We see that it has the form of a matrix of the full rank plus a combination of rank-1 operators.
This can also be written as follows:
\begin{multline}
\mathcal{X}_1=\Pi_+J^{-1}\Pi_+J\Pi_++\\+
\begin{pmatrix}
v_0 \\
\Pi_+J^{-1}v_0 \\
\Pi_+J^{-1}\Pi_+ Jv_0
\end{pmatrix}^T
\begin{pmatrix}
1+ \bar{v}_0 J^{-1} \Pi_+ J v_0 - \bar{v}_0J^{-1}v_0 \cdot \bar{v}_0J v_0 & \bar{v}_0J^{-1}v_0 & -1\\
\bar{v}_0J v_0 & -1 & 0\\
-1 & 0 & 0
\end{pmatrix}
\begin{pmatrix}
\bar{v}_0 \\
\bar{v}_0 J\Pi_+ \\
\bar{v}_0 J^{-1}\Pi_+ J\Pi_+
\end{pmatrix}.
\end{multline}
Now, using the relation
\begin{equation}
\label{eq:abcdDet}
\det(A+BCD)=\det A \cdot \det C \cdot \det(C^{-1}+DA^{-1}B),
\end{equation}
which holds for the rectangular matrices, we can write the tau functions ratio as
\begin{equation}
\frac{\tilde{\tau}}{\tau}=\det \mathcal{X}_2,
\end{equation}
where \(\mathcal{X}_2\) is a \(3\times 3\) matrix given explicitly by
\begin{multline}
\mathcal{X}_2=\begin{pmatrix}
1+ \bar{v}_0 J^{-1} \Pi_+ J v_0 - \bar{v}_0J^{-1}v_0 \cdot \bar{v}_0J v_0 & \bar{v}_0J^{-1}v_0 & -1\\
\bar{v}_0 \cdot \bar{v}_0J v_0 & -1 & 0\\
-1 & 0 & 0
\end{pmatrix}^{-1}
+\\+
\begin{pmatrix}
\bar{v}_0 \left(\Pi_+J^{-1}\Pi_+J\Pi_+ \right)^{-1}v_0 & \bar{v}_0 \left(\Pi_+J\Pi_+ \right)^{-1}v_0 & 1 \\
\bar{v}_0 \left(\Pi_+J^{-1}\Pi_+\right)^{-1}v_0 & 1 & \bar{v}_0 J \Pi_+ v_0\\
1 & \bar{v}_0 \Pi_+ J^{-1} v_0 & \bar{v}_0 \left(\Pi_+J^{-1}\Pi_+J\Pi_+ \right)v_0
\end{pmatrix}.
\end{multline}
Explicit computation of the finite determinant gives us
\begin{equation}
\mathcal{X}_2=
\begin{pmatrix}
\bar{v}_0 \left(\Pi_+J^{-1}\Pi_+J\Pi_+ \right)^{-1}v_0 & \bar{v}_0 \left(\Pi_+J\Pi_+ \right)^{-1}v_0 & 0 \\
\bar{v}_0 \left(\Pi_+J^{-1}\Pi_+\right)^{-1}v_0 & 0 & 0\\
0 & 0 & -1
\end{pmatrix},
\end{equation}
and therefore
\begin{equation}
\frac{\tilde{\tau}}{\tau}=\det \mathcal{X}_2= \bar{v}_0 \left(\Pi_+J\Pi_+ \right)^{-1} \cdot \bar{v}_0 \left(\Pi_+J^{-1}\Pi_+\right)^{-1}v_0.
\end{equation}
Now, using \eqref{eq:Jinverse}, we can rewrite the last relation as
\begin{equation}
\label{eq:tauRatio1}
\frac{\tilde{\tau}}{\tau}= \bar{v}_0\Psi_+^{-1}\Pi_+\Psi_- v_0\cdot
\bar{v}_0\Phi_+\Pi_+ \Phi_-^{-1} v_0.
\end{equation}
Recalling the analytic properties of \(\Phi_{\pm}\) and \(\Psi_{\pm}\), namely, that the functions with the ``\(+\)'' index are analytic in a neighborhood of \(0\), and the functions with the ``\(-\)'' index analytic in a neighborhood of infinity,  we can rewrite \eqref{eq:tauRatio1} as
\begin{equation}
\frac{\tilde{\tau}}{\tau}=\bar{v}_0\Psi_+(t,0)^{-1} \Psi_-(t,\infty)v_0
\cdot \bar{v}_0 \Phi_+(t,0) \Phi_-(t,\infty)^{-1}v_0.
\end{equation}
Using \eqref{eq:psipval}, we get
\begin{multline}
\bar{v}_0\Psi_+(t,0)^{-1} \Psi_-(t,\infty)v_0=\\=
\frac{r_0}{r_{\infty}}
\begin{pmatrix}
0 & 1
\end{pmatrix}
q^{\frac1{4}\boldsymbol{\sigma}_{\mathbf{3}}}
\begin{pmatrix}
1 & \frac{1}{q^{\sigma}-q^{-\sigma}}\\
0 & 1
\end{pmatrix}(s_\infty)^{-\boldsymbol{\sigma}_{\mathbf{3}}}
(qt)^{-\boldsymbol{\sigma}_{\mathbf{3}}\sigma}
(s_0)^{\boldsymbol{\sigma}_{\mathbf{3}}}
\begin{pmatrix}
1 & 0\\
\frac{-1}{q^{\sigma}-q^{-\sigma}} & 1
\end{pmatrix}t^{\frac1{4}\boldsymbol{\sigma}_{\mathbf{3}}}
\begin{pmatrix}
0 \\
1 
\end{pmatrix}=\\=
\frac{r_0}{r_{\infty}}\frac{s_{\infty}}{s_0} (qt)^{\sigma}/\sqrt[4]{qt}.
\end{multline}
The second factor can be found with the help of \eqref{eq:phipval} and \eqref{eq:phimval}:
\begin{multline}
\bar{v}_0 \Phi_+(t,0) \Phi_-(t,\infty)^{-1}v_0=\\=
\frac{r_{\infty}}{r_0}\begin{pmatrix}
0 & 1
\end{pmatrix}
\begin{pmatrix}
1 & c(t)/\sqrt{t}\\
0 & 1
\end{pmatrix} t^{-\frac1{4}\boldsymbol{\sigma}_{\mathbf{3}}}\mathsf{g}(t)^{\boldsymbol{\sigma}_{\mathbf{3}}}
\begin{pmatrix}
1 & 0\\
c(t) & 1
\end{pmatrix}
\begin{pmatrix}
0 \\
1 
\end{pmatrix}
=
\frac{r_{\infty}}{r_0}\mathsf{g}(t)^{-1}\sqrt[4]{t}.
\end{multline}
Combining the previous two formulas, we get\footnote{Notice that this formula is consistent with the B\"acklund transformation \eqref{eq:backlundg}, \eqref{eq:backlundsigma}.} \(\frac{\tilde{\tau}}{\tau}=(qt)^{\sigma}q^{-1/4}\frac{s_{\infty}}{s_0}\mathsf{g}(t)^{-1}\).
In this way, we have proved the following
\begin{theorem}
The \(q\)-Painlev\'e \(A_7^{(1)'}\) transcendent solving \eqref{eq:Painleve} can be expressed as a ratio of two Widom determinants \eqref{eq:widom}
\begin{equation}
\label{eq:gformula}
\mathsf{g}(t)=\frac{s_{\infty}}{s_0}t^{\sigma}q^{\sigma-1/4}\frac{\tau}{\tilde{\tau}},
\end{equation}
where the B\"acklund transformed tau function \(\tilde{\tau}\) is obtained from \(\tau\) the change of parameters \eqref{eq:backlundsigma}.
\end{theorem}
In principle, this is already sufficient to write down the solution of the \(q\)-Painlev\'e III\(_3\) equation.
It is however also interesting to find bilinear relations satisfied by \(\tau(t)\).
This task is accomplished in the next subsection.

\subsection{Isomonodromic evolution of the determinant}
\label{sec:detEvol}
We start from a transformation of the jump matrix that follows from \eqref{eq:Bmatrix}, \eqref{eq:linearzero}, \eqref{eq:Y0def}:
\begin{equation}
J(qt,z)=q^{-\frac1{4}\boldsymbol{\sigma}_{\mathbf{3}}}L_0(t,z/q)J(t,z),\qquad
J(t/q,z)=L_0(t/q,z/q)^{-1}q^{\frac1{4}\boldsymbol{\sigma}_{\mathbf{3}}}J(t,z).
\end{equation}
It implies the following transformation of the tau function \eqref{eq:widom}:
\begin{equation}
\tau(qt)=\det_{\mathcal{H}_+}\Pi_+J^{-1} L_0(t,z/q)^{-1}\Pi_+L_0(t,z/q) J \Pi_+
\end{equation}
The main ingredient of this formula is the conjugated projection operator
\begin{equation}
L_0(t,z/q)^{-1}\Pi_+L_0(t,z/q)=
\frac{1}{1-qt/z}\begin{pmatrix}
q^{-\sigma} & -q\sqrt{t}/z\\
-\sqrt{t} & q^{\sigma}
\end{pmatrix}
\Pi_+
\begin{pmatrix}
q^{\sigma} & q\sqrt{t}/z\\
\sqrt{t} & q^{-\sigma}
\end{pmatrix}
\end{equation}
To find it explicitly, let us act with this operator on a test function \(f\):
\begin{equation}
L_0(t,z/q)^{-1}\Pi_+L_0(t,z/q)f = \Pi_+ f -
L_0(t,z/q)^{-1}
\begin{pmatrix}
0 & q\sqrt{t}/z\\
0 & 0
\end{pmatrix} (\Pi_+f)(0),
\end{equation}
or equivalently
\begin{equation}
L_0(t,z/q)^{-1}\Pi_+L_0(t,z/q) = \Pi_+ + \frac{qt/z}{1-qt/z}
\begin{pmatrix}
-q^{-\sigma}/\sqrt{t} \\ 1
\end{pmatrix}
\otimes
\begin{pmatrix}
0 & 1
\end{pmatrix}
\operatorname{ev}_0\Pi_+ =
\Pi_+ + u_+\otimes \bar{u}_+.
\end{equation}
Therefore, it is the original projector plus an explicit rank-1 operator.
Next, let us compute the ratio of the tau functions:
\begin{multline}
\frac{\tau(qt)}{\tau(t)} = \det_{\mathcal{H}_+}
\left(\mathbb{I}+ \left(\Pi_+J^{-1}\Pi_+J\Pi_+\right)^{-1} \Pi_+J^{-1} u_+\otimes \bar{u}_+J\Pi_+ \right)
=\\=
1+ \bar{u}_+J\Pi_+\left(\Pi_+J^{-1}\Pi_+J\Pi_+\right)^{-1} \Pi_+J^{-1} u_+
=\\=
1 + \bar{u}_+\left(\Pi_+J^{-1}\Pi_+\right)^{-1} \Pi_+J^{-1} u_+
=\\=
1 + \begin{pmatrix}
0 & 1
\end{pmatrix}
\mathrm{ev}_0 \Phi_+\Pi_+\Phi_-^{-1}\Pi_+ \Phi_-\Phi_+^{-1}\frac{qt}{z-qt}
\begin{pmatrix}
-q^{-\sigma}/\sqrt{t} \\ 1
\end{pmatrix}
=\\=
1 + \begin{pmatrix}
0 & 1
\end{pmatrix}
\mathrm{ev}_0 \Phi_+\Pi_+\Phi_+^{-1}\frac{qt}{z-qt}
\begin{pmatrix}
-q^{-\sigma} /\sqrt{t}\\ 1
\end{pmatrix}
=\\=
1 + \begin{pmatrix}
0 & 1
\end{pmatrix}
\Phi_+(t,0)\left(\Phi_+^{-1}(t,qt) - \Phi_+^{-1}(t,0)\right)
\begin{pmatrix}
-q^{-\sigma}/ \sqrt{t}\\ 1
\end{pmatrix}
=\\=
\begin{pmatrix}
0 & 1
\end{pmatrix}
\Phi_+(t,0)\Phi_+(t,qt)^{-1}
\begin{pmatrix}
-q^{-\sigma}/\sqrt{t} \\ 1
\end{pmatrix},
\end{multline}
where we used a simplified version of the formula \eqref{eq:abcdDet}, which reads \(\det(\mathbb{I}+u\otimes \bar{u})=1+\bar{u}u\).
Now, use \eqref{eq:xidef} and \eqref{eq:phipval} to write
\begin{multline}
\label{eq:taushift}
\frac{\tau(qt)}{\tau(t)}=
\frac{r}{r_0}
\begin{pmatrix}
0&1
\end{pmatrix}
\begin{pmatrix}
1 & c(t)/\sqrt{t}\\
0 & 1
\end{pmatrix} t^{-\frac1{4}\boldsymbol{\sigma}_{\mathbf{3}}}\mathsf{g}(t)^{\boldsymbol{\sigma}_{\mathbf{3}}}
\Phi_+(t,qt)^{-1}
\begin{pmatrix}
-q^{-\sigma}/\sqrt{t} \\ 1
\end{pmatrix}=\\=
\frac{r}{r_0} \mathsf{g}(t)^{-1}
\begin{pmatrix}
0&1
\end{pmatrix}
\Phi_+(t,qt)^{-1}
\begin{pmatrix}
-q^{-\sigma}/\sqrt[4]{t} \\ \sqrt[4]{t}
\end{pmatrix}=\beta(t).
\end{multline}
In this way, one shows that the function \(\beta(t)\) defined in \eqref{eq:xidef} through the solutions of the linear problems has the meaning of the first logarithmic \(q\)-difference derivative of the tau function.
The counterpart of this function in the differential case was the Hamiltonian.
To get rid of this new function, one needs to compute the second \(q\)-difference derivative using \eqref{eq:betaeq}:
\begin{equation}
\label{eq:bilin1}
\frac{\tau(q^2t)\tau(t)}{\tau(qt)^2}=\frac{\beta(qt)}{\beta(t)}=
\frac{1+\mathsf{g}(qt)^{-2}qt}{1-qt}=\frac{1+qt(s_0/s_{\infty})^2 (qt)^{-2\sigma}q^{-2\sigma+1/2}\tilde{\tau}(qt)^2/\tau(qt)^2}{1-qt}.
\end{equation}
Shifting the variable in this equation by \(q\), we prove the following
\begin{theorem}
The tau functions \(\tau\) and \(\tilde{\tau}\) defined by \eqref{eq:widom} satisfy the following \(q\)-difference bilinear relation
\begin{equation}
(1-t)\tau(qt)\tau(q^{-1}t)=\tau(t)^2+(qt)^{1-2\sigma} (s_0/s_{\infty})^2 q^{-1/2} \tilde{\tau}(t)^2,
\end{equation}
as well as its B\"acklund-transformed version.
\end{theorem}
Let us also introduce another tau function
\begin{equation}
\label{eq:bigtau}
\mathcal{T}_{ren.}(t)=\frac{s_0}{s_{\infty}}(qt)^{\sigma^2}(qt;q,q)_{\infty} \tau(t),
\end{equation}
where \((z,q_1,q_2)_{\infty}\) is the double Pochhammer symbol
\begin{equation}
(z;q_1,q_2)_{\infty}=\prod_{i,k=0}^{\infty}\left(1-zq_1^iq_2^k\right),
\end{equation}
or
\begin{equation}
(z;q_1,q_2)_{\infty}=
\exp \left(-\sum_{n=1}^{\infty}\frac{z^n}{n(1-q_1^n)(1-q_2^n)} \right)
=\frac1{(z/q_1;1/q_1,q_2)_{\infty}}.
\end{equation}
The bilinear relation \eqref{eq:bilin1} can then be rewritten as
\begin{equation}
\label{eq:bilin2}
\mathcal{T}_{ren.}(qt)\mathcal{T}_{ren.}(q^{-1}t)=\mathcal{T}_{ren.}(t)^2+ \sqrt{t} \tilde{\mathcal{T}}_{ren.}(t)^2,
\end{equation}
while the B\"acklund-transformed relation becomes
\begin{equation}
\label{eq:bilin3}
\tilde{\mathcal{T}}_{ren.}(qt)\tilde{\mathcal{T}}_{ren.}(q^{-1}t)=\tilde{\mathcal{T}}_{ren.}(t)^2+ \sqrt{t} \mathcal{T}_{ren.}(t)^2.
\end{equation}

\section{Combinatorial expansion of the determinant}
\label{sec:combinatorics}
To find explicit combinatorial expansions, we follow a strategy similar to \cite{Cafasso:2017xgn}, namely, first transform the Widom determinant to the form of the Fredholm determinant
\begin{multline}
\label{eq:Fredholm}
\tau=\det_{\mathcal{H}_+}\Pi_+\Psi_+^{-1}\Psi_-\Pi_+ \Psi_-^{-1}\Psi_+\Pi_+=
\det_{\mathcal{H}_+}\left(\Pi_+-\Pi_+\Psi_+^{-1}\Psi_-\Pi_- \Psi_-^{-1}\Psi_+\Pi_+\right)=\\=
\det_{\mathcal{H}_+}\left(\Pi_+-\Psi_+\Pi_+\Psi_+^{-1}\Pi_-\cdot
\Psi_-\Pi_- \Psi_-^{-1}\Pi_+\right)=
\det_{\mathcal{H}_+}\left(\mathbb{I}-a\cdot d\right).
\end{multline}
The operators \(a\) and \(d\) can be rewritten as
\begin{align}
&a=\Psi_+\Pi_+\Psi_+^{-1}\Pi_-=
\Psi_+\Pi_+\Psi_+^{-1}(\mathbb{I}-\Pi_+)=
\Psi_+\Pi_+\Psi_+^{-1}-\Pi_+=
\Pi_--\Psi_+\Pi_-\Psi_+^{-1},\\
&d=\Psi_-\Pi_- \Psi_-^{-1}\Pi_+=
\Psi_-\Pi_- \Psi_-^{-1}(\mathbb{I}-\Pi_-)=
\Psi_-\Pi_- \Psi_-^{-1}-\Pi_-=
\Pi_+-\Psi_-\Pi_+ \Psi_-^{-1}.
\end{align}
The kernels of these operators are given explicitly by
\begin{equation}
\label{eq:adexpansion}
a(z,z')= \frac{\Psi_+(z) \Psi_+(z')^{-1}-\mathbb{I}}{z-z'},\qquad
d(z,z')=\frac{\mathbb{I}-\Psi_-(z)\Psi_-(z')^{-1}}{z-z'},
\end{equation}
which leads to
\begin{lemma}
The tau function \eqref{eq:widom} can be written as a block Fredholm determinant with an integrable kernel,
\begin{equation}
\label{eq:matrixFredholm}
\tau = \det_{\mathcal{H}_+\oplus \mathcal{H}_-}
\begin{pmatrix}
\mathbb{I} & a\\
d & \mathbb{I}
\end{pmatrix},
\end{equation}
where the blocks of the kernel are given by \eqref{eq:adexpansion}.
\end{lemma}

In this section, we find explicit factorized expressions for the elements of \(a\) and \(d\) in the Laurent basis, as well as factorized expressions for their minors.
This allows us to write the minor expansion of \eqref{eq:Fredholm} explicitly in \eqref{eq:tauminors}.
Then, we further simplify these minors using combinatorics and get the expansion \eqref{eq:tauExpansion} in terms of the Nekrasov functions.
This expansion also has a slightly cleaner renormalized version \eqref{eq:tauk}.

\subsection{Matrix elements of the operators}
\label{sec:matElements}
Let us start by finding series expansions of the kernels \(a\) and \(d\) \eqref{eq:adexpansion}.
We write these expansions as
\begin{equation}
a(z,z')=\sum_{m,n\in \mathbb{Z}'_+}^{\infty} a_{mm} z^{m-1/2}z'^{n-1/2},\qquad
d(z,z')=\sum_{m,n\in \mathbb{Z}'_+}^{\infty}d_{mn}z^{-m-1/2}z'^{-n-1/2},
\end{equation}
where
\begin{align}
&\mathbb{Z}'_+ = 1/2 + \mathbb{Z}_{\ge0}=\{1/2, 3/2, 5/2, \ldots \},\\
\nonumber
&\mathbb{Z}'_- = -\mathbb{Z}'_+ = \{-1/2, -3/2, -5/2, \ldots \}.
\end{align}
\subsubsection*{Operator \(a(z,z')\)}
The coefficients of expansion of \(a\) can be computed using the following trick.
First, we consider
\begin{multline}
a(z,z')-q^{1+\boldsymbol{\sigma}_{\mathbf{3}}\sigma}a(qz,qz')q^{-\boldsymbol{\sigma}_{\mathbf{3}}\sigma}=\frac{\Psi_+(z)\left(\mathbb{I}-L_\infty(z)L_{\infty}(z')^{-1}\right)\Psi_+(z')^{-1}}{z-z'}=\\=
\Psi_+(z)
\begin{pmatrix}
0 \\
1 
\end{pmatrix}\otimes
\begin{pmatrix}
 q^{-\sigma}&-1
\end{pmatrix}
\frac{\Psi_+(z')^{-1}}{z'-1}=\\=
\sum_{m,n\in \mathbb{Z}'_+} 
\begin{pmatrix} f_{a,m,+} \\ f_{a,m,-} \end{pmatrix}\otimes
\begin{pmatrix} g_{a,n,+} & g_{a,n,-} \end{pmatrix} z^{m-1/2}z'^{n-1/2}.
\end{multline}
Combining this formula with \eqref{eq:adexpansion} we get
\begin{equation}
a_{mn,\alpha\beta}=\frac{f_{a,m,\alpha}g_{a,n,\beta}}{1-q^{m+n+\sigma(\alpha-\beta)}}.
\end{equation}
The formulas for \(f\)'s and \(g\)'s can be obtained by explicit computation:
\begin{equation}
\Psi_+(z)
\begin{pmatrix}
0 \\
1 
\end{pmatrix}=
\begin{pmatrix}
\frac{-q^{-1/4}s_{\infty}}{q^{\sigma}-q^{-\sigma}}j_0(q^{1+2\sigma},z)\\
q^{1/4}s^{-\infty}j_0(q^{1-2\sigma},z)\end{pmatrix}
=\sum_{m\in \mathbb{Z}_+'}
\begin{pmatrix}
\frac{q^{-1/4}q^{\sigma}s_{\infty} }{(q^{2\sigma};q)_{m+1/2}(q;q)_{m-1/2}}  \\
\frac{q^{1/4}s^{-1}_{\infty} }{(q^{1-2\sigma};q)_{m-1/2}(q;q)_{m-1/2}}
\end{pmatrix}z^{m-1/2},
\end{equation}
and for the other factor
\begin{multline}
\begin{pmatrix}
 q^{-\sigma}&-1
\end{pmatrix}
\frac{\Psi_+(z')^{-1}}{z'-1}=
\begin{pmatrix}
-s_\infty^{-1}q^{1/4-\sigma} j_2(q^{1-2\sigma},qz)&
\frac{-s_{\infty}q^{2\sigma-1/4}}{1-q^{2\sigma}} j_2(q^{1+2\sigma},qz)
\end{pmatrix}=\\=
\sum_{n\in \mathbb{Z}'_+}
\begin{pmatrix}
\frac{-s_{\infty}^{-1} q^{1/4-\sigma}q^{(n-1/2)^2+(1-2\sigma)(n-1/2)}}{(q^{1-2\sigma};q)_{n-1/2}(q;q)_{n-1/2}} &
\frac{-s_{\infty} q^{2\sigma-1/4}q^{(n-1/2)^2+(1+2\sigma)(n-1/2)}}{(q^{2\sigma};q)_{n+1/2}(q;q)_{n-1/2}}
\end{pmatrix}z^{n-1/2}.
\end{multline}
Therefore,
\begin{equation}
\label{eq:fga}
f_{a,m}=\begin{pmatrix}
\frac{-q^{-1/4}q^{\sigma}s_{\infty} }{(q^{2\sigma};q)_{m+1/2}(q;q)_{m-1/2}}  \\
\frac{q^{1/4}s^{-1}_{\infty} }{(q^{1-2\sigma};q)_{m-1/2}(q;q)_{m-1/2}}
\end{pmatrix},
\qquad
g_{a,n}^T=
\begin{pmatrix}
\frac{-s_{\infty}^{-1} q^{n^2-2n\sigma}}{(q^{1-2\sigma};q)_{n-1/2}(q;q)_{n-1/2}} \\
\frac{-s_{\infty} q^{n^2+(2n+1)\sigma-1/2}}{(q^{2\sigma};q)_{n+1/2}(q;q)_{n-1/2}}
\end{pmatrix}
\end{equation}

\subsubsection*{Operator \(d(z,z')\)}
The same trick can be applied to find the expansion of \(d\):
\begin{multline}
d(z',z)-q^{1+\boldsymbol{\sigma}_{\mathbf{3}}\sigma}d(qz',qz)q^{-\boldsymbol{\sigma}_{\mathbf{3}}\sigma}=
\frac{\Psi_-(z')\left(L_0(z')L_0(z)^{-1}-\mathbb{I}\right)\Psi_-(z)^{-1}}{z'-z}
=\\=
\frac{\Psi_-(z')}{z'}
\begin{pmatrix} 1 \\ 0 \end{pmatrix}\otimes
\begin{pmatrix}-1 & q^{\sigma}t^{-1/2}\end{pmatrix}
\frac{\Psi_-(z)^{-1}}{1-z/t}
=\\=
\sum_{n,m\in \mathbb{Z}'_+} 
\begin{pmatrix} g_{d,n,+} \\ g_{d,n,-} \end{pmatrix}\otimes
\begin{pmatrix} f_{d,m,+} & f_{d,m,-} \end{pmatrix} z'^{-n-1/2}z^{-m-1/2}.
\end{multline}
The factors in the above tensor product read:
\begin{multline}
\label{eq:psimvec}
\frac{\Psi_-(z')}{z'}
\begin{pmatrix} 1 \\ 0 \end{pmatrix}=
\begin{pmatrix}
s_0t^{1/4-\sigma} q^{-1/4} z'^{-1}j_2(q^{1-2\sigma},qt/z')\\
\frac{s_0^{-1}(qt)^{1/4+\sigma}q^{-1/4+\sigma}}{1-q^{2\sigma}} z'^{-1} j_2(q^{1+2\sigma},qt/z')
\end{pmatrix}=\\=
\sum_{n\in \mathbb{Z}'_+}
\begin{pmatrix}
\frac{s_0t^{1/4-\sigma}q^{-1/4}(qt)^{n-1/2}q^{(n-1/2)(n-3/2)+(n-1/2)(1-2\sigma)}}{(q^{1-2\sigma};q)_{n-1/2}(q;q)_{n-1/2}} \\
\frac{s_0^{-1}(qt)^{1/4+\sigma}q^{-1/4+\sigma}(qt)^{n-1/2}q^{(n-1/2)(n-3/2)+(n-1/2)(1+2\sigma)}}{(q^{2\sigma};q)_{n+1/2}(q;q)_{n-1/2}}
\end{pmatrix}z'^{-1/2-n},
\end{multline}
and
\begin{multline}
\label{eq:psimcovec}
\begin{pmatrix}-1 & q^{1/2+\sigma}(qt)^{-1/2}\end{pmatrix}\frac{\Psi_-(z)^{-1}}{1-z/t}
=\\=
\begin{pmatrix}
\frac{s_0^{-1}(qt)^{3/4+\sigma}q^{-3/4}}{1-q^{2\sigma}}z^{-1}j_0(q^{1+2\sigma},t/z) &
-s_0t^{3/4-\sigma}q^{\sigma-3/4}z^{-1}j_0(q^{1-2\sigma},t/z)
\end{pmatrix}
=\\=
\begin{pmatrix}
\frac{s_0^{-1}(qt)^{3/4+\sigma}q^{-3/4}(qt)^{m-1/2}q^{1/2-m}}{(q^{2\sigma};q)_{m+1/2}(q;q)_{m-1/2}} &
\frac{-s_0t^{3/4-\sigma}q^{\sigma-3/4}(qt)^{m-1/2}q^{1/2-m}}{(q^{1-2\sigma};q)_{m-1/2}(q;q)_{m-1/2}}
\end{pmatrix}z^{-1/2-m}.
\end{multline}
It follows that
\begin{equation}
\label{eq:fgd}
g_{d,n}=
\begin{pmatrix}
\frac{s_0(qt)^{n-\sigma-1/4}q^{n^2-n-2n\sigma+\sigma}}{(q^{1-2\sigma};q)_{n-1/2}(q;q)_{n-1/2}} \\
\frac{s_0^{-1}(qt)^{n+\sigma-1/4}q^{n^2-n+2n\sigma}}{(q^{2\sigma};q)_{n+1/2}(q;q)_{n-1/2}}
\end{pmatrix},\qquad
f_{d,m}^T=
\begin{pmatrix}
\frac{-s_0^{-1}(qt)^{m+\sigma+1/4}q^{-1/4-m}}{(q^{2\sigma};q)_{m+1/2}(q;q)_{m-1/2}} \\
\frac{s_0 (qt)^{m-\sigma+1/4}q^{\sigma-m-1/4}}{(q^{1-2\sigma};q)_{m-1/2}(q;q)_{m-1/2}}
\end{pmatrix},
\end{equation}
and
\begin{equation}
\label{eq:dfg}
d_{nm,\beta\alpha}=\frac{g_{d,n,\beta}f_{d,m,\alpha}}{1-q^{-m-n-\sigma(\alpha-\beta)}}.
\end{equation}
The main outcome of these computations is that both operators \(a\) and \(d\) are essentially given by the Cauchy matrices \(\frac1{x_i-y_j}\) multiplied on the left and right sides by explicit diagonal factors.

\subsection{Minor expansion}
Now, we are ready to compute the minor expansion of \eqref{eq:Fredholm}.
We need the following identity involving Cauchy determinants:
\begin{multline}
\det \frac1{1-x_iy_j} \det \frac{1}{1-x^{-1}_iy^{-1}_j}=
\det \frac1{x_i^{-1}-y_j} \det \frac{1}{x_i-y^{-1}_j}=\\=
\frac{\prod_{i<j}(x_i^{-1}-x_j^{-1})(x_i-x_j)\prod_{i>j}(y_i-y_j)(y_i^{-1}-y_j^{-1})}{\prod_{ij}(x_i^{-1}-y_j)(x_i-y^{-1}_j)}
=\\=\frac{\prod_{i\neq j}(1-x_i/x_j)(1-y_i/y_j)}{\prod_{ij}(1-x_iy_j)(1-x_i^{-1}y_j^{-1})}
=\prod_{\substack{z,z'\in \{x_i\}\sqcup\{y_i^{-1}\}\\z\neq z'}}(1-z/z')^{\deg z \deg z'},
\end{multline}
where the degrees of the variables are fixed by \(\deg x_i=\deg x_i^{-1}=1\), \(\deg y_i=\deg y_i^{-1}=-1\).
Using this relation, we write
\begin{multline}
\label{eq:tauminors}
\tau=\det_{\mathcal{H}_+}(\mathbb{I}-a\cdot d)=\sum_{\substack{I,J\\|I|=|J|}}
\det_{\substack{(m,\alpha)\in I\\(-n,\beta)\in J}}
\frac{f_{a,m,\alpha}g_{a,n,\beta}}{1-q^{m+n+\sigma(\alpha-\beta)}}
\det_{\substack{(m,\alpha)\in I\\(-n,\beta)\in J}}
\frac{-g_{d,n,\beta}f_{d,m,\alpha}}{1-q^{-m-n-\sigma(\alpha-\beta)}}
=\\=
\sum_{\substack{I,J\\|I|=|J|}}\prod_{(m,\alpha)\in I}(-f_{a,m,\alpha}f_{d,m,\alpha})
\prod_{(-n,\beta)\in J}(g_{a,n,\beta}g_{d,n,\alpha})
\prod_{\substack{(k,\alpha),(k',\alpha')\in I\sqcup J\\(k,\alpha)\neq(k',\alpha')}}
(1-q^{k-k'+\sigma(\alpha-\alpha')})^{\operatorname{sign}(kk')}
=\\=\sum_{\mathtt{M}\in \mathbb{M}^2_0}\prod_{(k,\alpha)\in \mathtt{M}}g\!f_{k,\alpha}
\prod_{\substack{(k,\alpha),(k',\alpha')\in \mathtt{M}\\(k,\alpha)\neq(k',\alpha')}}
(1-q^{k-k'+\sigma(\alpha-\alpha')})^{\operatorname{sign}(kk')}.
\end{multline}
In this formula, \(I\) and \(J\) are the two subsets of the set of pairs ``(half-integer number, \(\pm1\))'', where the first element labels the Laurent mode, and the second one is a matrix index.
The ordering of rows and columns in the graphical representation of matrices in this paper is ``\(+1,-1\)''.
We also denote by \(\mathtt{M}\) the union
\begin{equation}
\mathtt{M}=I\sqcup J \in \mathbb{M}^2_0,
\end{equation}
where the set \(\mathbb{M}^2_0\) is defined by
\begin{equation}
\mathbb{M}^2_0 = \left\{I\sqcup J\middle|I \subset 2^{\mathbb{Z}_+'\times \{+,-\}}, J \subset 2^{\mathbb{Z}_-'\times \{+,-\}}, |I|=|J|\right\}
\end{equation}
and can be identified with the set of two-component Maya diagrams with the zero total charge, see, e.g., \cite{Gavrylenko:2016zlf}, \cite{Cafasso:2017xgn} for the pictures; they are the same in the difference and differential case.

We also define the products of the diagonal prefactors
\begin{equation}
g\!f_{m,\alpha}=f_{a,m,\alpha}f_{d,m,\alpha}, \quad m>0, \qquad \qquad
g\!f_{-n,\alpha} =-g_{a,n,\beta}g_{d,n,\alpha}, \quad n>0.
\end{equation}
Explicit evaluations of these coefficients follow from \eqref{eq:fga} and \eqref{eq:fgd} and are given by
\begin{equation}
g\!f_m=\begin{pmatrix}
\frac{-s_{\infty}/s_0t^{m+\sigma+1/4}q^{2\sigma-1/4}}{(q^{2\sigma};q)_{m+1/2}^2(q;q)_{m-1/2}^2}\\
\frac{-s_0/s_{\infty} t^{m-\sigma+1/4}q^{1/4}}{(q^{1-2\sigma};q)_{m-1/2}^2(q;q)_{m-1/2}^2}
\end{pmatrix},\qquad
g\!f_{-n}=\begin{pmatrix}
\frac{s_0/s_{\infty}t^{n-\sigma-1/4}q^{2n^2-4n\sigma-1/4}}{(q^{1-2\sigma};q)_{n-1/2}^2(q;q)_{n-1/2}^2}
\\
\frac{s_{\infty}/s_0t^{n+\sigma-1/4}q^{2n^2+(4n+2)\sigma-3/4}}{(q^{2\sigma};q)_{n+1/2}^2(q;q)_{n-1/2}^2}
\end{pmatrix}.
\end{equation}

\subsection{Nekrasov functions}
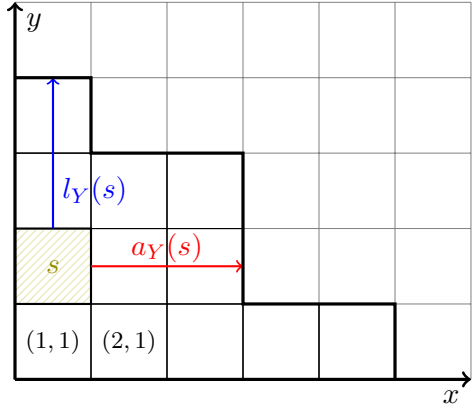
\begin{figure}[ht!]
\begin{center}
\begin{tikzpicture}[scale=1,
style1/.style={},
style2/.style={},
style3/.style={pattern color=olive!20!white,pattern={north east lines},thick},
style4/.style={red,thick,->},
style5/.style={blue, thick,->},]
\draw [help lines, step=1cm] (0,0) grid (6,5);

\draw[style1](0,2)rectangle(1,3);
\draw[style1](0,3)rectangle(1,4);

\draw[style2](0,1)rectangle(1,2);
\draw[style2](1,2)rectangle(2,3);

\draw[style2](0,0)rectangle(1,1) node[xshift=-0.5cm,yshift=-0.5cm]{\footnotesize\((1,1)\)};
\draw[style2](1,1)rectangle(2,2);
\draw[style2](2,2)rectangle(3,3);
\draw[style2](1,0)rectangle(2,1) node[xshift=-0.5cm,yshift=-0.5cm]{\footnotesize\((2,1)\)};
\draw[style2](2,0)rectangle(3,1);
\draw[style2](3,0)rectangle(4,1);
\draw[style2](4,0)rectangle(5,1);
\draw[style2](2,1)rectangle(3,2);

\draw[style3](0,1)rectangle(1,2) node[xshift=-0.5cm,yshift=-0.5cm] {\color{olive} \(s\)};

\draw[style5](1/2,2)-- node[xshift=0.55cm,yshift=-0.5cm] {\(l_Y(s)\)} (1/2,4);
\draw[style4](1,3/2)-- node[yshift=0.25cm]{\(a_Y(s)\)}(3,3/2);

\draw[->,very thick](0,0)--(6,0) node[below left] {\(x\)};
\draw[->,very thick](0,0)--(0,5) node[below right] {\(y\)};
\draw[very thick](0,4)--(1,4)--(1,3)--(3,3)--(3,1)--(5,1)--(5,0);
\end{tikzpicture}
\end{center}
\caption{\label{fig:armslegs} Example of the Young diagram. \(x(s)=1\), \(y(s)=2\), \(a_Y(s)=2\), \(l_Y(s)=2\).}
\end{figure}

There is a combinatorial theorem that was used in \cite{Nakajima:2003pg} to rewrite Nekrasov partition functions in more familiar terms by canceling negative contributions to the character of the tangent space near the fixed point on the instanton moduli space.
To formulate it, we first define
\begin{equation}
\label{eq:vchar}
V_{\alpha}=\sum_{s\in Y_{\alpha}}e^{\tau\epsilon_1(1-x(s))}e^{\tau\epsilon_2(1-y(s))},\quad
V_{\alpha}=\sum_{\alpha} e^{-\tau a_{\alpha}}V_{\alpha},\quad
W=\sum_{\alpha} e^{-\tau a_{\alpha}},\quad
\tau^{*}=-\tau,
\end{equation}
where \(Y_{\alpha}\) are the two Young diagrams, \((x(s), y(s))\) are coordinates of the box \(s\) in the plane starting from \((1,1)\), see Figure~\ref{fig:armslegs}.

Let us also define
\begin{equation}
\label{eq:chiNek}
\chi_{Nek}=-(1-e^{\tau\epsilon_1})(1-e^{\tau\epsilon_2})VV^{*}+VW^{*}+WV^{*}
\end{equation}
and
\begin{equation}
\chi_{NY}=\sum_{\alpha\beta}e^{\tau(a_{\alpha}-a_{\beta})}N_{\alpha\beta}(e^{\tau\epsilon_1},e^{\tau\epsilon_2}),
\end{equation}
where
\begin{equation}
\label{eq:chiNY}
N_{\alpha\beta}(e^{\tau\epsilon_1},e^{\tau\epsilon_2})=
\sum_{s\in Y_{\alpha}}e^{-\tau\epsilon_1l_{Y_{\beta}}(s)}e^{\tau\epsilon_2(a_{Y_{\alpha}}(s)+1)}+
\sum_{s\in Y_{\beta}}e^{\tau\epsilon_1(l_{Y_{\alpha}}(s)+1)}e^{-\tau\epsilon_2a_{Y_{\beta}}(s)}.
\end{equation}
Arm/leg lengths \(a_{Y_{\alpha}}(s)\), \(l_{Y_{\alpha}}(s)\) of the box \(s\) with respect to the diagram \(Y\)\footnote{\(s\) should not necessary lie in \(Y\), in this case arm and leg lengths become negative.} are shown in Figure~\ref{fig:armslegs}.

Now, we are ready to formulate the following
\begin{theorem}[Nakajima-Yoshioka]
\label{thm:NY}
The two formulas for the character are equivalent
\begin{equation}
\label{eq:chiequality}
\chi_{Nek}=\chi_{NY}.
\end{equation}
\end{theorem}
This theorem is useful because it allows us to rewrite a more complicated expression, growing quadratically with the number of boxes, as an expression of linear complexity.
It is also useful because it allows one to rewrite different seemingly inequivalent expressions for \(\chi\) in some unified form.
Namely, we are going to use it to rewrite \(\chi\) in the charged Frobenius coordinates and, in this way, relate it to the combinatorics of minors.

\subsubsection*{Young and Maya diagrams, Frobenius coordinates}
\begin{figure}[ht!]
\begin{center}
\begin{tikzpicture}[scale=1,
style1/.style={pattern color=blue!30!white,thick,pattern={north east lines}},
style2/.style={pattern color=red!30!white,thick,pattern={north west lines}},
style3/.style={pattern color=magenta!50!white,pattern={dots}, dashed},
style4/.style={red, thick,->},
style5/.style={blue, thick,->},]
\draw [help lines, step=1cm] (-3,0) grid (6,5);

\draw[style2](0,2)rectangle(1,3);
\draw[style1](0,3)rectangle(1,4);

\draw[style2](0,1)rectangle(1,2);
\draw[style2](1,2)rectangle(2,3);

\draw[style2](0,0)rectangle(1,1);
\draw[style2](1,1)rectangle(2,2);
\draw[style2](2,2)rectangle(3,3);
\draw[style2](1,0)rectangle(2,1);
\draw[style2](2,0)rectangle(3,1);
\draw[style2](3,0)rectangle(4,1);
\draw[style2](4,0)rectangle(5,1);
\draw[style2](2,1)rectangle(3,2);

\draw[style3](-1,0)rectangle(0,1);
\draw[style3](-2,0)rectangle(-1,1);
\draw[style3](-1,1)rectangle(0,2);

\draw[magenta,thick](1,3)--(-2,0) node [below] {\(-Q_{\alpha}\)};

\draw[style4](-3/2,1/2)--(5,1/2) node[above left]{\(m_1\)};
\draw[style4](-1/2,3/2)--(3,3/2) node[above left]{\(m_2\)};
\draw[style4](1/2,5/2)--(3,5/2) node[above left]{\(m_3\)};
\draw[style5](1/2,5/2)--(1/2,4) node[below right, xshift=-0.1cm]{\(n_1\)};

\draw[->,very thick](-3,0)--(6,0) node[yshift=-0.2cm] {\(x\)};
\draw[->,very thick](0,0)--(0,5) node[xshift=-0.25cm,yshift=-0.2cm] {\(y\)};

\draw[very thick](0,4)--(1,4)--(1,3)--(3,3)--(3,1)--(5,1)--(5,0);

\end{tikzpicture}
\end{center}
\caption{\label{fig:Frobenius} Charged Frobenius coordinates.}
\end{figure}
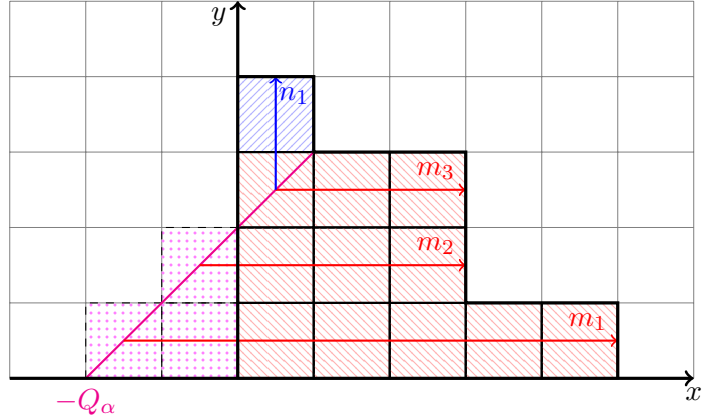

\newcommand\boxr{\tikz[scale=0.3,
style2/.style={pattern color=red!30!white,pattern={north west lines}},
]{\draw[style2](0,0)rectangle(1,1);}}

\newcommand\boxb{\tikz[scale=0.3,
style1/.style={pattern color=blue!30!white,pattern={north east lines}},
]{\draw[style1](0,0)rectangle(1,1);}}

\newcommand\boxm{\tikz[scale=0.3,
style3/.style={pattern color=magenta!50!white,pattern={dots}, dashed},
]{\draw[style3](0,0)rectangle(1,1);}}

We start by introducing the isomorphism between the charged Maya diagrams and charged Young diagrams:
\begin{equation}
\label{eq:isoYoung}
\mathbb{M}^2_0 = \mathbb{Y}^2\times \mathbb{Z}^2_0 = \mathbb{Y}^2\times \mathbb{Z},
\end{equation}
where \(\mathbb{Y}\) is the set of all Young diagrams, and \(\mathbb{Z}^2_0\) is the set of pairs \((Q_+,Q_-) = (Q,-Q)\).

The isomorphism is shown explicitly in Figure~\ref{fig:Frobenius}.
Namely, given a pair \((Y_{\alpha}, Q_{\alpha})\) we can define \(\#_{\alpha} n\) half-integer numbers \(n_i\) and \(\#_{\alpha} m = \#_{\alpha} n + Q_{\alpha}\) half-integer numbers \(m_j\).
We call these numbers charged Frobenius coordinates.
It is clear from this construction that
\begin{enumerate}
\item For any Young diagram \(Y_{\alpha}\) and any charge \(Q_{\alpha}\), all \(n_i\)'s are defined and are given by distinct half-integer numbers.
The same is true for \(m_j\)'s.
\item Given a fixed \(Q_{\alpha}\), any pair of sets of non-coinciding half-integer numbers \(n_i\) and \(m_i\), such that \(\#_{\alpha} m = \#_{\alpha} n+ Q_{\alpha}\), defines a Young diagram.
\end{enumerate}
The coordinates \(-n_i\) will be identified with the negative part of the Maya diagram (\(J\)), and \(m_i\) with the positive part (\(I\)).

Taking into account this explicit construction of the Frobenius coordinates, the isomorphism \eqref{eq:isoYoung} can be formulated as follows.
Given an element of \(\mathbb{M}^2_0\), we first compute the charges \(Q_{\alpha}\) as the differences \(\#_{\alpha} m - \#_{\alpha}n\), and then construct the corresponding Young diagrams from their charged Frobenius coordinates.

The inverse map is constructed simply by computing the charged Frobenius coordinates of the Young diagrams \(Y_{\alpha}\) with respect to the charges \(Q_{\alpha}\).

\subsubsection*{Characters in the Frobenius coordinates}
Now we want to express the character \(\chi_{Nek}\) \eqref{eq:chiNek} in the Frobenius coordinates.

In our computations, we put from the beginning
\begin{equation}
\epsilon_1=1, \quad \epsilon_2=-1.
\end{equation}
After this redefinition
\begin{equation}
\label{eq:Vaxy}
V_{\alpha}=\sum_{s\in Y_{\alpha}}e^{\tau(y(s)-x(s))}.
\end{equation}
The character \(V_{\alpha}\) can be written using a version of inclusion/exclusion formula (see Figure~\ref{fig:Frobenius}):
\begin{equation}
V_{\alpha}=V^{\boxb}_{\alpha}+V^{\boxm\boxr}_{\alpha}-V_{\alpha}^{\boxm}
\end{equation}
The values of the summands can be obtained from the computations of geometric progressions and are equal to
\begin{equation}
V^{\boxm\boxr}_{\alpha}=e^{\tau Q_{\alpha}}\sum_{i=1}^{\#_{\alpha}m}\frac{1-e^{-\tau(m_i+1/2)}}{1-e^{-\tau}},
\quad
V^{\boxb}_{\alpha}=e^{\tau Q_{\alpha}}\sum_{i=1}^{\#_{\alpha}n} \frac{e^{\tau}-e^{\tau(n_i+1/2)}}{1-e^{\tau}},
\end{equation}

\begin{equation}
V^{\boxm}_{\alpha}=\sum_{y=1}^{Q_{\alpha}}\frac{e^{\tau Q_{\alpha}}-e^{\tau(y-1)}}{1-e^{-\tau}}=\frac{Q_{\alpha}e^{\tau Q_{\alpha}}}{1-e^{-\tau}}-\frac{1-e^{\tau Q_{\alpha}}}{(1-e^{-\tau})(1-e^{\tau})}.
\end{equation}
The resulting expression can be slightly simplified:
\begin{equation}
\label{eq:Vamn}
V_{\alpha}=e^{\tau Q_{\alpha}-\tau/2}\left(\sum_i \frac{e^{\tau n_i}}{1-e^{-\tau}}-
\sum_i \frac{e^{-\tau m_i}}{1-e^{-\tau}}\right)+
\frac{e^{\tau Q_{\alpha}}-1}{e^{\tau}(1-e^{-\tau})^2}.
\end{equation}
Notice that it does not have a pole at \(e^{-\tau}=1\) due to the identity
\begin{equation}
Q_{\alpha}=\#_{\alpha}m-\#_{\alpha}n.
\end{equation}
The full character of \(V\) \eqref{eq:vchar} can be written as
\begin{equation}
V=\frac{e^{-\tau/2}}{1-e^{-\tau}} \mathcal{V} + \frac{e^{-\tau}}{(1-e^{-\tau})^2}(\mathcal{W}-W),
\end{equation}
where
\begin{equation}
\mathcal{V}=\sum_{(k,\alpha)\in \mathtt{M}} \operatorname{sign}(-k) e^{-\tau(k+\sigma_{\alpha})}, \qquad
\mathcal{W}=\sum_{\alpha} e^{-\tau\sigma_{\alpha}},
\qquad
\sigma_{\alpha} = a_{\alpha}-Q_{\alpha}.
\end{equation}
In this way, the character of the tangent space becomes
\begin{equation}
\chi_{Nek}=e^{\tau}(1-e^{-\tau})^2 VV^{*}+VW^{*}+WV^{*},
\end{equation}
or explicitly
\begin{equation}
\chi_{Nek}=-\mathcal{V}\mathcal{V}^{*}+\frac{e^{-\tau/2}}{1-e^{-\tau/2}}
\left(\mathcal{V} \mathcal{W}^{*}-\mathcal{W}\mathcal{V}^{*}\right) +
\frac{e^{-\tau}}{(1-e^{-\tau})^2}\left(\mathcal{W}\mathcal{W}^{*}-WW^{*}\right).
\end{equation}
Another expression for \(\chi\) following from Theorem~\ref{thm:NY} is 
\begin{equation}
\chi_{NY}=\sum_{\alpha\beta}e^{\tau(\sigma_{\alpha}-\sigma_{\beta}+Q_{\alpha}-Q_{\beta})}\left(
\sum_{s\in Y_{\alpha}}e^{-\tau(a_{Y_{\alpha}}(s)+l_{Y_{\beta}}(s)+1)}+
\sum_{s\in Y_{\beta}}e^{\tau(a_{Y_{\beta}}(s)+l_{Y_{\alpha}}(s)+1)}
\right).
\end{equation}
Now, we define the Nekrasov factor
\begin{multline}
\label{eq:zvecdef}
Z_{vec}=\operatorname{PE}(-\chi_{NY})=\prod_{\alpha\beta}
\prod_{s\in Y_{\alpha}}\left(1-q^{-\sigma_{\alpha}+\sigma_{\beta}-Q_{\alpha}+Q_{\beta} + a_{Y_{\alpha}}(s)+l_{Y_{\beta}}(s)+1}\right)^{-1}\\\times
\prod_{s\in Y_{\beta}}\left(1-q^{-\sigma_{\alpha}+\sigma_{\beta}-Q_{\alpha}+Q_{\beta}-a_{Y_{\beta}}(s)-l_{Y_{\alpha}}(s)-1}\right)^{-1},
\end{multline}
where the plethystic exponential \(\operatorname{PE}\) is defined by
\begin{equation}
\operatorname{PE}\left(\sum_i e^{-\tau v_i} - \sum_j e^{-\tau w_j}\right):=
\frac{\prod_i(1-q^{v_i})}{\prod_j(1-q^{w_j})}.
\end{equation}
It has the obvious property
\begin{equation}
\operatorname{PE}\left(\sum k_i A_i\right) = \prod_i \operatorname{PE}(A_i)^{k_i}, \qquad k_i \in \mathbb{Z}.
\end{equation}
Thanks to Theorem~\ref{thm:NY}, we can rewrite the Nekrasov factor \eqref{eq:zvecdef} equivalently as
\begin{equation}
\label{eq:Z2Z1Z0}
Z_{vec}=\operatorname{PE}(-\chi_{Nek})=Z_2Z_1Z_0,
\end{equation}
where
\begin{align}
&Z_2=\operatorname{PE}(\mathcal{V}\mathcal{V}^{*}-|\mathcal{V}|),\\
&Z_1=\operatorname{PE}\left(\frac{e^{-\tau/2}}{1-e^{-\tau}}
\left(\mathcal{W}\mathcal{V}^{*} - \mathcal{V} \mathcal{W}^{*}\right)
+|\mathcal{V}|\right),\\
&Z_0=\operatorname{PE}\left(\frac{e^{-\tau}}{(1-e^{-\tau})^2}
\left(WW^{*} - \mathcal{W}\mathcal{W}^{*}\right)\right).
\end{align}
Here \(|\mathcal{V}|\) is the number of summands in \(\mathcal{V}\).
It should be added and subtracted during the separation of different terms to avoid the appearance of the zero weights since \(\operatorname{PE}(1)=0\).

Explicit expressions for \(Z_{2,1,0}\) read
\begin{align}
&Z_2=\prod_{\substack{(k,\alpha), (k',\alpha') \in \mathtt{M}\\(k,\alpha)\neq(k',\alpha')}}(1-q^{\sigma_{\alpha}-\sigma_{\alpha'}+k-k'})^{\operatorname{sign}(kk')},\\
&Z_1=\prod_{(k,\alpha)\in \mathtt{M}} h_{k,\alpha},\\
\label{eq:Z0}
&Z_0=\prod_{\alpha\beta}\frac
{(q^{1+\sigma_{\alpha}-\sigma_{\beta}+Q_{\alpha}-Q_{\beta}};q,q)_{\infty}}
{(q^{1+\sigma_{\alpha}-\sigma_{\beta}};q,q)_{\infty}},
\end{align}
where
\begin{equation}
h_{k,\alpha}=\left(
\prod_{\beta\neq \alpha}
\frac{(q^{1/2+k+\sigma_{\alpha}-\sigma_{\beta}};q)_{\infty}}
{(q^{1/2-k+\sigma_{\beta}-\sigma_{\alpha}};q)_{\infty}}\right)^{\operatorname{sign}(k)}
\times \lim_{\epsilon\to 0} (1-q^{\epsilon})
\left(\frac{(q^{1/2+k+\epsilon};q)_{\infty}}
{(q^{1/2-k+\epsilon};q)_{\infty}}\right)^{\operatorname{sign}(k)}.
\end{equation}
At the moment, we are able to express the Nekrasov factor in the charged Frobenius coordinates.
It remains to identify it with the terms of expansion \eqref{eq:tauminors}, which is done in the next subsection.

\subsection{Relation between minors and Nekrasov functions}
\begin{remark}
Partially, this subsection is a \(q\)-generalization of the proof from \cite[Appendix A]{Gavrylenko:2016zlf}. However, this time, it is significantly simplified and made much less technical using the ideas from \cite{Nakajima:2003pg}.
\end{remark}
Combining \eqref{eq:tauminors} and \eqref{eq:Z2Z1Z0} we can write
\begin{equation}
\label{eq:tauNek1}
\tau=\sum_{\mathtt{M}\in \mathbb{M}^2_0}Z_{vec}Z_0^{-1}\prod_{(k,\alpha)\in \mathtt{M}}(g\!f_{k,\alpha}/h_{k,\alpha}) = 
\sum_{\mathtt{M}\in \mathbb{M}^2_0}Z_{vec}Z_0^{-1}\prod_{(k,\alpha)\in \mathtt{M}}\overline{g\!f}_{k,\alpha}=
\sum_{\mathtt{M}\in \mathbb{M}^2_0}Z_{vec}Z_0^{-1}\overline{Z}_1^{-1},
\end{equation}
where
\begin{equation}
\overline{Z}_1 = \prod_{(k,\alpha)\in \mathtt{M}}(g\!f_{k,\alpha}/h_{k,\alpha}) =
\prod_{(k,\alpha)\in \mathtt{M}}\overline{g\!f}_{k,\alpha}.
\end{equation}
The explicit formulas for \(\overline{g\!f}_{k,\alpha}\) are given by
\begin{align}
\overline{g\!f}_m=&\begin{pmatrix}
\frac{-s_{\infty}/s_0t^{m+\sigma+1/4}q^{2\sigma-1/4}}{(q^{2\sigma};q)_{m+1/2}^2(q;q)_{m-1/2}^2}
\frac{(q^{1/2-m+\epsilon};q)_{\infty}(q^{1/2-m-2\sigma})_{\infty}}{(1-q^{\epsilon})(q^{1/2+m+\epsilon};q)_{\infty}(q^{1/2+m+2\sigma};q)_{\infty}}
\\
\frac{-s_0/s_{\infty} t^{m-\sigma+1/4}q^{1/4}}{(q^{1-2\sigma};q)_{m-1/2}^2(q;q)_{m-1/2}^2}
\frac{(q^{1/2-m+\epsilon};q)_{\infty}(q^{1/2-m+2\sigma};q)_{\infty}}{(1-q^{\epsilon})(q^{1/2+m+\epsilon};q)_{\infty}(q^{1/2+m-2\sigma};q)_{\infty}}
\end{pmatrix},\quad m>0,\\
\overline{g\!f}_{-n}=&\begin{pmatrix}
\frac{s_0/s_{\infty}t^{n-\sigma-1/4}q^{2n^2-4n\sigma-1/4}}{(q^{1-2\sigma};q)_{n-1/2}^2(q;q)_{n-1/2}^2}
\frac{(q^{1/2-n+\epsilon};q)_{\infty}(q^{1/2-n+2\sigma};q)_{\infty}}{(1-q^{\epsilon})(q^{1/2+n+\epsilon};q)_{\infty}(q^{1/2+n-2\sigma};q)_{\infty}}
\\
\frac{s_{\infty}/s_0t^{n+\sigma-1/4}q^{2n^2-n+(4n+2)\sigma-3/4}}{(q^{2\sigma};q)_{n+1/2}^2(q;q)_{n-1/2}^2}
\frac{(q^{1/2-n+\epsilon};q)_{\infty}(q^{1/2-n-2\sigma})_{\infty}}{(1-q^{\epsilon})(1-q^{1/2+n+\epsilon};q)_{\infty}(q^{1/2+n+2\sigma};q)_{\infty}}
\end{pmatrix},\quad n>0.
\end{align}
These expressions can be simplified using the following identities:
\begin{equation}
(u;q)_n=(u;q)_{\infty}/(uq^n;q)_{\infty},\qquad
(u;q)_n=q^{n(n-1)/2}u^n(-1)^n(u^{-1}q^{1-n};q)_{\infty}/(u^{-1}q;q)_{\infty}.
\end{equation}
The result is
\begin{equation}
\overline{g\!f}_m=
\begin{pmatrix}
\frac{s_{\infty}/s_0t^{1/4+m+\sigma}(q^{1-2\sigma};q)_{\infty}}{q^{m^2+2m\sigma-\sigma}(q^{2\sigma};q)_{\infty}} \\
-\frac{s_0/s_{\infty}t^{1/4+m-\sigma}(q^{2\sigma};q)_{\infty}}{q^{-1/2+m^2-2m\sigma+\sigma}(q^{1-2\sigma};q)_{\infty}}
\end{pmatrix}, \quad
\overline{g\!f}_{-n}=
\begin{pmatrix}
\frac{s_0/s_{\infty}t^{-1/4+n-\sigma}(q^{2\sigma};q)_{\infty}}{q^{-n^2+2n\sigma+\sigma}(q^{1-2\sigma};q)_{\infty}} \\
-\frac{s_{\infty}/s_0t^{-1/4+n+\sigma}(q^{1-2\sigma};q)_{\infty}}{q^{1/2-n^2-2n\sigma-\sigma}(q^{2\sigma};q)_{\infty}} 
\end{pmatrix},\quad m,n>0.
\end{equation}
Now, we introduce a new variable
\begin{equation}
\label{eq:Sdef}
S=i\frac{s_{\infty}}{s_0}q^{-1/4+\sigma}\frac{(q^{1-2\sigma};q)_{\infty}}{(q^{2\sigma};q)_{\infty}}.
\end{equation}
It has a simple transformation property under the B\"acklund transformation, namely, \(\tilde{S}=-S^{-1}\).
Using this property, \(h_{k,\alpha}\) can be written as
\begin{align}
\label{eq:gfS}
\overline{g\!f}_{m,\alpha} & = -i S^{\alpha}t^{1/4+m+\alpha\sigma}q^{1/4-m^2-2m\alpha\sigma},\quad m>0,\\
\overline{g\!f}_{-n,\alpha} & = i S^{-\alpha}t^{-1/4+n-\alpha\sigma}q^{-1/4+n^2-2n\alpha\sigma},\quad n>0.
\end{align}

Now we can compare the formulas \eqref{eq:Vaxy} and \eqref{eq:Vamn}:
\begin{equation}
\label{eq:diagramIdentity}
e^{-\tau Q_{\alpha}}\sum_{s\in Y_{\alpha}}e^{\tau(y(s)-x(s))} =
\left(\sum_{i=1}^{\#_{\alpha}n} \frac{e^{\tau n_i}}{e^{\tau/2}-e^{-\tau/2}}-
\sum_{i=1}^{\#_{\alpha}m} \frac{e^{-\tau m_i}}{e^{\tau/2}-e^{-\tau/2}}\right)+
\frac{1-e^{-\tau Q_{\alpha}}}{(e^{\tau/2}-e^{-\tau/2})^2}.
\end{equation}
Introducing the following functions of the Young diagram
\begin{equation}
\label{eq:NTdef}
N_{\alpha}=|Y_{\alpha}|=\sum_{s\in Y_{\alpha}} 1,\qquad
T_{\alpha}=||Y_{\alpha}||=\sum_{s\in Y_{\alpha}}(x(s)-y(s))
\end{equation}
and computing the first few coefficients of expansions of \eqref{eq:diagramIdentity} in \(\tau\), we get
\begin{equation}
\sum_{i=1}^{\#_{\alpha}m} 1-\sum_{i=1}^{\#_{\alpha}n} 1=Q_{\alpha},\quad
\sum_{i=1}^{\#_{\alpha}m} m_i + \sum_{i=1}^{\#_{\alpha}n} n_i=\frac{Q_{\alpha}^2}{2}+N_{\alpha},\quad
\sum_{i=1}^{\#_{\alpha}m} m_i^2 - \sum_{i=1}^{\#_{\alpha}n} n_i^2=\frac{4Q_{\alpha}^3-Q_{\alpha}}{12}+2Q_{\alpha}N_{\alpha}+2T_{\alpha}.
\end{equation}
Using the above identities, we can compute \(\overline{Z}_1^{-1}\) as
\begin{multline}
\label{eq:Z1}
\overline{Z}_1^{-1}=\prod_{(k,\alpha)\in \mathtt{M}}\overline{g\!f}_{k,\alpha}=
\prod_{\alpha=\pm1}S^{\alpha Q_{\alpha}}t^{Q_{\alpha}^2/2+N_{\alpha}+\alpha\sigma Q_{\alpha}}q^{-2Q_{\alpha}N_{\alpha}-2T_{\alpha}-2(N_{\alpha}+Q_{\alpha}^2/2)\alpha\sigma}
=\\=
S^{2Q}t^{N_++N_-+(Q+\sigma)^2-\sigma^2}q^{2(Q+\sigma)(N_--N_+)}q^{-2T_--2T_+}.
\end{multline}
Combining together \eqref{eq:Z0}, \eqref{eq:tauNek1}, \eqref{eq:Z1}, we can now prove the following
\begin{theorem}
Explicit expression of the tau function \eqref{eq:widom} is given by
\begin{multline}
\label{eq:tauExpansion}
\tau(t)=t^{-\sigma^2}\prod_{\epsilon=\pm}(q^{1+2\epsilon\sigma};q,q)_{\infty}\\
\times\sum_{Q\in \mathbb{Z}}S^{2Q}\sum_{Y_{\pm}\in \mathbb{Y}}
\frac{t^{|\vec{Y}|+(\sigma+Q)^2}q^{2(Q+\sigma)(N_--N_+)}q^{-2T_--2T_+}}
{\prod_{\epsilon=\pm1}(q^{1+2\epsilon(\sigma+Q)};q,q)_{\infty}}Z_{vec}(\sigma+Q|\vec{Y}),
\end{multline}
where \(S\) is defined by \eqref{eq:Sdef}, \(Z_{vec}(\sigma+Q|\vec{Y})\) is given by \eqref{eq:zvecdef}, and \(N_{\alpha}\), \(T_{\alpha}\) are determined by \eqref{eq:NTdef}.
\end{theorem}
The B\"acklund-transformed tau function is given by
\begin{multline}
\tilde{\tau}(t)=t^{-(\sigma-1/2)^2}\prod_{\epsilon=\pm}(q^{1-\epsilon+2\epsilon\sigma};q,q)_{\infty}\\
\times\sum_{Q\in \mathbb{Z}}(-S)^{2Q}\sum_{Y_{\pm}\in \mathbb{Y}}
\frac{t^{|\vec{Y}|+(\sigma-1/2+Q)^2}q^{2(Q-1/2+\sigma)(N_--N_+)}q^{-2T_--2T_+}}
{\prod_{\epsilon=\pm1}(q^{1-\epsilon+2\epsilon(\sigma+Q)};q,q)_{\infty}}
Z_{vec}(\sigma-1/2+Q|\vec{Y})
=\\=
t^{-(\sigma-1/2)^2}\prod_{\epsilon=\pm}(q^{1-\epsilon+2\epsilon\sigma};q,q)_{\infty}
\frac{(q^{2\sigma};q)_{\infty}}{(q^{1-2\sigma};q)_{\infty}}S\\
\times\sum_{Q\in \frac12+\mathbb{Z}}S^{2Q}\sum_{Y_{\pm}\in \mathbb{Y}}
\frac{t^{|\vec{Y}|+(\sigma+Q)^2}q^{2(Q+\sigma)(N_--N_+)}q^{-2T_--2T_+}}
{\prod_{\epsilon=\pm1}(q^{1+2\epsilon(\sigma+Q)};q,q)_{\infty}}
Z_{vec}(\sigma+Q|\vec{Y}).
\end{multline}
Let us now introduce the renormalized tau functions analogous to \eqref{eq:bigtau}:
\begin{equation}
\label{eq:tauRenormalizationM2}
\mathcal{T}_0^{[-2]}(t)=\frac{q^{-\sigma^2}(qt)^{\sigma^2}}{\prod_{\epsilon=\pm1}(q^{1+2\epsilon\sigma};q,q)_{\infty}}\tau(t),\quad
\mathcal{T}_{\frac12}^{[-2]}(t)=-i \frac{s_0}{s_{\infty}}\frac{q^{-\sigma^2}(qt)^{(\sigma-1/2)^2}}{\prod_{\epsilon=\pm1}(q^{1+2\epsilon\sigma};q,q)_{\infty}}\tilde{\tau}(t).
\end{equation}
These tau functions satisfy the bilinear identities
\begin{equation}
(1-t)\mathcal{T}^{[-2]}_\mu(tq)\mathcal{T}^{[-2]}_\mu(t/q)=
\mathcal{T}^{[-2]}_\mu(t)^2-\sqrt{t}\,\mathcal{T}^{[-2]}_{\frac12-\mu}(t)^2,\quad \mu=0, \frac12,
\end{equation}
and their explicit expressions read
\begin{multline}
\label{eq:tauk}
\mathcal{T}^{[k]}_\mu(t)=\mathcal{T}_{\mu}^{[k]}\left(S^2,u;t,q\right)
=\\=
\sum_{Q\in \mu+\mathbb{Z}}S^{2Q}\sum_{Y_{\pm}\in \mathbb{Y}}
\frac{t^{|\vec{Y}|+(\sigma+Q)^2}q^{k(Q+\sigma)(|Y_+|-|Y_-|)}q^{k(||Y_+||+||Y_-||)}}
{\prod_{\epsilon=\pm1}(q^{1+2\epsilon(\sigma+Q)};q,q)_{\infty}}Z_{vec}(\sigma+Q|\vec{Y})
=\\=
\sum_{Q\in \mu+\mathbb{Z}}S^{2Q} \mathcal{Z}^{[k]}\left(uq^Q;t,q\right).
\end{multline}

The Painlev\'e transcendent \eqref{eq:gformula} can be expressed in terms of these tau functions as
\begin{equation}
\label{eq:gT}
\mathsf{g}(t)=-i t^{1/4} \frac{\mathcal{T}_0^{[-2]}(t)}{\mathcal{T}_{\frac12}^{[-2]}(t)}.
\end{equation}

The tau functions \(\mathcal{T}_{\mu}^{[k]}\) for \(k=0\) were first obtained in \cite{Bershtein:2016aef}, their generalizations for \(k=1, 2\) were obtained in \cite{Bershtein:2018srt}.
The gauge theory expressions for the instanton partition functions originate from \cite{Iqbal:2003ix,Tachikawa:2004ur,Gottsche:2006bm}.
At the moment, there exist two proofs of the fact that the tau functions \eqref{eq:tauk} solve the \(q\)-Painlev\'e equations, given in \cite{2019SIGMA..15..074M,Shchechkin:2020ryb}.

One may also introduce the quantity
\begin{equation}
\label{eq:tauRenorm}
\mathcal{T}_{\mu}=(qt;q,q)_{\infty} \mathcal{T}^{[-2]}_{\mu}
\end{equation}
analogous to \eqref{eq:bigtau}.
It satisfies the standard bilinear identities for \(k=0\):
\begin{equation}
\mathcal{T}^{[-2]}_\mu(tq)\mathcal{T}^{[-2]}_\mu(t/q)=
\mathcal{T}^{[-2]}_\mu(t)^2-\sqrt{t}\,\mathcal{T}^{[-2]}_{\frac12-\mu}(t)^2.
\end{equation}
We show in the next section that, indeed, \(\mathcal{T}_{\mu}(t)\) can be identified with \(\mathcal{T}^{[0]}_{\mu}(t)\), and therefore it also has an explicit combinatorial representation.
This relation is also known from \cite{Bershtein:2018srt}.

\section{Alternative Fredholm determinants and combinatorial expansions}
\label{sec:alternativeDet}
In this section, we obtain Fredholm determinant representations for the tau functions \eqref{eq:tauk} for \(k=-1,0,1,2\) and prove the following
\begin{theorem}
\label{thm:tauIdentity}
The tau functions \(\mathcal{T}_{\mu}^{[0]}\) and \(\mathcal{T}_{\mu}^{[-2]}\) are related by
\begin{equation}
\mathcal{T}_{\mu}^{[0]}=(qt;q,q)_{\infty}\mathcal{T}_{\mu}^{[-2]}.
\end{equation}
\end{theorem}

As for the \(k=\pm 1\) case, it was found in \cite{Bershtein:2018srt} that this case corresponds to the \(q\)-Painlev\'e \(A_7^{(1)}\) equation.
Since the tau function with \(k=\pm 1\) also appears in our setting, we have also obtained its Fredholm determinant representation.
Nevertheless, we did not study its Riemann-Hilbert problem, which is needed to provide a rigorous proof of this representation.

\subsection{Alternative linear systems}
The original factorization of the jump matrix is given by \eqref{eq:Y0def}, \eqref{eq:Yinfdef}:
\begin{align}
\Psi_-(z)=\Psi^{[2]}_-(z)=
r_0\left(s_0/(qt)^{\sigma}\right)^{\boldsymbol{\sigma}_{\mathbf{3}}}
&\begin{pmatrix}
j_2(q^{1-2\sigma},qt/z) & \frac{qt/z}{q^{\sigma}-q^{1-\sigma}}j_2(q^{2-2\sigma},qt/z)\\
\frac{-1}{q^{\sigma}-q^{-\sigma}}j_2(q^{1+2\sigma},qt/z) & j_2(q^{2\sigma},qt/z)
\end{pmatrix}t^{\frac1{4}\boldsymbol{\sigma}_{\mathbf{3}}},
\\
\Psi_+(z)=\Psi_+^{[0]}(z)=
r_\infty
(s_\infty)^{\boldsymbol{\sigma}_{\mathbf{3}}}
&\begin{pmatrix}
j_0(q^{2\sigma},z) & \frac{-1}{q^{\sigma}-q^{-\sigma}}j_0(q^{1+2\sigma},z)\\
\frac{-z}{q^{\sigma}-q^{1-\sigma}}j_0(q^{2-2\sigma},z) & j_0(q^{1-2\sigma},z)
\end{pmatrix}
q^{-\frac1{4}\boldsymbol{\sigma}_{\mathbf{3}}}.
\end{align}
Let us denote the tau function defined for such a factorization by \(\tau^{[-2]}\).
With this notation, the formula \eqref{eq:Fredholm} can be written more explicitly as
\begin{equation}
\tau^{[-2]}=\tau^{[0,2]} = \tau_W\left[J^{[0,2]}\right] = \det(\mathbb{I}-a^{[0]}\cdot d^{[2]})
\end{equation}
The inverse matrices to \(\Psi^{[2]}_-\), \(\Psi^{[0]}_+\) are
\begin{align}
\Psi_-^{[2]}(z)^{-1}=
t^{-\frac1{4}\boldsymbol{\sigma}_{\mathbf{3}}}&\begin{pmatrix}
j_0(q^{2\sigma},qt/z) & \frac{-qt/z}{q^{\sigma}-q^{1-\sigma}}j_0(q^{2-2\sigma},qt/z)\\
\frac{1}{q^{\sigma}-q^{-\sigma}}j_0(q^{1+2\sigma},qt/z) & j_0(q^{1-2\sigma},qt/z)
\end{pmatrix}
r_0^{-1}\left((qt)^{\sigma}/s_0\right)^{\boldsymbol{\sigma}_{\mathbf{3}}},
\\
\Psi_+^{[0]}(z)^{-1}=
q^{\frac1{4}\boldsymbol{\sigma}_{\mathbf{3}}}&\begin{pmatrix}
j_2(q^{1-2\sigma},z) & \frac{1}{q^{\sigma}-q^{-\sigma}}j_2(q^{1+2\sigma},z)\\
\frac{z}{q^{\sigma}-q^{1-\sigma}}j_2(q^{2-2\sigma},z) & j_2(q^{2\sigma},z)
\end{pmatrix}
r_\infty^{-1} (s_\infty)^{-\boldsymbol{\sigma}_{\mathbf{3}}}.
\end{align}
We are also interested in the following matrices:
\begin{align}
\Psi^{[0]}_-(z)=
r_0\left(s_0/(qt)^{\sigma}\right)^{\boldsymbol{\sigma}_{\mathbf{3}}}
&\begin{pmatrix}
j_0(q^{1-2\sigma},qt/z) & \frac{qt/z}{q^{\sigma}-q^{1-\sigma}}j_0(q^{2-2\sigma},qt/z)\\
\frac{-1}{q^{\sigma}-q^{-\sigma}}j_0(q^{1+2\sigma},qt/z) & j_0(q^{2\sigma},qt/z)
\end{pmatrix}t^{\frac1{4}\boldsymbol{\sigma}_{\mathbf{3}}},
\\
\Psi^{[1]}_-(z)=
r_0(s_0/(qt)^{\sigma})^{\boldsymbol{\sigma}_{\boldsymbol{3}}}
&\begin{pmatrix}
j_1(q^{1-2\sigma},q^{3/2}t/z) & j_1(q^{1-2\sigma},q^{1/2}t/z)\\
\frac{1}{q^{-2\sigma}-1}j_1(q^{1+2\sigma},q^{3/2}t/z) &
\frac{1}{1-q^{2\sigma}} j_1(q^{1+2\sigma},q^{1/2}t/z)
\end{pmatrix},
\end{align}
These matrices solve the following linear systems
\begin{equation}
Y_0^{[k]}=z^{\sigma\boldsymbol{\sigma}_{\mathbf{3}}}\Psi_-^{[k]},\qquad
Y_0^{[k]}(qz)=Y_0^{[k]}(z)L_0^{[k]},
\end{equation}
where
\begin{align}
&L_0^{[0]}(z)=\frac1{1-\frac{t}{z}}
\begin{pmatrix}
q^{\sigma} & \frac{\sqrt{t}}{z}\\
\sqrt{t} & q^{-\sigma}
\end{pmatrix}
=
\begin{pmatrix}
q^{-\sigma} & -\frac{\sqrt{t}}{z}\\
-\sqrt{t} & q^{\sigma}
\end{pmatrix}^{-1},
\\
&L_0^{[1]}(z)=
\begin{pmatrix}
0 & -q^{\sigma}\\
q^{-\sigma} & q^{-\sigma}+q^{\sigma}+\frac{t}{z}
\end{pmatrix},
\\
&L_0^{[2]}(z)=
\begin{pmatrix}
q^{\sigma} & \frac{\sqrt{t}}{z}\\
\sqrt{t} & q^{-\sigma}
\end{pmatrix}.
\end{align}
The determinants of these solutions admit simple evaluations
\begin{equation}
\det \Psi_-^{[0]}(z)=r_0^2 \frac1{(qt/z)_{\infty}},\qquad
\det \Psi_-^{[1]}(z)=r_0^2,\qquad
\det \Psi_-^{[2]}(z)=r_0^2 (qt/z)_{\infty}.
\end{equation}

Explicit computations, analogous to Section \ref{sec:matElements}, give the following relations between the generalized matrix elements:
\begin{equation}
d^{[k]}_{nm,\beta\alpha}=\frac{g^{[k]}_{d,n,\beta}f^{[k]}_{d,m,\alpha}}{1-q^{-m-n-\sigma(\alpha-\beta)}},
\qquad
g^{[k]}_{d,n,\beta}=q^{kn^2/2-kn\sigma}g^{[0]}_{d,n,\beta},
\qquad
f^{[k]}_{d,m,\alpha}=q^{-km^2/2-km\sigma}f^{[0]}_{d,m,\alpha}.
\end{equation}
The original matrix elements \eqref{eq:dfg} were given by \(d_{nm,\beta\alpha}=d_{nm,\beta\alpha}^{[2]}\).
Let us consider more general tau function
\begin{equation}
\tau^{[k_1,k_2]} = \det(\mathbb{I}-a^{[k_1]}\cdot d^{[k_2]}).
\end{equation}
In the case \(k_1=0\), such a modification changes the combinatorial expansion \eqref{eq:tauExpansion} corresponding to \(k_2=2\) in the following way:
\begin{multline}
\tau^{[0,k]}(t)=t^{-\sigma^2}\prod_{\epsilon=\pm}(q^{1+2\epsilon\sigma};q,q)_{\infty}\\
\times\sum_{Q\in \mathbb{Z}}S^{2Q}\sum_{Y_{\pm}\in \mathbb{Y}}
\frac{t^{|\vec{Y}|+(\sigma+Q)^2}q^{-k(Q+\sigma)(N_+-N_-)}q^{-k(T_-+T_+)}}
{\prod_{\epsilon=\pm1}(q^{1+2\epsilon(\sigma+Q)};q,q)_{\infty}}Z_{vec}(\sigma+Q|\vec{Y}).
\end{multline}
Moreover, the modification of both matrix elements should give
\begin{equation}
\tau^{[k_1,k_2]}(t)=\tau^{[k_1-k_2]}(t),
\end{equation}
so that in total we have \(k=-2,-1,0,1,2\).
The cases \(k\) and \(-k\) are related by the transformation \((Y_+,Y_-)\mapsto (Y_-^T,Y_+^T)\), \((Q,\sigma)\mapsto (-Q,-\sigma)\).
In this way, we reduce the number of different tau functions to \(k=0,1,2\).
In the next sections, we relate \(k=0\) and \(k=-2\) cases.

\subsection{Generalized Szeg\H{o} formula}
Here, we study the dependence of the Widom determinant on the scalar factor \(\mathfrak{J}(z)\).
Namely, let
\begin{equation}
J_{\mathfrak{t}}=J(\mathfrak{t},z)=\mathfrak{J}(\mathfrak{t},z)J(z)=\mathfrak{J}_+(\mathfrak{t},z)\mathfrak{J}_-(\mathfrak{t},z)^{-1}J(z),
\end{equation}
where \(\mathfrak{t}\) is some parameter that describes an interpolation between \(J(z)\) and \(\mathfrak{J}(z)J(z)\).
We can choose it such that, for example,
\begin{equation}
\mathfrak{J}(0,z)=\mathbb{I}, \quad \mathfrak{J}(1,z)=\mathfrak{J}(z).
\end{equation}

The two factorizations of the jump \(J_{\mathfrak{t}}\) are
\begin{equation}
J(\mathfrak{t},z)=(\mathfrak{J}_+(\mathfrak{t},z)\Phi_+(z))(\mathfrak{J}_-(\mathfrak{t},z)\Phi_-(z))^{-1}=
(\mathfrak{J}_-(\mathfrak{t},z)\Psi_-(z))^{-1}(\mathfrak{J}_+(\mathfrak{t},z)\Psi_+(z)).
\end{equation}
Let us recall Widom's variational formula \cite{WIDOM19761,Cafasso:2017xgn}, which reads
\begin{equation}
\partial_{\mathfrak{t}}\log\tau[J_{\mathfrak{t}}]=\oint \frac{dz}{2\pi i} \tr J(\mathfrak{t},z)^{-1}\partial_{\mathfrak{t}} J(\mathfrak{t},z)
\left(\partial_z\Phi_-(\mathfrak{t},z) \Phi_-(\mathfrak{t},z)^{-1}+ \Psi_+(\mathfrak{t},z)^{-1}\partial_z\Psi_+(\mathfrak{t},z)\right).
\end{equation}
In our case, it can be simplified to
\begin{equation}
\partial_{\mathfrak{t}}\log\tau[J_{\mathfrak{t}}]=\oint \frac{dz}{2\pi i} \partial_{\mathfrak{t}}\log \frac{\mathfrak{J}_+}{\mathfrak{J}_-}\partial_z\left(\log\det \Phi_-+N \log \mathfrak{J}_-+\log\det\Psi_++N\log \mathfrak{J}_+ \right).
\end{equation}
This formula can be integrated back to
\begin{equation}
\label{eq:genSzego}
\log\frac{\tau[\mathfrak{J}J]}{\tau[J]}=\oint \frac{dz}{2\pi i} \left(N\log \mathfrak{J}_+ \partial_z\log \mathfrak{J}_- + \log \mathfrak{J}_+ \partial_z\log\Phi_- - \log \mathfrak{J}_-\partial_z\log\Psi_+\right).
\end{equation}
The first part, quadratic in \(\log \mathfrak{J}\), is the \(N\times N\) Szeg\H{o} formula for the jump \(\mathfrak{J}\), and two other terms are some linear corrections to it.

\subsection{Modification of the determinant}
Now, we notice that using the identity
\begin{equation}
j_2(u,z)=(z;q)_{\infty}j_0(u,z)
\end{equation}
we can relate
\begin{equation}
\Psi_-^{[0]}(t,z)=(qt/z;q)_{\infty}\Psi_-^{[2]}(t,z),
\end{equation}
and therefore the relation between \(k=0\) and \(k=-2\) tau functions is described by the generalized Szeg\H{o} formula \eqref{eq:genSzego}.
To find it, we need to compute
\begin{equation}
\det\Phi_-=r_{\infty}^{-2}r^2(qt/z;q)_{\infty},\qquad \det\Psi_+=r_{\infty}^{2}\frac1{(z;q)_{\infty}}
\end{equation}
and to use \eqref{eq:genSzego} together with
\begin{equation}
\mathfrak{J}_-(z)=\frac1{(qt/z;q)_{\infty}},\qquad \mathfrak{J}_+(z)=1.
\end{equation}
This finally gives
\begin{multline}
\label{eq:SzegoSum}
\log\frac{\tau^{[0]}}{\tau^{[-2]}}=\log\frac{\tau\left[\mathfrak{J}J^{[0,2]}\right]}{\tau\left[J^{[0,2]}\right]}
=
\oint\frac{dz}{2\pi i}\log (qt/z;q)_{\infty}\partial_z\log(z;q)_{\infty}
=\\=
\sum_{n=1}^{\infty}\oint \frac{dz}{2\pi i} \frac{(qt)^n/z^n}{n(1-q^n)}\partial_z \frac{z^n}{n(1-q^n)}
=
\sum_{n=1}^{\infty}\frac{(qt)^n}{n(1-q^n)^2} = \log(qt;q,q)_{\infty},
\end{multline}
which is the desired prefactor from \eqref{eq:bigtau} and from \eqref{eq:tauRenorm}.
Switching to slightly redefined tau functions \eqref{eq:tauk}, we get the statement of Theorem~\ref{thm:tauIdentity}.

\section{Meaning of the tau function and its non-linear connection problem}
\label{sec:connection}
In this section, we find the relation between the local asymptotics of solutions of the linear system and the isomonodromic tau functions \eqref{eq:omegadef}, \eqref{eq:Omegat}, \eqref{eq:Omega1}.
Then, we use these local asymptotics to find \(t\to 1/t\) transformation of the \(q\)-Painlev\'e transcendent \eqref{eq:gduality} and of the isomonodromic tau function \eqref{eq:UpsilonFormula}.
We have also derived the fusion kernels for \(c=\infty\) \eqref{eq:upsilonFormula} and \(c=1\) \eqref{eq:fusionFormula} \(q\)-deformed conformal blocks.

\subsection{Asymptotics of the linear system around singularities}
\label{sec:Yasymp}
We would like to introduce the following vectors
\begin{equation}
\label{eq:omegadef}
\Omega^{(t)}(t)=
r^{-1}Y(t,t)
\begin{pmatrix}
\frac{\mathsf{g}(t)^2+1}{\mathsf{g}(t)} \\
\frac{\mathsf{g}(t)^2+t}{\mathsf{g}(qt)}
\end{pmatrix},\qquad
\Omega^{(1)}(t)=
r^{-1}Y(t,q)
\begin{pmatrix}
-\frac{\mathsf{g}(t)^2+t}{\mathsf{g}(qt)\mathsf{g}(t)} \\ \mathsf{g}(t)^2+1
\end{pmatrix}.
\end{equation}
The evolution equations for these vectors read
\begin{align}
&\Omega^{(t)}(qt)=
r^{-1}Y(t,t)\begin{pmatrix}
\frac{\mathsf{g}(qt)}{\mathsf{g}(t)} & 1\\
t & \frac{\mathsf{g}(t)}{\mathsf{g}(qt)}
\end{pmatrix}
\begin{pmatrix}
\frac{\mathsf{g}(qt)^2+1}{\mathsf{g}(qt)} \\
\mathsf{g}(t)(\mathsf{g}(qt)^2+1)
\end{pmatrix}=
(1+\mathsf{g}(qt)^2)\Omega^{(t)}(t),\\
&\Omega^{(1)}(qt)=
r^{-1}Y(t,q)
\begin{pmatrix}
1 & \frac{qt}{z \mathsf{g}(qt)\mathsf{g}(t)}\\
\mathsf{g}(qt)\mathsf{g}(t) & 1
\end{pmatrix}^{-1}
\begin{pmatrix}
-\frac{\mathsf{g}(t)(\mathsf{g}(qt)^2+1)}{\mathsf{g}(qt)} \\ \mathsf{g}(qt)^2+1
\end{pmatrix}
=\frac{1+\mathsf{g}(qt)^2}{1-t} \Omega^{(1)}(t).
\end{align}
In this way, we see that the ratios \(\Omega_1^{(t)}/\Omega_2^{(t)}\) and \(\Omega_1^{(1)}/\Omega_2^{(2)}\) are invariant under the isomonodromic evolution.

Let us now obtain an explicit expression for these vectors.
The formulas \eqref{eq:PhiPlusOmegat} and \eqref{eq:Y0def} give
\begin{equation}
\Omega^{(t)}(t)=
r^{-1}\alpha(t) Y_0(t,t) \begin{pmatrix}
q^{\sigma} /\sqrt[4]{t}\\
\sqrt[4]{t}
\end{pmatrix}
= \frac{r_0}{r} \alpha(t) (s_0)^{\boldsymbol{\sigma}_{\mathbf{3}}} \mathcal{Y}_0(q)
\begin{pmatrix}q^{\sigma} \\ 1\end{pmatrix}.
\end{equation}
To find \(\alpha(t)\), we use the equations \eqref{eq:PhiPlusOmegat}, \eqref{eq:xidef}:
\begin{equation}
\label{eq:alphabeta1}
\frac{r}{r_0}\Phi_+(t,qt)^{-1}
\begin{pmatrix}-q^{-\sigma} /\sqrt[4]{t}\\\sqrt[4]{t}\end{pmatrix}=
\beta(t)\begin{pmatrix}\frac{-1}{\mathsf{g}(qt)}\\ \mathsf{g}(t)\end{pmatrix},
\qquad
\Phi_+(t,qt)^T\begin{pmatrix}\sqrt[4]{t} \\ q^{-\sigma}/\sqrt[4]{t}\end{pmatrix}=
\alpha(t) \begin{pmatrix}\mathsf{g}(qt) \\ \frac{1}{\mathsf{g}(t)}\end{pmatrix}.
\end{equation}
The second equation can be transformed as
\begin{equation}
\Phi_+(t,qt)^T \boldsymbol{\sigma}_{\mathbf{2}}
\begin{pmatrix}-q^{-\sigma} /\sqrt[4]{t}\\\sqrt[4]{t}\end{pmatrix}
=
\alpha(t)\frac{\mathsf{g}(qt)}{\mathsf{g}(t)}\boldsymbol{\sigma}_{\mathbf{2}}\begin{pmatrix}\frac{-1}{\mathsf{g}(qt)}\\ \mathsf{g}(t)\end{pmatrix},
\end{equation}
and then
\begin{equation}
\label{eq:alpha1}
\det\Phi_+(t,qt) \cdot \Phi_+(t,qt)^{-1}
\begin{pmatrix}-q^{-\sigma} /\sqrt[4]{t}\\\sqrt[4]{t}\end{pmatrix}
=
\alpha(t)\frac{\mathsf{g}(qt)}{\mathsf{g}(t)}\begin{pmatrix}\frac{-1}{\mathsf{g}(qt)}\\ \mathsf{g}(t)\end{pmatrix}.
\end{equation}
Now, we compare \eqref{eq:alpha1} with \eqref{eq:alphabeta1}, use \eqref{eq:detsolution}, \eqref{eq:detY0inf} to find that \(\det\Phi_+(t,qt)=\frac{r^2}{r_0^2}\frac1{(qt;q)_{\infty}}\) and finally get
\begin{equation}
\beta(t)=\frac{\mathsf{g}(qt)}{\mathsf{g}(t)}\frac{r_0}{r}(qt;q)_{\infty}\alpha(t).
\end{equation}

To complete the computation, it remains to find \(\mathcal{Y}_0(q)\) \eqref{eq:Y0def}.
The values of the \(q\)-Bessel functions are given by
\begin{equation}
j_0(u,q)=\frac{\theta_p(u;q)}{(u;q)_{\infty}(q;q)_{\infty}}, \qquad
j_2(u,q)=\frac{\theta_p(u;q)}{(u;q)_{\infty}}.
\end{equation}
Here, \(\theta_p\) denotes the partial theta function, see, e.g., \cite{Warnaar2003}:
\begin{equation}
\theta_p(u;q)=\sum_{n=0}^{\infty}(-1)^nq^{n(n-1)/2}u^n,
\end{equation}
which solves the difference relation
\begin{equation}
u\theta_p(qu;q)+\theta_p(u;q)=1.
\end{equation}
It is related to the full theta function by
\begin{equation}
\theta_p(u;q)+\theta_p(q/u;q)-1=\theta_1(u;q),
\end{equation}
where
\begin{equation}
\theta_1(u;q)=
(u;q)_{\infty}(q/u;q)_{\infty}(q;q)_{\infty}=(q;q)_{\infty}\theta(u;q)=
\sum_{n\in \mathbb{Z}}(-1)^nq^{n(n-1)/2}u^n.
\end{equation}
In this way, we obtain
\begin{multline}
\mathcal{Y}_0(q) = q^{-\sigma \boldsymbol{\sigma}_{\mathbf{3}}}
\begin{pmatrix}
\frac{\theta_p(q^{1-2\sigma};q)}{(q^{1-2\sigma};q)_{\infty}} & \frac{q}{q^{\sigma}-q^{1-\sigma}}\frac{\theta_p(q^{2-2\sigma};q)}{(q^{2-2\sigma};q)_{\infty}}\\
\frac{-1}{q^{\sigma}-q^{-\sigma}}\frac{\theta_p(q^{1+2\sigma};q)}{(q^{1+2\sigma};q)_{\infty}} & \frac{\theta_p(q^{2\sigma};q)}{(q^{2\sigma};q)_{\infty}}
\end{pmatrix}= \\ =
q^{-\sigma \boldsymbol{\sigma}_{\mathbf{3}}}
\begin{pmatrix}
\frac{\theta_p(q^{1-2\sigma};q)}{(q^{1-2\sigma};q)_{\infty}} & q^{1-\sigma}\frac{\theta_p(q^{2-2\sigma};q)}{(q^{1-2\sigma};q)_{\infty}}\\
q^{\sigma}\frac{\theta_p(q^{1+2\sigma};q)}{(q^{2\sigma};q)_{\infty}} & \frac{\theta_p(q^{2\sigma};q)}{(q^{2\sigma};q)_{\infty}}
\end{pmatrix}.
\end{multline}
The action on the vector of our interest simplifies this expression to
\begin{equation}
\mathcal{Y}_0(q)
\begin{pmatrix}
q^{\sigma} \\ 1
\end{pmatrix}=
\begin{pmatrix}
\frac{1}{(q^{1-2\sigma};q)_{\infty}} \\ \frac{q^{\sigma}}{(q^{2\sigma};q)_{\infty}}
\end{pmatrix},
\end{equation}
so that we finally get
\begin{equation}
\label{eq:Omegat}
\Omega^{(t)}(t)=
\frac{\tau(qt)}{\tau(t)}\frac{\mathsf{g}(t)}{\mathsf{g}(qt)}\frac1{(qt;q)_{\infty}}
\begin{pmatrix}
\frac{s_0}{(q^{1-2\sigma};q)_{\infty}} \\ \frac{s_0^{-1} q^{\sigma}}{(q^{2\sigma};q)_{\infty}}
\end{pmatrix}.
\end{equation}
\subsubsection*{Singularity at \(z=1\)}
To find the other vector, \(\Omega^{(1)}(t)\), we use the following trick.
First, we notice that the equation defining \(s_{\infty}\) \eqref{eq:fors2} can be written as
\begin{equation}
\begin{pmatrix}
1 & q^{-\sigma}
\end{pmatrix}
\mathcal{Y}_{\infty}(q)^{-1}
\begin{pmatrix}
s_{\infty}(t)^{-1} & 0\\
0 & s_{\infty}(t)
\end{pmatrix}
\Omega^{(1)}(t)=0.
\end{equation}
The explicit expression for \(\mathcal{Y}_\infty(q)^{-1}\) reads
\begin{equation}
\mathcal{Y}_{\infty}(q)^{-1}=
\begin{pmatrix}
\frac{\theta_p(q^{1-2\sigma};q)}{(q^{1-2\sigma};q)_{\infty}} & \frac{q^{\sigma} \theta_p(q^{1+2\sigma};q)}{(q^{2\sigma};q)_{\infty}}\\
\frac{q^{1-\sigma}\theta_p(q^{2-2\sigma};q)}{(q^{1-2\sigma};q)_{\infty}} & \frac{\theta_p(q^{2\sigma};q)}{(q^{2\sigma};q)_{\infty}}
\end{pmatrix} q^{(1/4-\sigma)\boldsymbol{\sigma}_{\mathbf{3}}}.
\end{equation}
Its action on the co-vector is given by
\begin{equation}
\begin{pmatrix}1 & q^{-\sigma}\end{pmatrix}\mathcal{Y}_{\infty}(q)^{-1}=
\begin{pmatrix}\frac{q^{1/4-\sigma}}{(q^{1-2\sigma};q)_{\infty}} &
\frac{q^{-1/4}}{(q^{2\sigma};q)_{\infty}}\end{pmatrix}.
\end{equation}
Therefore, the vectors \(\Omega^{(1)}(t)\), \(\Omega^{(t)}(t)\) satisfy the following equations:
\begin{equation}
\label{eq:omegaequations}
\frac{\Omega^{(1)}_1(t)}{\Omega^{(1)}_2(t)}=-s_{\infty}^2q^{\sigma-1/2}\frac{(q^{1-2\sigma};q)_{\infty}}{(q^{2\sigma};q)_{\infty}},\qquad
\frac{\Omega^{(t)}_1(t)}{\Omega^{(t)}_2(t)}=s_0^2q^{-\sigma}\frac{(q^{2\sigma};q)_{\infty}}{(q^{1-2\sigma};q)_{\infty}}.
\end{equation}
\subsubsection*{Symmetry of the linear system}
Now, we notice that if \(Y(t,z)\) solves the system \eqref{eq:linear}, then the expression
\begin{equation}
Y^d(t,z)=\frac{(qt/z;q)_{\infty}}{(z;q)_{\infty}}
\begin{pmatrix} 0 & -i/b\\ ib & 0 \end{pmatrix}
Y(t,qt/z)
\begin{pmatrix} 0 & -i/\mathsf{g}(t)\\ i\mathsf{g}(t) & 0 \end{pmatrix}
\end{equation}
solves the same system:
\begin{equation}
Y^d(t,q z)=Y^d(t,z)L(t,z).
\end{equation}
Since it has the same determinant and the same diagonal monodromy as \(Y(t,z)\), we can use 1-parametric freedom to choose \(b\) such that \(Y^d\) becomes equal to \(Y\).
This can be done by applying this symmetry to vectors defined in \eqref{eq:omegadef}:
\begin{multline}
\Omega^{(1)}(t)=\frac{(t;q)_{\infty}}{(q;q)_{\infty}}
\begin{pmatrix} 0 & -i/b\\ ib & 0 \end{pmatrix}
Y(t,t)
\begin{pmatrix} 0 & -i/\mathsf{g}(t)\\ i\mathsf{g}(t) & 0 \end{pmatrix}
\begin{pmatrix}
-\frac{\mathsf{g}(t)^2+t}{\mathsf{g}(qt)\mathsf{g}(t)} \\ \mathsf{g}(t)^2+1
\end{pmatrix}=
\\=
\frac{(t;q)_{\infty}}{(q;q)_{\infty}}
\begin{pmatrix} 0 & -1/b\\ b & 0 \end{pmatrix} \Omega^{(t)}(t)
=
\frac{(t;q)_{\infty}}{(q;q)_{\infty}}
\begin{pmatrix} -\Omega_2^{(t)}(t)/b\\ \Omega_1^{(t)}(t)b \end{pmatrix}.
\end{multline}
Comparing the last expression with \eqref{eq:omegaequations}, we find
\begin{equation}
-s_{\infty}^2q^{\sigma-1/2}\frac{(q^{1-2\sigma};q)_{\infty}}{(q^{2\sigma};q)_{\infty}}=
\frac{\Omega^{(1)}_1(t)}{\Omega^{(1)}_2(t)}=
-b^{-2}\frac{\Omega^{(t)}_2(t)}{\Omega^{(t)}_1(t)}=
-b^{-2}s_0^{-2}q^{\sigma}\frac{(q^{1-2\sigma};q)_{\infty}}{(q^{2\sigma};q)_{\infty}},
\end{equation}
which gives us an equation for \(b\).
We can choose its solution to be consistent with the exactly solvable case \eqref{eq:gAlgebraic}:
\begin{equation}
b=\frac{q^{1/4}}{s_0s_{\infty}}.
\end{equation}
Taking this into account, we finally get
\begin{equation}
\label{eq:Omega1}
\Omega^{(1)}(t)=
\frac{\tau(qt)}{\tau(t)}\frac{\mathsf{g}(t)}{\mathsf{g}(qt)}\frac{1-t}{(q;q)_{\infty}}
\begin{pmatrix}
-\frac{s_\infty q^{\sigma-1/4}}{(q^{2\sigma};q)_{\infty}} \\ \frac{s_\infty^{-1}q^{1/4}}{(q^{1-2\sigma};q)_{\infty}}
\end{pmatrix}.
\end{equation}

\subsection{Tau functions of the algebraic solution}
Comparing \eqref{eq:Omegat}, \eqref{eq:Omega1} with \eqref{eq:Y0exact}, \eqref{eq:Yinfexact} using \eqref{eq:omegadef}, and also the fact that for the algebraic solution \(\tau(t)=\tilde{\tau}(t)\), we find that
\begin{equation}
\tau^{alg.}(t)=\tilde{\tau}^{alg.}(t)=
\frac{(-\sqrt{qt};\sqrt{q},\sqrt{q})_{\infty}}{(qt;q,q)_{\infty}}=
\frac1{(\sqrt{qt};\sqrt{q},\sqrt{q})_{\infty}},
\end{equation}
which was also found in \cite{Bershtein:2018srt}.

Differently normalized versions of these tau functions \eqref{eq:tauRenormalizationM2} are
\begin{equation}
\mathcal{T}_0^{\substack{alg.\\ [-2]}}(t)=\frac{1}{(q^{1/2};q,q)_{\infty}(q^{3/2};q,q)_{\infty}}\frac{t^{\frac1{16}}}{(\sqrt{qt};\sqrt{q},\sqrt{q})_{\infty}} =
\frac{C^{alg.}(q)t^{\frac1{16}}}{(\sqrt{qt};\sqrt{q},\sqrt{q})_{\infty}} =
i\mathcal{T}_{\frac12}^{\substack{alg.\\ [-2]}}(t).
\end{equation}
Likewise, for zero Chern-Simons level we have
\begin{equation}
\mathcal{T}_0^{\substack{alg.\\ [0]}}(t)=
i\mathcal{T}_{\frac12}^{\substack{alg.\\ [0]}}(t)=
C^{alg.}(q)t^{\frac1{16}}(-\sqrt{qt};\sqrt{q},\sqrt{q})_{\infty}.
\end{equation}
The above solutions were written for the case \(S^2=-1, \sigma=\frac{1}{4}\).
It is more convenient, however, to consider their analytic continuation that corresponds to \(S^2=1, \sigma=\frac{1}{4}\):
\begin{equation}
\label{eq:TauExact}
\mathcal{T}_0^{\substack{alg.'\\ [0]}}(t)=
\mathcal{T}_{\frac12}^{\substack{alg.'\\ [0]}}(t)=
C^{alg.}(q)t^{\frac1{16}}(\sqrt{qt};\sqrt{q},\sqrt{q})_{\infty}.
\end{equation}

Such tau functions were also obtained in the context of representation theory in \cite{2020SIGMA..16..077B}.

\subsection{Connection problem}
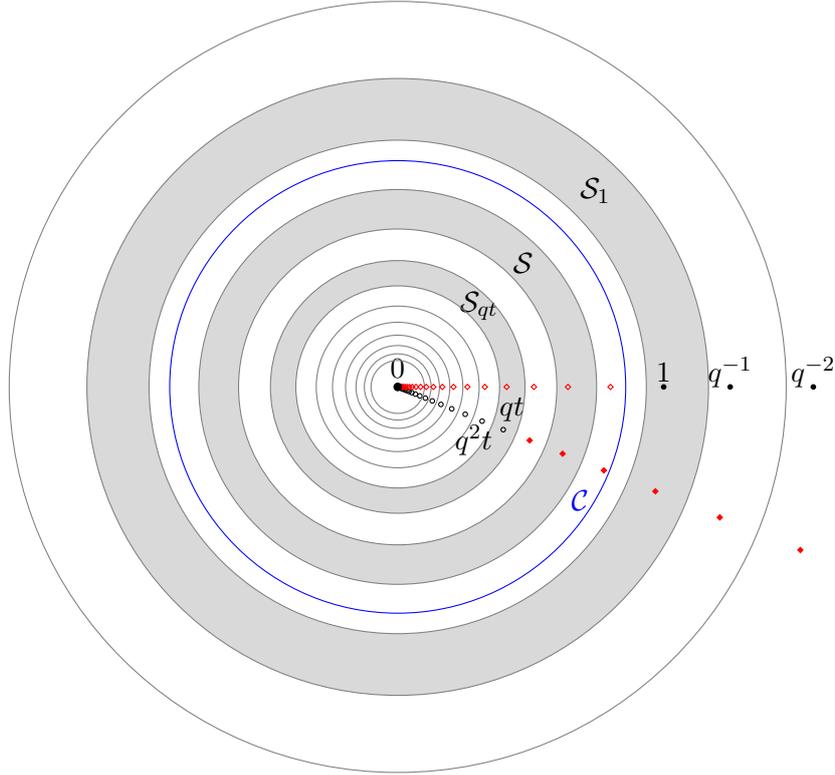
\begin{figure}[ht!]
\begin{center}
\begin{tikzpicture}[scale=1]
\draw[radius=3,start angle=0,end angle=-360,color=blue](3,0)arc;
\path[fill=gray!30!white] (2.61719,0) arc [radius=2.61719,start angle=0,end angle=360] -- (2.09375,0) arc[radius=2.09375,start angle=0,end angle=-360] -- cycle;
\node at (1.65,1.65) {\(\mathcal{S}\)};
\path[fill=gray!30!white] (1.34,0) arc [radius=1.34,start angle=0,end angle=360] -- (1.675,0) arc[radius=1.675,start angle=0,end angle=-360] -- cycle;
\node at (1.07,1.07) {\(\mathcal{S}_{qt}\)};
\path[fill=gray!30!white] (3.27148,0) arc [radius=3.27148,start angle=0,end angle=360] -- (4.08936,0) arc[radius=4.08936,start angle=0,end angle=-360] -- cycle;
\node at (2.6,2.6) {\(\mathcal{S}_1\)};
\node at (2.4,-1.5){\color{blue}\(\mathcal{C}\)};
\node at (0,0.25){\(0\)};
\foreach \x in {1.5, 1.2, 0.96, 0.768, 0.6144, 0.49152, 0.393216, 0.314573, 0.251658, 0.201327, 0.161061, 0.128849, 0.103079, 0.0824634, 0.0659707, 0.0527766} {\draw[radius=0.03] ({\x*sqrt(6/7)},{-\x*sqrt(1/7)}) circle;};
\foreach \x in {5.11169, 4.08936, 3.27148, 2.61719, 2.09375, 1.675, 1.34, 1.072, 0.8576, 0.68608, 0.548864, 0.439091, 0.351273} {\draw[radius=\x, thin,gray](0,0)circle;};
\node at (1.5,-0.3) {\(qt\)};
\node at (1,-0.7) {\(q^2t\)};
\foreach \x in {3.5, 4.375, 5.46875} {\draw[fill, radius=0.03] (\x,0) circle;};
\foreach \x in {2.8, 2.24, 1.792, 1.4336, 1.14688, 0.917504, 0.734003, 0.587203, 0.469762, 0.37581, 0.300648, 0.240518, 0.192415, 0.153932, 0.123145, 0.0985162, 0.078813, 0.0630504, 0.0504403, 0.0403523} {\draw (\x,0) pic{zeror};};
\foreach \x in {1.875, 2.34375, 2.92969, 3.66211, 4.57764, 5.72205} {\draw ({\x*sqrt(6/7)},{-\x*sqrt(1/7)}) pic{poler};};
\node at (3.5,0.2) {\(1\)};
\node at (4.375,0.2) {\(q^{-1}\)};
\node at (5.46875,0.2) {\(q^{-2}\)};
\draw[fill, radius=0.05](0,0) circle;
\end{tikzpicture}
\end{center}
\caption{\label{fig:duality} Zeros (\zerob,\,\zeror) and poles (\poleb,\,\poler) of different solutions.}
\end{figure}

The original solution \(Y(t,z)\) of \eqref{eq:linear} degenerates at \(z=tq^n, n\in \mathbb{Z}_{>0}\) and has poles at \(z=q^n, n\in \mathbb{Z}_{\le 0}\).
These points are represented by {\zerob} and {\poleb}, respectively, in Figures \ref{fig:contours}, \ref{fig:duality}.
This solution serves our purposes well in the region \(|t|<1\).
If we consider \(|t|\ge1\), then any domain \(\mathcal{S}_R^{\epsilon}\) \eqref{eq:SRepsdef} contains either singularity or degeneration point, and \(Y(t,z)\) does not represent a solution of a Riemann-Hilbert problem defining Krichever's monodromy.

\subsubsection*{Complementary solution of the linear system}
To study the region \(|t|>1\), we can consider a complementary solution given by
\begin{equation}
\check{Y}(t,z)=z^{\frac12 \log_q t}T(z)Y(t,z),
\end{equation}
where
\begin{equation}
T(z)=
z^{\check{\sigma} \boldsymbol{\sigma}_{\mathbf{3}}}\check{\lambda}^{\boldsymbol{\sigma}_{\mathbf{3}}}
\begin{pmatrix}
\frac{\theta\left(q^{\check{\sigma}+\sigma}\sqrt{t};q\right)\theta\left(z q^{\check{\sigma}-\sigma}/\sqrt{t};q\right)}{\theta\left(q^{2\sigma};q\right)\theta\left(z/t;q\right)}
& \frac{\theta\left(q^{-\check{\sigma}+\sigma}/\sqrt{t};q\right)\theta\left(z q^{\check{\sigma}+\sigma}/\sqrt{t};q\right)}{\theta\left(q^{2\sigma};q\right)\theta\left(z/t;q\right)}\\
-\frac{\theta\left(q^{\sigma-\check{\sigma}}\sqrt{t};q\right)\theta\left(z q^{-\check{\sigma}-\sigma}/\sqrt{t};q\right)}{\theta\left(q^{-2\check{\sigma}};q\right)\theta\left(z/t;q\right)}
& \frac{\theta\left(q^{\check{\sigma}+\sigma}/\sqrt{t};q\right)\theta\left(z q^{-\check{\sigma}+\sigma}/\sqrt{t};q\right)}{\theta\left(q^{2\check{\sigma}};q\right)\theta\left(z/t;q\right)}
\end{pmatrix}\lambda^{-\boldsymbol{\sigma}_{\mathbf{3}}}z^{-\sigma \boldsymbol{\sigma}_{\mathbf{3}}}.
\end{equation}
One can easily check that
\begin{equation}
T(qz)=T(z)/\sqrt{t},\qquad \det T(z)=\frac{\theta(z;q)}{\theta(z/t;q)}=\frac{(z;q)_{\infty}(q/z;q)_{\infty}}{(tq/z;q)_{\infty}(z/t;q)_{\infty}},
\end{equation}
and therefore \(\check{Y}(t,z)\) solves the same system \eqref{eq:linear}.
Moreover, its determinant is
\begin{equation}
\det \check{Y}(t,z)=r^2 \frac{(q/z;q)_{\infty}}{(z/t;q)_{\infty}}.
\end{equation}
The structure of zeros (\zeror) and poles (\poler) of this determinant is complementary to the one of \(Y(t,z)\), see Figure~\ref{fig:duality}.
This makes \(\check{Y}(t,z)\) a candidate for the second solution that can be used for \(|t|>1\).
To actually show this, we need to ensure that \(\check{Y}(t,z)\) does not have other singularities.
This is achieved by canceling the two potential singularities at \(z=1\) and \(z=qt\):
\begin{equation}
0=\Res_{z=t}\check{Y}(t,qz)=\Res_{z=t}T(qz)Y(t,z)L(t,z),
\qquad T(q)Y(t,q)=T(q)Y(t,1)L(t,1).
\end{equation}

Now, we use \eqref{eq:Ltt} and \eqref{eq:intertwinerInf} to write
\begin{equation}
\Res_{z=t}T(z)Y(t,t)
\begin{pmatrix}
\frac{\mathsf{g}(t)^2+1}{\mathsf{g}(t)}\\
\frac{\mathsf{g}(t)^2+t}{\mathsf{g}(qt)}
\end{pmatrix}=0,\qquad
T(q)Y(t,q)\begin{pmatrix}-\frac{\mathsf{g}(t)^2+t}{\mathsf{g}(qt)\mathsf{g}(t)} \\ \mathsf{g}(t)^2+1\end{pmatrix}=0.
\end{equation}
Using \eqref{eq:omegadef}, these equations can be rewritten as
\begin{equation}
\label{eq:TtT1equation}
\Res_{z=t}T(z)\Omega^{(t)}(t)=0,\qquad
T(1)\Omega^{(1)}(t)=0.
\end{equation}
The first equation from \eqref{eq:TtT1equation} can be written explicitly using \eqref{eq:Omegat}:
\begin{equation}
\label{eq:TtEquation}
\lambda^{-1}t^{-\sigma}\frac{s_0}{(q^{1-2\sigma};q)_{\infty}} \theta\left(q^{\check{\sigma}-\sigma}\sqrt{t};q\right)+
\lambda t^{\sigma} \frac{s_0^{-1} q^{\sigma}}{(q^{2\sigma};q)_{\infty}} \theta\left(q^{-\check{\sigma}+\sigma}/\sqrt{t};q\right)=0,
\end{equation}
whereas the second equation can be written explicitly using \eqref{eq:Omega1}:
\begin{multline}
\label{eq:T1Equation}
-\lambda^{-1}\frac{s_\infty q^{\sigma-1/4}}{(q^{2\sigma};q)_{\infty}}
\theta\left(q^{\check{\sigma}+\sigma}\sqrt{t};q\right)\theta\left(q^{\check{\sigma}-\sigma}/\sqrt{t};q\right)
\\+\lambda  \frac{s_\infty^{-1}q^{1/4}}{(q^{1-2\sigma};q)_{\infty}}
\theta\left(q^{-\check{\sigma}+\sigma}/\sqrt{t};q\right)\theta\left(q^{\check{\sigma}+\sigma}/\sqrt{t};q\right)=0.
\end{multline}
The parameter \(\lambda\) can be found explicitly from the above equations, and the result is given by
\begin{equation}
\lambda^2=-t^{-2\sigma}q^{-\sigma}\frac{s_0^2(q^{2\sigma};q)_{\infty}}{(q^{1-2\sigma};q)_{\infty}}
\frac{\theta\left(q^{\check{\sigma}-\sigma}\sqrt{t};q\right)}{\theta\left(q^{-\check{\sigma}+\sigma}/\sqrt{t};q\right)}.
\end{equation} 
Alternatively, we can exclude \(\lambda\) from \eqref{eq:TtEquation} and \eqref{eq:T1Equation} and get
\begin{equation}
-t^{-2\sigma}\frac{s_0^2(q^{2\sigma};q)_{\infty}}{(q^{1-2\sigma};q)_{\infty}}
\frac{\theta\left(q^{\check{\sigma}-\sigma}\sqrt{t};q\right)}{\theta\left(q^{-\check{\sigma}+\sigma}/\sqrt{t};q\right)}=
\frac{s_\infty^2 q^{2\sigma-1/2}(q^{1-2\sigma};q)_{\infty}}{(q^{2\sigma};q)_{\infty}}
\frac{\theta\left(q^{\check{\sigma}+\sigma}\sqrt{t};q\right)\theta\left(q^{\check{\sigma}-\sigma}/\sqrt{t};q\right)}{\theta\left(q^{-\check{\sigma}+\sigma}/\sqrt{t};q\right)\theta\left(q^{\check{\sigma}+\sigma}/\sqrt{t};q\right)}.
\end{equation}

Using the definition \eqref{eq:Sdef} and simplifying the theta functions therein, we can rewrite the last relation as
\begin{multline}
S^2=(q/t)^{2\sigma}\frac{
\theta\left(q^{\check{\sigma}(t)-\sigma}\sqrt{t};q\right)
\theta\left(q^{-\check{\sigma}(t)-\sigma}\sqrt{t};q\right)}
{\theta\left(q^{\check{\sigma}(t)+\sigma}\sqrt{t};q\right)
\theta\left(q^{-\check{\sigma}(t)+\sigma}\sqrt{t};q\right)}
=\\=
t^{-2\log_q u} u^2 \frac{
\theta\left(\check{u}(t)u^{-1}\sqrt{t};q\right)
\theta\left(\check{u}(t)^{-1}u^{-1}\sqrt{t};q\right)}
{\theta\left(\check{u}(t)u\sqrt{t};q\right)
\theta\left(\check{u}(t)^{-1}u\sqrt{t};q\right)}.
\end{multline}
This equation defines \(\check{\sigma}(t)\) as a function of \(S\), \(\sigma\), \(t\).
Its r.h.s. is \(q^2\)-periodic, but not \(q\)-periodic, which induces the following periodicity property of \(\check{\sigma}\):
\begin{equation}
\check{\sigma}(qt)=\frac12-\check{\sigma}(t),
\end{equation}
which is the B\"acklund transformation \eqref{eq:backlundsigma}.
We can also define another function \(\sigma^i(t)\), which is \(q\)-periodic:
\begin{equation}
\sigma^i(t)=\frac{1-\phi(t)}{4}+\phi(t)\check{\sigma}(t),
\end{equation}
where \(\phi(t)\) is defined everywhere except \(t=q^{n+1/2}, n\in \mathbb{Z}\), and satisfies the conditions
\begin{equation}
\phi(t)^2=1, \qquad \phi(qt)=-\phi(t),\qquad \phi(1/t)=\phi(t).
\end{equation}

\subsubsection*{Dual \(q\)-Painlev\'e transcendent}
We can notice that the equation \eqref{eq:Painleve} has the following involution symmetry that maps solutions to solutions:
\begin{equation}
\mathsf{g}(t)\mapsto \mathsf{g}^i(t)=t^{\frac{1+\phi(t)}{4}}\mathsf{g}(1/t)^{\phi(t)}.
\end{equation}
This mapping has a counterpart acting on the linear system \eqref{eq:Lax}:
\begin{equation}
L(t,z)=t^{\frac12}t^{-\frac1{4}\boldsymbol{\sigma}_{\mathbf{3}}} B^b(z/t)^{\frac{1-\phi(t)}{2}}L^i(1/t,z/t)B^b(qz/t)^{\frac{1-\phi(t)}{2}}t^{\frac1{4}\boldsymbol{\sigma}_{\mathbf{3}}},
\end{equation}
where \(L^i(t,z)\) stands for the matrix \(L(t,z)\) \eqref{eq:Lax} with \(\mathsf{g}(t)\) replaced by \(\mathsf{g}^i(t)\).
One consequence of such a symmetry is that
\begin{equation}
\check{Y}'(t,z)=z^{\frac12 \log_q t} B^b(q)^{\frac{1-\phi(t)}{2}}Y^i(1/t,z/t)B^b(qz/t)^{\frac{1-\phi(t)}{2}}t^{\frac1{4}\boldsymbol{\sigma}_{\mathbf{3}}}
\end{equation}
also solves the system \eqref{eq:linear}.
It has the same determinant as \(\check{Y}(t,z)\), therefore we can use the diagonal freedom in \(\check{Y}(t,z)\) to ensure that
\begin{equation}
\check{Y}'(t,z)=\check{Y}(t,z),
\end{equation}
or equivalently,
\begin{equation}
\label{eq:dualityRelation}
T(z)Y(t,z)=B^b(q)^{\frac{1-\phi(t)}{2}}Y^i(1/t,z/t)B^b(qz/t)^{\frac{1-\phi(t)}{2}}t^{\frac1{4}\boldsymbol{\sigma}_{\mathbf{3}}}.
\end{equation}

This allows us to identify \(\sigma^i(t)\) with one of the two parameters of the solution \(\mathsf{g}^i(t)\). The other parameter is \(S^i(t)\):
\begin{equation}
\mathsf{g}^i(\sigma,S;1/t)=t^{-\frac{1+\phi(t)}{4}}\mathsf{g}(\sigma,S,t)^{\phi(t)}=\mathsf{g}(\sigma^i(t),S^i(t);1/t).
\end{equation}
This equation relates two solutions of \(q\)-Painlev\'e III\(_3\).
However, these solutions contain a discontinuous function \(\phi(t)\).
We can perform a time-dependent B\"acklund transformation
\begin{equation}
t^{\frac{\phi(t)-1}{4}}\mathsf{g}(\sigma^i(t),S^i(t);1/t)^{\phi(t)}=\mathsf{g}(\check{\sigma}(t),\check{S}(t);1/t)=t^{\frac12}\mathsf{g}(\sigma,S;t)\eqcolon\check{\mathsf{g}}(\sigma,S;1/t).
\end{equation}
Then the function \(\check{\mathsf{g}}(t)\) no longer solves the \(q\)-Painlev\'e III\(_3\) equation, but in contrast to \(\mathsf{g}^i(t)\) it becomes analytic.

The mapping between \((t,\sigma, S)\) and \((1/t,\sigma^i, S^i)\) is an involution.
Using this fact, we can show that\footnote{Another option is to study the relation \eqref{eq:dualityRelation} in more detail, but this will mostly reproduce the previous computations.}
\begin{equation}
\check{S}(t)^2=e^{2 \log \check{u}(t)\log(t/q)/\log q}\frac{
\theta\left(u\check{u}(t)\sqrt{t};q\right)
\theta\left(u^{-1}\check{u}(t)\sqrt{t};q\right)}
{\theta\left(u\check{u}(t)^{-1}\sqrt{t};q\right)
\theta\left(u^{-1}\check{u}(t)^{-1}\sqrt{t};q\right)}.
\end{equation}
This expression can be rewritten as
\begin{equation}
\check{S}^2=e^{u\partial_u \mathfrak{s}(u,\check{u};t,q)},
\end{equation}
where we used the notation 
\begin{equation}
\label{eq:sDef}
\mathfrak{s}(u,\check{u};t,q)=\left((\log u)^2 + (\log\check{u})^2\right)\log_q(t/q)
+\sum_{\epsilon,\epsilon'=\pm1}\gamma \left(u^{\epsilon}\check{u}^{\epsilon'}\sqrt{t};q\right),
\end{equation}
and
\begin{equation}
\label{eq:gammaDef}
\log\theta(z;q) = z \partial_z \gamma(z;q)
\end{equation}
is so-called elliptic dilogarithm, which we define and study later; see also \cite{Zagier2000} for a similar function.
The same formula exists for \(S\).
In this way, we have proved the following
\begin{theorem}
The solutions of the \(q\)-Painlev\'e III\(_3\) equation \eqref{eq:gT} evaluated at \(t\) and at \(1/t\) are related by
\begin{equation}
\label{eq:gduality}
g\left(\log_qu,S;t\right) = \sqrt{t} g\left(\log_q\check{u},\check{S};1/t\right).
\end{equation}
The relation between parameters is given by
\begin{equation}
\check{S}^2=e^{\check{u}\partial_{\check{u}} \mathfrak{s}(u,\check{u};t,q)},
\qquad
S^2=e^{-u\partial_{u} \mathfrak{s}(u,\check{u};t,q)}.
\end{equation}
It preserves the symplectic form
\begin{equation}
\omega=d\log S^2\wedge d\log u=d\log \check{S}^2\wedge d\log \check{u}.
\end{equation}
\end{theorem}
The generating function \(\mathfrak{s}(\sigma,\check{\sigma},t)\) of this transformation describes the difference between the Liouville forms:
\begin{equation}
\label{eq:LiouvilleDiff}
\log \check{S}^2d\log\check{u} - \log S^2d\log u = d_{\mathcal{M}} \mathfrak{s}(u,\check{u};t,q),
\end{equation}
where \(d_{\mathcal{M}}\) stands for the external differential with fixed \(t\).

\subsection{Connection constants}
\subsubsection*{Special functions}
In order to describe the function \(\gamma(z;q)\), we first introduce the elliptic gamma function, see, e.g., \cite{FELDER200044}:
\begin{equation}
\label{eq:GammaRecurrence}
\Gamma(z;p,q)=\frac{(z;p,q)_{\infty}}{(pq/z;p,q)_{\infty}}.
\end{equation}
It satisfies the recurrence relation
\begin{equation}
\Gamma(pz;p,q)=\theta(z;q)\Gamma(z;p,q)
\end{equation}
and has a representation valid for \(pq<|z|<1\)
\begin{equation}
\label{eq:GammaSeries}
\Gamma(z;p,q)=\exp \left(-\sum_{n=1}^{\infty}\frac{z^n-\left(pq/z\right)^n}{n(1-p^n)(1-q^n)} \right).
\end{equation}
It follows from this representation that the \(p\to 1\) limit of \(\Gamma\) is described by
\begin{equation}
\label{eq:gammaLim}
\Gamma(z;p,q)=e^{\frac1{p-1}\gamma(z;q)+O(1)},
\end{equation}
where
\begin{equation}
\label{eq:gammaSeries}
\gamma(z;q)=\sum_{n=1}^{\infty}\frac{z^n-\left(q/z\right)^n}{n^2(1-q^n)}.
\end{equation}
Taking the \(p\to 1\) limit of \eqref{eq:GammaRecurrence}, we obtain \eqref{eq:gammaDef}.

The elliptic gamma function satisfies the following simple identity:
\begin{equation}
\Gamma(qz;pq,q)\Gamma(z;p,qp)=\Gamma(z;p,q).
\end{equation}
Taking the \(p\to 1\) limit of this identity, we get
\begin{equation}
\label{eq:gammaGamma}
\Gamma(qz;q,q)=e^{-q\partial_q \gamma(z;q)},
\end{equation}
which can also be verified independently using \eqref{eq:GammaSeries} and \eqref{eq:gammaSeries}.

\subsubsection*{Blow-up relations}
The works \cite{nakajima2005instanton,Gottsche:2006bm} study the equivariant 5d instanton partition functions on the blow-up of \(\mathbb{C}^2\), denoted by \(\widehat{\mathcal{Z}}\), and relate them to the instanton partition functions on \(\mathbb{C}^2\):
\begin{equation}
\widehat{\mathcal{Z}}_{k,d}(\varepsilon_1,\varepsilon_2,a;t,\boldsymbol{\beta})=
\sum_{Q\in \frac{k}{2}+\mathbb{Z}}
\mathcal{Z}\left(\varepsilon_1,\varepsilon_2-\varepsilon_1,a+\varepsilon_1Q;q_1^{d-1}t,\boldsymbol{\beta}\right)
\mathcal{Z}\left(\varepsilon_2,\varepsilon_1-\varepsilon_2,a+\varepsilon_2Q;q_2^{d-1}t,\boldsymbol{\beta}\right),
\end{equation}
where
\begin{equation}
q_i=e^{\boldsymbol{\beta}\varepsilon_i}.
\end{equation}
Partition functions \(\mathcal{Z}\) can also be identified with the \(q\)-deformed conformal blocks \cite{Awata:2009ur} with central charge \(c=1+6\frac{(\varepsilon_1+\varepsilon_2)^2}{\varepsilon_1\varepsilon_2}\).
They are generalizations to arbitrary value of \(\varepsilon_1/\varepsilon_2\) of our \(\mathcal{Z}^{[0]}\) \eqref{eq:tauk}:
\begin{equation}
\mathcal{Z}(\varepsilon,-\varepsilon,a;t,\boldsymbol{\beta})
 = \mathcal{Z}^{[0]}(u;t,q),\qquad q=e^{\boldsymbol{\beta}\varepsilon}, \quad a=\sigma\varepsilon, \quad q^{\boldsymbol{\beta}a}=u.
\end{equation}

Another limit of this partition function is \(\varepsilon_2\to 0\), or the so-called Nekrasov-Shatashvili limit \cite{Braverman:2004cr,Nekrasov:2009uh}, also called the quasiclassical limit:
\begin{equation}
\mathcal{Z}(\varepsilon_1,\varepsilon_2,a;t,\boldsymbol{\beta})
=e^{\frac1{\boldsymbol{\beta}\varepsilon_2}\mathcal{F}(\varepsilon_1,a;t,\boldsymbol{\beta})}
= e^{\frac1{\boldsymbol{\beta} \varepsilon_2}\mathcal{F}^{[0]}(u;t,q)}.
\end{equation}

The blow-up relations from \cite{nakajima2005instanton} read
\begin{equation}
\mathcal{Z}_{0,1}(\varepsilon_1,\varepsilon_2,a;t,\boldsymbol{\beta})=
\mathcal{Z}(\varepsilon_1,\varepsilon_2,a;t,\boldsymbol{\beta}),
\qquad
\mathcal{Z}_{1,1}(\varepsilon_1,\varepsilon_2,a;t,\boldsymbol{\beta})=0.
\end{equation}
We can also take the \(\varepsilon_2\to 0\) limit of these equations:
\begin{equation}
1=\sum_{Q\in \mathbb{Z}} \mathcal{Z}\left(\varepsilon_1,-\varepsilon_1,a;t,\boldsymbol{\beta}\right)
e^{\frac1{\boldsymbol{\beta}\varepsilon_2}\left(\mathcal{F}(\varepsilon_1-\varepsilon_2,a+\varepsilon_2Q;t,\boldsymbol{\beta})-\mathcal{F}(\varepsilon_1,a;t,\boldsymbol{\beta})\right)+O(\varepsilon_2)},
\end{equation}
or, in the notation of the present paper,
\begin{multline}
1=\sum_{Q\in \mathbb{Z}} \mathcal{Z}^{[0]}\left(uq^Q;t,q\right)
e^{\frac1{\boldsymbol{\beta}\varepsilon_2}\left(\mathcal{F}^{[0]}(ue^{\boldsymbol{\beta}\varepsilon_2Q};t,qe^{-\boldsymbol{\beta}\varepsilon_2})-\mathcal{F}^{[0]}(u;t,q) \right) + O(\varepsilon_2)}
=\\=
\sum_{Q\in \mathbb{Z}} \mathcal{Z}^{[0]}(uq^Q;t,q)
e^{Q u\partial_u \mathcal{F}^{[0]}(u;t,q) - q\partial_q \mathcal{F}^{[0]}(u;t,q)}
=\\=
\mathcal{T}^{[0]}_0 \left(
e^{u\partial_u \mathcal{F}^{[0]}(u;t,q)},u;t,q\right) e^{-q\partial_q \mathcal{F}^{[0]}(u;t,q)}.
\end{multline}
Similar of computations can also be found in \cite{Nekrasov:2020qcq,OlegMovie}.
The same can be done for the second blow-up relation.
In this way, we see that their \(\varepsilon_2\to 0\) limits read
\begin{equation}
\label{eq:tauBlowup}
\mathcal{T}_{\frac12}^{[0]}\left(S_{\star}(u;t,q)^2,u;t,q\right)=0,
\qquad
\mathcal{T}_0^{[0]}\left(S_{\star}(u;t,q)^2,u;t,q\right)=e^{q\partial_q \mathcal{F}^{[0]}(u;t,q)},
\end{equation}
where
\begin{equation}
\log S_{\star}(u;t,q)^2=u\partial_u \mathcal{F}^{[0]}(u;t,q).
\end{equation}
This describes the so-called Malgrange divisor, a sub-manifold on which the Riemann-Hilbert problem does not have a solution.

\subsubsection*{Connection constants from blow-up relations}
We would like to find a relation between the tau functions, analogous to \eqref{eq:gduality}:
\begin{equation}
\label{eq:UpsilonDef}
\mathcal{T}_\mu^{[0]}\left(S^2,u;t,q\right)=
\Upsilon(S,\check{S},u,\check{u};t,q)\mathcal{T}_\mu^{[0]}\left(\check{S}^2,\check{u};1/t,q\right).
\end{equation}
To do this, we first use the blow-up relations \eqref{eq:tauBlowup} to show that every zero of the tau function \(\mathcal{T}^{[0]}_{\frac12}\) has two equivalent descriptions:
\begin{equation}
\log S_{\star}\left(u;t,q\right)^2=u\partial_u \mathcal{F}^{[0]}\left(u;t,q\right),
\qquad
\log S_{\star}\left(\check{u};1/t,q\right)^2=\check{u}\partial_{\check{u}} \mathcal{F}^{[0]}\left(\check{u};1/t,q\right).
\end{equation}
Now, using \eqref{eq:LiouvilleDiff}, we get
\begin{equation}
\partial_{\check{u}}\mathcal{F}^{[0]}(\check{u};1/t,q) d\check{u}-
\partial_u\mathcal{F}^{[0]}\left(u;t,q\right) du=d_{\mathcal{M}}\mathfrak{s}(u,\check{u};t,q).
\end{equation}
Since the derivatives in the l.h.s. are only monodromy derivatives, we have an equality
\begin{equation}
d_{\mathcal{M}}\left(
\mathcal{F}^{[0]}\left(\check{u};1/t,q\right) -
\mathcal{F}^{[0]}\left(u;t,q\right)
- \mathfrak{s}(u,\check{u};t,q)
\right)=0
\end{equation}
that holds on the Malgrange divisor and implies the equality
\begin{equation}
\label{eq:upsilonDef}
\mathcal{F}^{[0]}\left(u;t,q\right)=
\mathcal{F}^{[0]}\left(\check{u};1/t,q\right)
- \mathfrak{s}(u,\check{u};t,q) - \mathfrak{s}_0(t,q) =
\mathcal{F}^{[0]}\left(\check{u};1/t,q\right) + \upsilon(u,\check{u};t,q)
\end{equation}
for an unknown function \(\mathfrak{s}_0\).

The function \(\upsilon(u,\check{u};t,q)\) can be called additive connection constant for the quasiclassical conformal block since the quasiclassical limit of equality
\begin{equation}
e^{\frac1{\boldsymbol{\beta} \varepsilon_2}\mathcal{F}^{[0]}\left(u;t,q\right)}
= \int e^{\frac1{\boldsymbol{\beta} \varepsilon_2}\upsilon(u,\check{u};t,q)}
e^{\frac1{\boldsymbol{\beta} \varepsilon_2} \mathcal{F}^{[0]}\left(\check{u};1/t,q\right)} du
\end{equation}
gives precisely the desired equality between the classical \(q\)-deformed conformal blocks.

To find a connection between the tau functions, we use the blow-up relation and rewrite the equation \eqref{eq:UpsilonDef}:
\begin{equation}
\Upsilon\left(S_{\star},\check{S}_{\star},u,\check{u};t,q\right) =
e^{q\partial_q \mathcal{F}^{[0]}\left(u;t,q\right)}/
e^{q\partial_q \mathcal{F}^{[0]}\left(\check{u};1/t,q\right)}=
e^{q \partial_q \upsilon(u,\check{u};t,q)}.
\end{equation}
Using \eqref{eq:sDef}, \eqref{eq:gammaGamma}, \eqref{eq:upsilonDef}, this can be written explicitly as
\begin{equation}
\Upsilon(u,\check{u};t,q)=t^{\sigma^2+\check{\sigma}^2}
e^{-q \partial_q \mathfrak{s}_0(t,q)}
\prod_{\epsilon,\epsilon'=\pm 1}
\Gamma(qu^{\epsilon}\check{u}^{\epsilon'}\sqrt{t};q,q).
\end{equation}

\subsubsection*{Comparison with the algebraic solution}
To find the unknown function \(\mathfrak{s}_0\), we first compute the connection constant for the algebraic solution  \(\Upsilon\) \eqref{eq:TauExact}:
\begin{multline}
\Upsilon^{alg.'}(t)=\frac{\mathcal{T}_0^{\substack{alg.'\\ [0]}}(t)}
{\mathcal{T}_0^{\substack{alg.'\\ [0]}}(1/t)}
= t^{\frac{1}{8}}\Gamma(\sqrt{qt},\sqrt{q},\sqrt{q})
= t^{\frac{1}{8}}\Gamma(q^{1/2}\sqrt{t};q,q)
\Gamma(q\sqrt{t};q,q)^2
\Gamma(q^{3/2}\sqrt{t};q,q).
\end{multline}
On the other hand, one has
\begin{equation}
\Upsilon^{alg.'}=
t^{\frac{1}{8}}\Gamma(q\sqrt{t};q,q)^2
\Gamma(q^{3/2}\sqrt{t};q,q)\Gamma(q^{1/2}\sqrt{t};q,q)
e^{q \partial_q \mathfrak{s}_0(t,q)}.
\end{equation}
Comparing the two formulas, we see that
\begin{equation}
q\partial_q \mathfrak{s}_0(t,q)=0,
\end{equation}
therefore \(\mathfrak{s}_0(t,q)\) can only depend on \(t\).
To fix the remaining freedom, we need some extra arguments.
However, it does not affect the connection constant for the tau functions.

In this way, we have proved the following
\begin{theorem}
The connection constant for the tau function of the \(q\)-Painlev\'e \(A_7^{(1)'}\) equation \eqref{eq:UpsilonDef} is given by
\begin{equation}
\label{eq:UpsilonFormula}
\Upsilon(u,\check{u};t,q)=t^{\sigma^2+\check{\sigma}^2}
\prod_{\epsilon,\epsilon'=\pm 1}
\Gamma(qu^{\epsilon}\check{u}^{\epsilon'}\sqrt{t};q,q).
\end{equation}
\end{theorem}
This result looks in a sense even simpler than connection constants for the differential tau functions computed in \cite{Its:2014lga,2015arXiv150607485I,Iorgov:2013uoa,TorusConnection}.

Our proof of this theorem strongly relies on the blow-up relations, and it looks to be the shortest way.
However, we believe that it can be done in a more rigorous way using Oleg Lisovyy's idea of differentiating the Fredholm determinants with respect to monodromy data and comparing the corresponding 1-forms for two different determinants.
Such computation for the case of toric isomonodromic deformations was done in \cite{DelMonte:2022nem}.

We have also proved another
\begin{theorem}
The transformation properties of the classical conformal block are described by
\begin{equation}
\mathcal{F}^{[0]}\left(u;t,q\right)=
\operatorname{Extr}_{\check{u}}\left(\upsilon(u,\check{u};t,q)+
\mathcal{F}^{[0]}\left(\check{u};1/t,q\right)\right),
\end{equation}
where \(\operatorname{Extr}\) stands for extremum,
and the additive connection constant \(\upsilon\) is given explicitly by
\begin{equation}
\label{eq:upsilonFormula}
\upsilon(u,\check{u};t,q)=-\left((\log u)^2 + (\log\check{u})^2\right)\log_q(t/q)
-\sum_{\epsilon,\epsilon'=\pm1}\gamma \left(u^{\epsilon}\check{u}^{\epsilon'}\sqrt{t};q\right)-\mathfrak{s}_0(t).
\end{equation}
\end{theorem}

\subsection{Fusion kernels}
Once we know the transformation of the isomonodromic tau function, we can also compute the transformation of the individual conformal block:
\begin{multline}
\mathcal{Z}^{[0]}(u;t,q)=\oint_{q^{-1/2}}^{q^{1/2}} \mathcal{T}_0^{[0]}(S^2,u;t,q) \frac{d\log S^2}{2\pi i}=
\\=
\oint_{q^{-1/2}}^{q^{1/2}}\Upsilon(u,\check{u};t,q)\mathcal{T}_\mu^{[0]}\left(\check{S}^2,\check{u};1/t,q\right) \frac{d\log S^2}{2\pi i}
=\\=
-\oint_{q^{1/2}}^{q^{-1/2}}\Upsilon(u,\check{u};t,q)\sum_{Q\in \mathbb{Z}}\mathcal{T}_\mu^{[0]}\left(\check{S}^2,\check{u}q^Q;1/t,q\right) \frac{d\log S^2}{2\pi i}=
\\=
-\int_{q^{\infty}}^{q^{-\infty}}\mathcal{T}_\mu^{[0]}\left(\check{S}^2,\check{u};1/t,q\right)\Upsilon(u,\check{u};t,q) \frac{d\log S^2}{2\pi i}
=\\=
-\int_0^{\infty}\mathcal{T}_\mu^{[0]}\left(\check{S}^2,\check{u};1/t,q\right)\frac{\partial\log\check{S}^2(u,\check{u},t)}{\partial\log\check{u}} \Upsilon(u,\check{u};t,q) \frac{d\log \check{u}}{2\pi i}=
\\=
\int\mathcal{T}_\mu^{[0]}\left(\check{S}^2,\check{u};1/t,q\right) \mathfrak{S}(u,\check{u};t,q) \frac{d\log \check{u}}{2\pi i},
\end{multline}
where
\begin{equation}
\mathfrak{S}(u,\check{u};t,q) = 
-\check{u}\frac{\partial\log\check{S}^2(u,\check{u},t)}{\partial \check{u}} \Upsilon(u,\check{u};t,q)=
\frac{\partial^2 \mathfrak{s}(u,\check{u};t,q)}{\partial\log u\,\partial\log \check{u}}\Upsilon(u,\check{u};t,q)
\end{equation}
is the fusion kernel that we want to find.
The above transformations are a bit schematic and are essentially analogous to the computations in \cite{Iorgov:2013uoa}.
First, we single out the conformal block from the tau function with the help of contour integration.
Second, we transform a combination of sum and integral into an integral from \(0\) to \(\infty\).
Finally, we switch from integration in \(\check{S}\) to integration in \(\check{u}\).
In principle, the latter step requires some analysis of the analytic properties of the integrand, but we skip it and do this formally.

The expression for the second derivative is
\begin{equation}
\frac{\partial^2 \mathfrak{s}(u,\check{u};t,q)}{\partial\log u\,\partial\log \check{u}} =
\frac{\sqrt{t}}{u\check{u}}\frac{\theta(u^2;q)\theta(\check{u}^2;q)\theta(t;q)(q;q)_{\infty}^2}
{\prod_{\epsilon,\epsilon'=\pm 1}\theta(u^{\epsilon}\check{u}^{\epsilon'}\sqrt{t};q)}.
\end{equation}
Combining everything and using the recurrence property of \(\Gamma\), we can formulate the following
\begin{conjecture}
The fusion kernel for \(c=1\) \(q\)-deformed conformal blocks is given by
\begin{equation}
\label{eq:fusionFormula}
\mathfrak{S}(u,\check{u};t,q)=\frac{t^{\sigma^2+\check{\sigma}^2-1/2}}{u\check{u}}\theta(u^2;q)\theta(\check{u}^2;q)\theta(t;q)(q;q)_{\infty}^2
\prod_{\epsilon,\epsilon'=\pm 1}\Gamma(u^{\epsilon}\check{u}^{\epsilon'}\sqrt{t};q,q).
\end{equation}
\end{conjecture}

\section{Discussion}
\label{sec:discussion}
The present paper provides an example of the general \(q\)-isomonodromic Fredholm determinant and the main technical tools to study it.
However, it opens many more problems than it solves.
We list these problems and possible approaches to their solutions below.

\subsection{Generalization to other \(q\)-isomonodromic problems}
The generalizations of the Fredholm determinant to different isomonodromic problems should still be worked out.
This will clearly include passing to higher rank matrices and to the operators acting between the functions on different circles in the spirit of \cite{Gavrylenko:2016zlf,Cafasso:2017xgn}.
It would also be nice to find a simplification of the computations from Section \ref{sec:determinantVariation} to make them more easily applicable in more complicated settings.
Some parallel work related to combinatorial expansion of the Fredholm determinant for \(q\)-Painlev\'e VI is being done in \cite{qPVI:unpublished}.

Also, we need an efficient description of all possible \(q\)-isomonodromic problems that allow for a Fredholm determinant representation.
For example, the papers \cite{Bershtein:2017swf,Bershtein:2018srt} provide a conjecture that all isomonodromic systems obtained by deautonomization of the Goncharov-Kenyon dimer integrable models can be solved by the topological string partition functions.
Such expressions can definitely be re-summed into some Fredholm determinants, but the rigorous derivation of such formulas is still an open problem.
Also, some systems, even in the \(q\)-Painlev\'e family, can be more complicated than just deautonomizations of the dimer models, see, e.g., \cite{Bershtein:2024lvd}.
The question is if we can generalize the Fredholm determinant also to this case.

\subsection{Relation to the Arinkin-Borodin tau functions}
The tau functions defined in the present paper are very similar to those in \cite{2007arXiv0706.3073A,2015arXiv150606718K}.
Namely, our tau functions cannot be expressed explicitly in terms of the Painlev\'e transcendents, and moreover, their first \(q\)-difference derivatives cannot be expressed in terms of the Painlev\'e transcendents either; only the second \(q\)-difference derivatives have explicit expressions.
This is also the case in \cite{2007arXiv0706.3073A,2015arXiv150606718K}, since their tau functions, as well as their first logarithmic derivatives, are sections of non-trivial line bundles on the moduli space of the difference flat connections.

Therefore, we conjecture that our tau functions should be equal to some ratios of the \(q\)-generalizations of the tau functions from \cite{2007arXiv0706.3073A,2015arXiv150606718K}:
\begin{equation}
\label{eq:tauRatio}
\tau(t)=\frac{\hat{\tau}_{AB}}{\prod_i\hat{\tau}^{(i)}_{ref.}},
\end{equation}
where \(\hat{\tau}^{(i)}_{ref.}\) are some reference tau functions, for example, the tau functions of simpler Riemann-Hilbert problems.
An avatar of this phenomenon can be the factor \(t^{-\sigma^2}\prod_{\epsilon=\pm}(q^{1+2\epsilon\sigma};q,q)_{\infty}\) in \eqref{eq:tauExpansion}.
At least, we know that such tau functions in the differential case are interpreted as the 3-point tau functions \cite{Bertola:2022jxj}.
The expressions like \eqref{eq:tauRatio} also appeared in \cite{Cafasso:2017xgn}.

If we manage to define the Arinkin-Borodin tau function for arbitrary jump \(J(z)\) in some equivalence class, for example, in the class of jumps that differ by rational modifications, then we can formulate the following
\begin{conjecture}
The Widom determinant has an expression
\begin{equation}
\tau_W[J]=\frac{\tau_{AB}[J]\tau_{AB}[\mathbb{I}]}{\tau_{AB}[\Phi_+]\tau_{AB}[\Phi_-^{-1}]},
\end{equation}
where \(\Phi_{\pm}\) are factorizations of \(J\) \eqref{eq:Jfactorizations}.
\end{conjecture}
This conjecture can also be supported by the study of \(1\times 1\) \(q\)-isomonodromic systems.
This study reduces to the usual Szeg\H{o} formula and technically only involves the computations like at the end of \eqref{eq:SzegoSum}.

\subsection{Relation to the Riemann-Hilbert problem on a torus}
It was noticed in a different context in \cite{DelMonte:2020wty} that
\begin{enumerate}
\item  Isomonodromic deformations on a torus can be described by the Riemann-Hilbert problem of the form
\begin{equation}
\label{eq:torusRHP}
\Psi(qz) = J_{tor}(z)\Psi(z)e^{2\pi i \boldsymbol{Q}}
\end{equation}
with the jump matrix \(J_{tor}(z)\) given by the ratio of solutions of the 3-point problems given by usual hypergeometric functions.
\item Isomonodromic tau function for the problem on a torus is given by some other determinant, different from \eqref{eq:widom}:
\begin{equation}
\mathcal{T}_{tor}[\varpi,J_{tor}]=\det
\begin{pmatrix}
\mathbb{I}-\varpi^{-1}q^{-z\partial_z}\Pi_-J_{tor}^{-1} & \Pi_-J_{tor} \\
\Pi_+J_{tor}^{-1} & \mathbb{I}-\varpi q^{z\partial_z}\Pi_+J_{tor}
\end{pmatrix}.
\end{equation}
\end{enumerate}

We notice that the Riemann-Hilbert problem \eqref{eq:torusRHP} with the rational jump coincides with the linear system \eqref{eq:linear} up to multiplication of solution by \(z^{\sigma \boldsymbol{\sigma}_{\mathbf{3}}}\).
This raises the following problems:
\begin{itemize}
\item To understand the role of the tau function \(\mathcal{T}_{tor}\) with the rational jump for the \(q\)-difference problem.
\item To develop similar Fredholm determinant formalism for the \emph{elliptic difference} isomonodromic deformations \cite{Krichever:2004bb}.
In this case, analogously to the elliptic differential isomonodromic deformations, the jump should be expressed in terms of the \(q\)-hypergeometric functions.
\end{itemize}

\subsection{Free-fermionic constructions of the vertex operators}
The original motivation for the Fredholm determinant describing isomonodromic deformations came from the free fermion construction \cite{Gavrylenko:2016moe}.
This likely generalizes to the \(q\)-different setting, allowing us to construct more general \(q\)-isomonodromic vertex operators for the \(q\)-W-algebras.
Such kinds of objects are already present in \cite{Jimbo:2017ael}, but they are constructed directly from Nekrasov functions.
Instead, we believe that there is another, purely analytic definition of such vertex operators that allows us to derive Nekrasov functions independently.

\subsection{Relation to the \(q\)-difference spectral problems}
It is known \cite{Bonelli:2017gdk,Gavrylenko:2023ewx} that the spectral problems for some \(q\)-difference Schr\"odinger operators are described by the tau function \eqref{eq:tauk} with \(S=S(t)\) given by some non-trivial \(q\)-periodic function.
We need to understand the meaning of such a tau function from the \(q\)-isomonodromic system point of view.
Probably, this can give us some derivation of the quantization conditions \cite{Grassi:2014uua} from the isomonodromic problems, in the spirit of \cite{Grassi:2019coc,Bershtein:2021uts}.

\subsection{Relation to discrete \(q\)-difference Fredholm determinants}
It is known that some \(q\)-isomonodromic tau functions arise in the context of combinatorial problems \cite{2007arXiv0706.3073A,2015arXiv150606718K,Borodin2003,2019arXiv190306159D,Borodin:2002pv} as discrete gap probabilities.
There is one example of reconstruction of such gap probability-like hypergeometric determinant for the differential Painlev\'e~VI in \cite{Gavrylenko:2016zlf}, so it would be natural to expect some similar phenomenon in the \(q\)-difference case.

There is also a more general question.
Our original problems are defined on the discrete sets \(\{z_0q^n\}\) and \(\{t_0 q^n\}\), although the definition of the Fredholm determinant includes the space of analytic functions, and discrete indices appear only in the Fourier space.
However, since the Fourier and the coordinate space look similar in the \(q\)-difference setting, it would be natural to expect some other dual determinant defined on a discrete set.

We saw an example of simplification of the matrix Fredholm determinant to the scalar one for some specific solution of the differential Painlev\'e VI equation \cite[Section 4.1]{Gavrylenko:2016zlf}.
In that case, the contour was deformed to the contour around the branch cut.
Since, in our case, the analog of the branch cut is the \(q\)-lattice, like \(tq^n, n>0\), the integral around it should become the sum of residues.
In this way, we can formulate the following
\begin{conjecture}
The determinant \eqref{eq:matrixFredholm} can be written as a determinant on the \(q\)-lattice
\begin{equation}
\mathcal{D}=\mathcal{D}^+_t\sqcup \mathcal{D}^-_1=
\{tq^n| n\in \mathbb{Z}_{>0}\} \sqcup \{q^n|n\in \mathbb{Z}_{<0}\}.
\end{equation}
\end{conjecture}
We are almost sure that this happens for the solution analogous to \cite[Section 4.1]{Gavrylenko:2016zlf}, and moreover, in that case 
\(\mathcal{D}=\mathcal{D}^+_t=
\{tq^n| n\in \mathbb{Z}_{>0}\}\).
However, some additional ideas might be needed to prove it in the general case.

It would also be interesting to understand if the path integral formalism from \cite{Tata:2022qtb} can be generalized to the \(q\)-difference case, if it gives any new determinants, and if these determinants become discrete in some situations.

\subsection{Fusion kernels and elliptic cluster algebras}
We have computed two fusion kernels: for \(c=\infty\) conformal blocks \eqref{eq:upsilonFormula}, for \(c=1\) conformal blocks \eqref{eq:fusionFormula}, and also the connection constant for the tau functions.
All these formulas have some similarities.
Let us define the following function:
\begin{equation}
\hat{\Upsilon}(u,\check{u};t,q_1,q_2)=\left(t q_1^{-1}q_2^{-1}\right)^{-\frac{(\log u)^2+(\log \check{u})^2}{\log q_1\log q_2}}
\prod_{\epsilon,\epsilon'=\pm 1}\Gamma(u^{\epsilon}\check{u}^{\epsilon'}\sqrt{t};q_1,q_2)^{-1}.
\end{equation}
Using the identity
\begin{equation}
\Gamma(t;q_1,q_2)=\frac1{\Gamma(t/q_2;q_1,q_2^{-1})},
\end{equation}
and also the identity \eqref{eq:gammaLim},
we can show that
\begin{equation}
\Upsilon(u,\check{u};t,q)=\hat{\Upsilon}(u,\check{u};t,q,q^{-1})
\end{equation}
and
\begin{equation}
\upsilon(u,\check{u};t,q)=\lim_{p\to 1}(p-1)\log \hat{\Upsilon}(u,\check{u};t,q,p).
\end{equation}
Therefore, both \(c=\infty\) and \(c=1\) functions are two different limits of the same function.

However, the formula for \(c=1\) fusion kernel is more complicated:
\begin{equation}
\mathfrak{S}(u,\check{u};t,q)=
\frac{t^{-\frac{(\log u)^2+(\log\check{u})^2}{\log q\log q^{-1}}-1/2}}{\prod_{\epsilon,\epsilon'=\pm 1}\Gamma(u^{\epsilon}\check{u}^{\epsilon'}\sqrt{t};q,q^{-1})}
\frac{\theta(u^2;q)\theta(\check{u}^2;q)\theta(t;q)(q;q)_{\infty}^2}
{u\check{u}\prod_{\epsilon,\epsilon'=\pm 1}\theta(u^{\epsilon}\check{u}^{\epsilon'}\sqrt{t};q)}.
\end{equation}
It is unclear whether it can be generalized for \(q_1q_2\neq 1\), or the situation in the general case is more complicated, and the corresponding fusion kernel is given by some Ponsot-Teschner integral \cite{Ponsot:1999uf,Eberhardt:2023mrq}.
It is possible that the correct way to find such an integral is to guess the difference relations satisfied by this kernel in the spirit of \cite{Roussillon:2020lyc,Nemkov:2016qxf}.

Independently from this, we see that symmetry transformations of monodromy data are described by the generating functions containing elliptic dilogarithms.
In the differential case, such transformations were described by the usual dilogarithms, and in many cases, they could be given by cluster mutations, e.g., \cite{2017JPhA...50y5202L}.
Therefore, we suggest the following
\begin{conjecture}
There should exist an elliptic generalization of cluster mutations, some special transformations with elliptic dilogarithms \eqref{eq:gammaSeries} playing the role of generating functions.
The algebra of functions connected by such transformations can be called elliptic cluster algebra.
Monodromy manifolds of the \(q\)-isomonodromic systems, like in \cite{AFST2020629511190,2023arXiv230104083R,joshi2024segresurfacesgeometrypainleve,Joshi:2022kud}, should have some elliptic cluster structure, and their automorphisms should be described by elliptic mutations.
\end{conjecture}

Cluster coordinates on monodromy manifolds usually appear in exact WKB descriptions of monodromies.
In this sense, it is interesting if computations from \cite{DelMonte:2024dcr} can produce any elliptic mutations for the \(q\)-difference case.

Another place where cluster algebras appear is the \(q\)-Painlev\'e evolutions themselves \cite{Bershtein:2024lvd}.
For example, equation \eqref{eq:Painleve} can be obtained as the composition of two mutations and permutations.
In these cases, cluster algebras are more complicated than for the monodromy manifolds.
In particular, their automorphism groups contain affine Weyl groups \cite{Bershtein:2017swf}.
Abelian subgroups of these groups generate \(q\)-isomonodromic flows.
We also state the following
\begin{conjecture}
Elliptic isomonodromic deformation flows \cite{Krichever:2004bb,Nijhoff} can be obtained as combinations of elliptic mutations in the corresponding elliptic cluster algebras.
\end{conjecture}

\printbibliography




\end{document}